\documentclass[aps,prx,longbibliography,notitlepage,reprint,superscriptaddress]{revtex4-2}

\usepackage[utf8]{inputenc}
\usepackage{amsmath}

\usepackage[colorlinks=true, allcolors=blue]{hyperref} %hyperrefs
\usepackage{graphicx,bm,siunitx,cleveref,url}

\begin{document}

\title{Spiderweb nanomechanical resonators via Bayesian optimization:  \\ inspired by nature and guided by machine learning}\thanks{This work was published in \href{https://doi.org/10.1002/adma.202106248}{Adv. Mater. 2106248} (2021).}

\author{Dongil Shin}\thanks{These authors contributed equally to this work.}
\affiliation{Delft University of Technology, Faculty of Mechanical, Maritime and Materials Engineering, Department of Materials Science and Engineering}
\affiliation{Delft University of Technology, Faculty of Mechanical, Maritime and Materials Engineering, Department of Precision and Microsystems Engineering}

\author{Andrea Cupertino}\thanks{These authors contributed equally to this work.}
\affiliation{Delft University of Technology, Faculty of Mechanical, Maritime and Materials Engineering, Department of Precision and Microsystems Engineering}

\author{Matthijs H. J. de Jong}
\affiliation{Delft University of Technology, Faculty of Mechanical, Maritime and Materials Engineering, Department of Precision and Microsystems Engineering}
\affiliation{Delft Unversity of Technology, Faculty of Applied Sciences, Department of Quantum Nanoscience, Kavli Institute of Nanoscience}

\author{Peter G. Steeneken}
\affiliation{Delft University of Technology, Faculty of Mechanical, Maritime and Materials Engineering, Department of Precision and Microsystems Engineering}
\affiliation{Delft Unversity of Technology, Faculty of Applied Sciences, Department of Quantum Nanoscience, Kavli Institute of Nanoscience}

\author{Miguel A. Bessa}
\email[]{M.A.Bessa@tudelft.nl}
\affiliation{Delft University of Technology, Faculty of Mechanical, Maritime and Materials Engineering, Department of Materials Science and Engineering}

\author{Richard A. Norte}
\email[]{R.A.Norte@tudelft.nl}
\affiliation{Delft University of Technology, Faculty of Mechanical, Maritime and Materials Engineering, Department of Precision and Microsystems Engineering}
\affiliation{Delft Unversity of Technology, Faculty of Applied Sciences, Department of Quantum Nanoscience, Kavli Institute of Nanoscience}

% Abstract should be written in the present tense and impersonal style (i.e., avoid we), and be at most 200 words long
\begin{abstract}

From ultra-sensitive detectors of fundamental forces to quantum networks and sensors, mechanical resonators are enabling next-generation technologies to operate in room temperature environments. Currently, silicon nitride nanoresonators stand as a leading microchip platform in these advances by allowing for mechanical resonators whose motion is remarkably isolated from ambient thermal noise. However, to date, human intuition has remained the driving force behind design processes. Here, inspired by nature and guided by machine learning, a spiderweb nanomechanical resonator is developed that exhibits vibration modes which are isolated from ambient thermal environments via a novel "torsional soft-clamping" mechanism discovered by the data-driven optimization algorithm. This bio-inspired resonator is then fabricated; experimentally confirming a new paradigm in mechanics with quality factors above 1 billion in room temperature environments. In contrast to other state-of-the-art resonators, this milestone is achieved with a compact design which does not require sub-micron lithographic features or complex phononic bandgaps, making it significantly easier and cheaper to manufacture at large scales. Here we demonstrate the ability of machine learning to work in tandem with human intuition to augment creative possibilities and uncover new strategies in computing and nanotechnology.
\end{abstract}

\maketitle

Major advances in nanotechnology have allowed mechanical resonators to improve dramatically over the last decades. One of the most sought after characteristics for a mechanical resonator is noise isolation from thermal environments, namely at room temperature conditions where thermomechanical noise can dominate. The degree of mechanical isolation is characterized by a resonator's mechanical quality factor, $Q_m$. Typically $Q_m$ is defined as the ratio of energy stored in a resonator over the energy dissipated over one cycle of oscillation. Inversely, mechanical quality factors can indicate the dissipation of mechanical noise into a resonator from ambient environments. For mechanical sensors, a resonator's isolation from ambient thermal noise can greatly enhance their ability to detect ultra-small forces, pressures, positions, masses, velocities, and accelerations. For quantum technologies, mechanical quality factor dictates the average number of coherent oscillations a nanomechanical resonator (in the quantum regime) can undergo before one phonon of thermal noise enters the resonator and causes decoherence of its quantum properties \cite{marquardt2007quantum}. From microchip sensing to quantum networks, cryogenics are conventionally required to counteract thermal noise but enabling these burgeoning technologies to operate in ambient temperatures would have a significant impact on their widespread use.

In room temperature environments, on-chip mechanical resonators with state-of-the-art quality factors have mostly consisted of high-aspect-ratio suspended nanostructures fabricated from tensile thin-films. Silicon nitride (Si\textsubscript{3}N\textsubscript{4}) films have been the material of choice for their high intrinsic stress, yield strength, temperature stability, chemical inertness and prevalence in nanotechnology. Over the years, researchers have developed improved design principles that manipulate the strain, bending and mode shape in nanomechanical resonators to improve quality factors, which are ultimately limited by bending losses as the resonator oscillates in vacuum. Another important characteristic is the mechanical frequency of a nanomechanical resonator's vibrational mode, $f_m$. For high-precision detectors of fundamental forces like gravity and dark matter \cite{halg2021membrane,metcalfe2014applications} and quantum-limited commercial sensors \cite{page2021gravitational, safavi2011proposal}, resonators with high $Q_m$ and low $f_m$ are a long-standing goal. Minimizing $f_m/Q_m$ is a key figure-of-merit towards quantum-limited force \cite{reinhardt2016ultralow} or acceleration \cite{krause2012high} sensitivities ($S_f,\, S_a \propto (f_m/Q_m)^{0.5}$) and for enabling quantum sensing of forces like dark matter \cite{carney2021mechanical,manley2021searching} and gravity \cite{schmole2016micromechanical, miao2020quantum} where low frequency and higher quality factor are advantageous. For phonon-based quantum technologies, a mechanical resonator's vibrational modes are initialized into the quantum regime, where their motion harbors less than one quanta of vibration (phonon) \cite{chan2011laser,Delic892}. Mechanical resonators in these quantum regimes must have sufficiently high $Q_m\times f_m > k_B T_{room}/h$ to suppress the effects of room temperature, $T_{room}$, thermal noise on their fragile quantum properties. While there are only a handful of platforms \cite{ghadimi2017radiation,ghadimi2018elastic,beccari2021hierarchical,hoj2021ultra,chakram2014dissipation,guo2019feedback,norte2016mechanical,tsaturyan2017ultracoherent} to overcome these stringent requirements on $Q_m\times f_m$ at room temperature, a general goal has been to achieve the highest $Q_m$ and lowest $f_m$ possible while still maintaining $Q_m \times f_m$ above $6 \times 10^{12}$~Hz. 

\begin{figure*}[t] % If you want the figure along the entire page width use \begin{figure*}
    \centering
	\includegraphics[width = 0.92\textwidth]{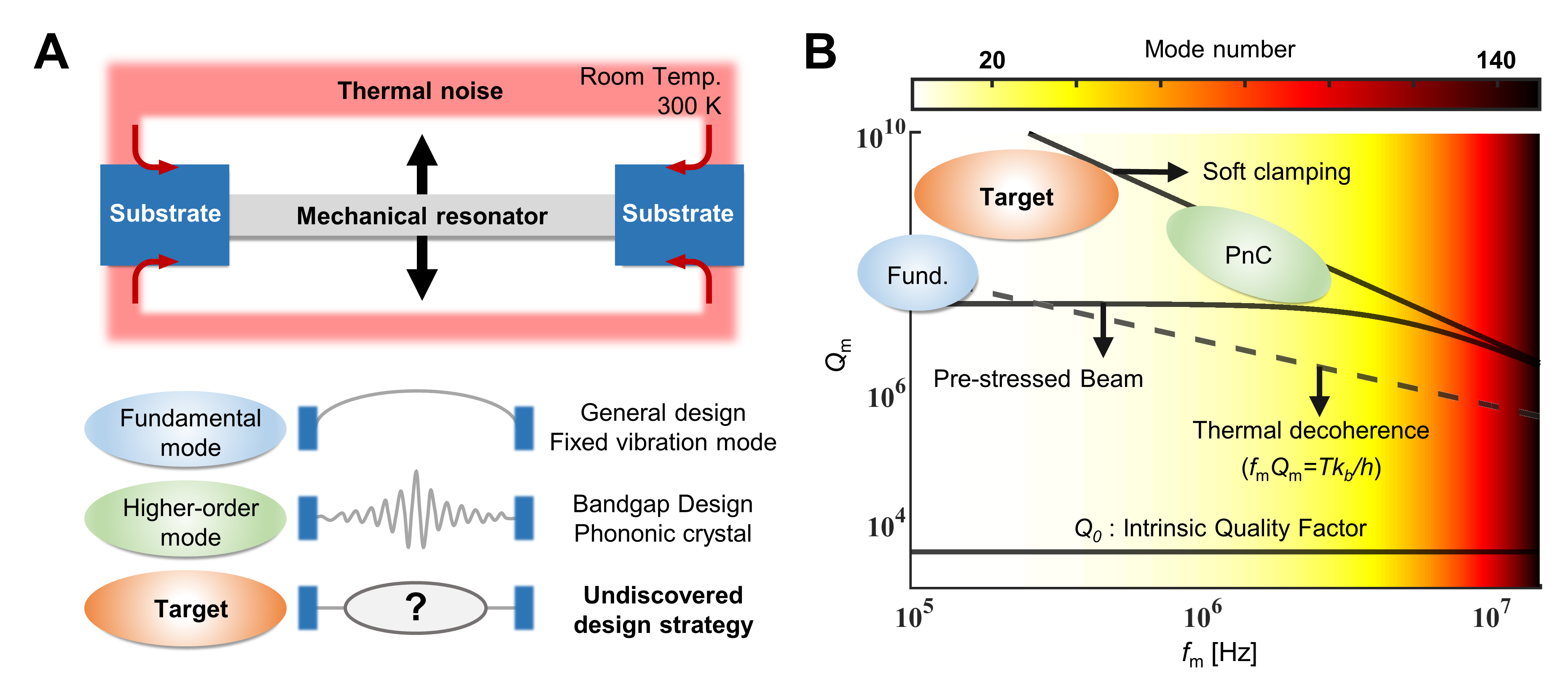} 
	\caption{\textbf{A} Illustration of the mechanical resonator and target vibration modes for a high quality factor resonator. Unlike fundamental modes \cite{norte2016mechanical,fedorov2020fractal}, or higher-order modes \cite{ghadimi2018elastic}, the target mode shape for lower-order mode hadn't been discovered. \textbf{B} Quality factor ($Q_m$) versus frequency ($f_m$) for a double clamped $50$~nm thick $3$~mm long silicon nitride beam. The bottom solid line corresponds to the intrinsic quality factor of Si\textsubscript{3}N\textsubscript{4}, the intermediate solid line to the effect of high pre-stress, and the upper solid line to the effect of complete elimination of clamp losses (perfect soft clamping) on the string resonator. The dashed line corresponds to the mechanical decoherence constraint, indicating that the resonator having the quality factor above can complete one full coherent oscillation without a thermal phonon entering the resonator. The figure highlights the target unexplored region of designs with high quality factors and lower-order modes.}
	\label{fig_1}
\end{figure*}

Previously proposed nanomechanical resonators follow  strategies largely motivated by one-dimensional analytical models of resonators \cite{schmid2011damping} because they provide easy-to-interpret design rules. It is important to note that while silicon nitride has been conventionally used for high-Q resonators, these design principles are valid for nearly any strained thin-film material. These analytical models show that higher strain (i.e. higher stress $\sigma$ and low Young's modulus $E$), longer ($L$) and thinner ($t$) geometries generally lead to higher quality factors in both nanomechanical membranes and strings. For example, $f_m/Q_m$ of the double clamped beam's fundamental mode is proportional to $\sqrt{E}/L^3$ $(L+1.4t\sqrt{E/\sigma})$, when assuming a thin long pre-stressed beam (explicit formulation can be found in the Supporting Information). While increasing the aspect ratio of resonators usually leads to smaller $f_m/Q_m$, it also makes them much more challenging to fabricate reliably. \textbf{Figure \ref{fig_1}}\textbf{A} illustrates conventional design strategies. When considering the fundamental mode of resonators, mechanical quality factors are typically improved by pre-stressing -- a form a strain engineering called dissipation dilution \cite{verbridge2006high,schmid2008damping,zwickl2008high,fedorov2019generalized} which increases stored energies and lowers dissipation compared to unstressed resonators. It overcomes the fundamental limit of the material's intrinsic damping from its bulk and surface, enabling a higher quality factor by orders of magnitude. Simultaneously, resonator models \cite{schmid2011damping} also explain that for high-aspect ratio pre-stressed resonators ($t<<L$) the dominating loss that decreases $Q_m$ is due to the mode's sharp curvature at the clamped boundary between the oscillating element and the substrate (on which the resonators are fabricated). This observation motivated the use of phononic crystals (i.e. phononic bandgap) which confine a higher-order mode from the clamping regions using a periodic pattern around the mode. Now rather than having large curvature near the edges, the phononic crystals enable a soft-clamping \cite{ghadimi2018elastic,tsaturyan2017ultracoherent,guo2019feedback} to reduce the mode's curvature close to the rigid clamp, and thereby eliminating this dissipation mechanism when operating at high order vibration modes as depicted in the schematic in \textbf{Figure \ref{fig_1}}\textbf{A}. Phononic crystals enable higher quality factors that approach hundreds of millions, at the cost of operating at higher frequencies and typically requiring higher aspect-ratio resonators which are more difficult to manufacture reliably. 

% \vspace{5mm}

To illustrate the design space, \textbf{Figure \ref{fig_1}}\textbf{B} shows the mechanical quality factor $Q_m$ versus vibration frequency $f_m$ for a $50$~nm thick and $3$~mm long Si\textsubscript{3}N\textsubscript{4} beam. The color scale indicates the various out-of-plane vibration modes, from fundamental to higher-order modes with shorter wavelengths. The Supporting Information ("Lessons from string resonators") provides additional information about the bounds shown in the figure. The considered pre-stressed double clamped beam shows the improvement of $Q_m$ caused by enhancement of stored energy (intermediate solid line) when compared to the unstressed beam (horizontal solid line at the bottom, $Q_0$) and the improvement caused if perfect soft clamping is achieved around the region shown in \textbf{Figure \ref{fig_1}}\textbf{A} (top solid line). These lines help illustrate the design region in the graph's top left corner that remains largely unexplored in current resonators that aim mainly for high quality factor. Previous works have focused on increasing $Q_m$ of the fundamental mode with strain engineering \cite{ norte2016mechanical,reinhardt2016ultralow}, including design strategies as topology optimization \cite{gao2020systematic, hoj2021ultra} or hierarchical designs \cite{fedorov2020fractal,beccari2021hierarchical}. Here we pursue high quality factors at lower frequencies by following a new approach inspired by nature and guided by machine learning.

% \vspace{5mm}

Spiderweb designs have unique geometries that make them one of the most well-known and fascinating classes of micro-mechanical structures found in nature. Despite their ubiquitous presence, experts from physics, materials science, and biology are still uncovering the elusive mechanics of spiderwebs that enable them to be remarkably robust vibration sensors \cite{cranford2012nonlinear,zaera2014uncovering}. Spider silk threads have high toughness and stiffness, reaching yield strengths on the order of a gigapascal -- about 5 times higher than steel \cite{jyoti2019structural} and about the same as Si\textsubscript{3}N\textsubscript{4}. They are used to create lightweight fibrous web structures which harbor an extraordinary strength-to-weight ratio rarely observed among other structures found in nature or science \cite{vollrath1992spider,gosline1999mechanical,boutry2009biomechanical,meyer2014compliant}.  Furthermore, in the case of spiders that sense their prey via the webs, these structures are designed to be most sensitive to vibrations emanating from the web and not from surrounding environmental vibrational disturbances \cite{masters1984vibrations,du2011structural,barrows1915reactions,mortimer2015unpicking}. Since their unique sensing capabilities have been relentlessly optimized over millions of years of complex evolutionary competition \cite{gao2003materials,aizenberg2005skeleton,kamat2000structural}, spiderwebs stand as a promising starting point for machine learning algorithms to design nanomechanical sensors \cite{miniaci2016spider,krushynska2017spider,liu2018spider}.

% \vspace{5mm}

\begin{figure}[t] % If you want the figure along the entire page width use \begin{figure*}
    \centering
	\includegraphics[width = 0.45\textwidth]{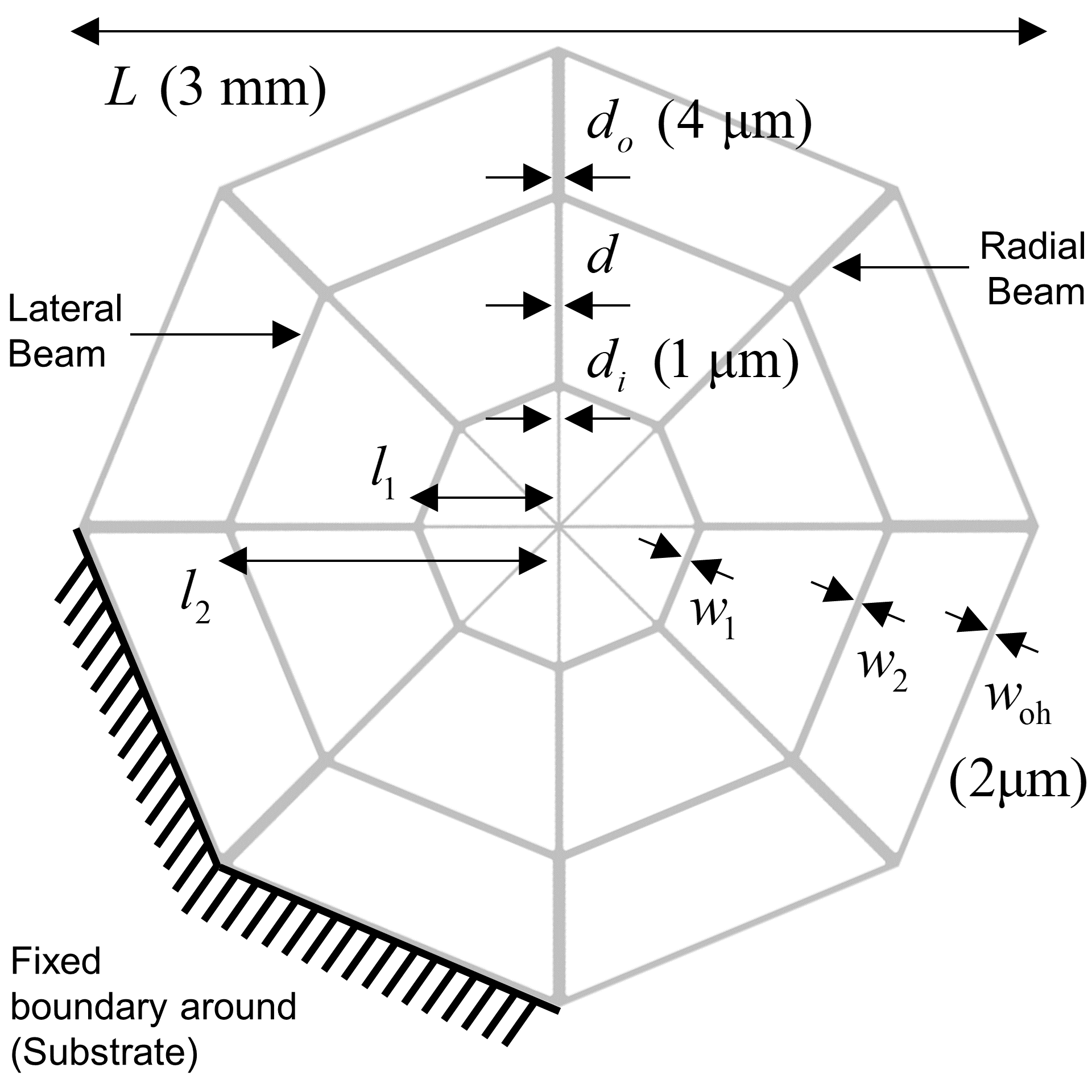} 
	\caption{Spiderweb nanomechanical resonator model with $N_r=8$. $d$, $w_1$, $w_2$, $l_1$, and $l_2$ are the design parameters, $d_i$, $d_o$, $w_{oh}$, and $L$ are set as 1 \textmu m, 4 \textmu m, 2 \textmu m, and 3 mm, respectively.}
	\label{fig_2}
\end{figure}

% \vspace{5mm}

Without making any assumptions about how a spiderweb functions as a vibrational sensor in nature, we propose a web-like structure composed of radial beams, lateral beams, and junctions between them, as shown in \textbf{Figure \ref{fig_2}}. Instead of spiderweb threads which are microns thick, we consider highly-stressed Si\textsubscript{3}N\textsubscript{4} that can be as thin as 20~nm while being suspended over several millimeters. The properties of Si\textsubscript{3}N\textsubscript{4} were considered to be $E$=250 GPa, $\nu$=0.23, $\rho$=3100 kg/m$^3$ with an initial released stress of 1.07 GPa based on measurements. The parameterized model shown in \textbf{Figure \ref{fig_2}} includes six design parameters, $d$, $w_1$, $w_2$, $l_1$, $l_2$ and $N_r$. The two inner rings (i.e. the rings formed by the lateral beams) were constrained to have at least a distance of 8 \textmu m between them. We also considered even numbers between 4 to 16 for the number of lateral beams per ring, $N_r$. Note that the even number of $N_r$ was considered to include not only the symmetric but also the anti-symmetric periodic boundary condition. The beam width at the outer and inner parts of the radial beams ($d_o$ and $d_i$) were taken to be 4 \textmu m and 1 \textmu m, respectively. The width of the structure at the resonator-substrate interface, $w_{oh}$, was set to 2 \textmu m (half the maximum beam width) in the finite element model to reflect the inevitable overhang originating from the fabrication process. The fixed boundary condition around the resonator is modeled to reflect the overhang attached to a fixed substrate. Additionally, we gave a 1 \textmu m radius fillet for every corner at the junction of lateral and radial beams. By limiting the model's features such as tether widths and fillets to micron scales (rather than sub-micron), it ensures that these structures can ultimately be defined using photolithography which allow for significantly easier, large-scale fabrication. Finally, the simulation model in the paper considers $L$, $t$ as 3 mm and 50 nm, respectively. From the simulation, we estimated the mechanical quality factor of the resonator by calculating the dissipation dilution \cite{fedorov2019generalized,yu2012control} of the out-of-plane vibration modes. The quality factors are calculated as, 

% \vspace{5mm}

\begin{equation} \label{eq_plate_quality_factor_1}
\frac{Q_m}{Q_0}=\frac{12(1-\nu^2)}{E t^2}\frac{\int \alpha dS}{\int \beta dS}
\end{equation}

\noindent with $\alpha$ and $\beta$ being defined as,

\begin{equation} \label{eq_plate_quality_factor_2}
\alpha= \sigma_{xx} u_{z,x}^2 + \sigma_{yy} u_{z,y}^2 + 2\sigma_{xy} u_{z,x}u_{z,y}  
\end{equation}

\setlength{\abovedisplayskip}{0pt} \setlength{\abovedisplayshortskip}{0pt}

\begin{equation} \label{eq_plate_quality_factor_3}
\beta= u_{z,xx}^2+u_{z,yy}^2+2\nu u_{z,xx}u_{z,yy}+2(1-\nu)u_{z,xy}^2 
\end{equation}

% \vspace{5mm}

\noindent with $u_z$ being the out-of-plane displacement during vibration, and $\sigma$ the stress distribution resulting from static analysis for the initial stress. The comma denotes a partial derivative with respect to that coordinate. $Q_0$ is the intrinsic quality factor defined as $Q_0^{-1}=Q_\mathrm{volume}^{-1}+Q_\mathrm{surface}^{-1}$, where $Q_\mathrm{volume}$ is the bulk material loss of Si\textsubscript{3}N\textsubscript{4}, and $Q_\mathrm{surface}$ is the surface loss that varies linearly with the resonator's thickness. For thin resonators in room temperature, we assume $Q_0$ $\approx$ 6900 $t$/100 nm \cite{villanueva2014evidence}. Note that $\alpha$ is proportional to the elastic energy in tension, which corresponds to the energy stored in the resonator, and $\beta$ is proportional to the bending loss. The detailed derivation is provided in the Supporting Information ("Derivation of the quality factor for two-dimensional structures"). With the parameterized model, we aimed to find the highest quality factor considering general types of mode shape below 1 MHz, which is the tenth mode frequency of the same size of the pre-stressed string in \textbf{Figure \ref{fig_1}}\textbf{B}. Given that the length $L$ of our spiderweb nanomechanical resonators are limited to 3~mm, this ensures a target frequency in the hundreds of kHz regime.

% \vspace{5mm}

The choice of optimization algorithm to guide the data-driven design process represents the most crucial part for solving real application problems and depends on the characteristics of the problem and data availability. For example, recently machine learning algorithms have proved their success in material design problems with abundant data \cite{gu2018bioinspired,carrasquilla2017machine,jiao2020evolutionary}. On the contrary in this case, a new resonator is designed based on a new spiderweb model shown in \textbf{Figure \ref{fig_2}} and therefore, no prior data is available. Trial-and-error experimentation is difficult because conducting a single experiment of a particular design takes several days of fabrication and testing. In addition, fast analytical predictions of the quality factors and vibrational mode frequencies of designs are also not possible due to the complexity of the two-dimensional geometry, which also requires to consider various vibration mode shapes. In fact, finite element analyses of these structures are computationally expensive; taking between $10$ and $30$ minutes using 20 CPU cores of our high performance computing cluster. Despite only considering periodic boundary conditions and simulating only a fraction of the total structure, these long simulation times arise from the fine mesh of elements required by the high aspect ratio structure such that the subtle shape curvatures are captured in small but crucial regions, such as joints. The geometry is meshed with 4 to 6 elements in the beam's width direction using shell elements, and the mesh resolution near the joints is about double that (detail of the finite element model information can be found in Supporting Information). Therefore, a week of computation can only generate data corresponding to less than 1000 design iterations. This is then classified as a data-scarce optimization problem where each new design iteration should be as informative to the design goal as possible. Under these conditions, using data-scarce machine learning to guide the optimization process is particularly effective, as achieved by the Bayesian Optimization method \cite{ding2018human,shalloo2020automation,shields2021bayesian}.

% \vspace{5mm}

Bayesian optimization \cite{pelikan1999boa} constructs a machine learning regression model usually from Gaussian processes \cite{bessa2019bayesian}, by predicting model uncertainty and seeking the optimum solution in fewer iterations than competing algorithms \cite{Frazier2016,shahriari2015taking}. Applying the algorithm to new problem domains, which requires new kinds of surrogate models without pre-domain knowledge as in our problem, is especially beneficial \cite{shahriari2015taking}. For readers unfamiliar with the topic, the Supporting Information includes a short introduction to the method. In the context of designing the spiderweb nanomechanical resonator, Bayesian optimization is expected to not only explore the design space to find new vibrational modes that induce soft clamping with a compact design, but also use them to reach high quality factors in the low frequency regime for a given resonator size. In this work we used the GPyOpt python implementation of the method \cite{gpyopt2016}, and MATLAB for the pre- and post-processing of our spider-web design. The finite element analysis was performed by COMSOL \cite{comsol}.

% \vspace{5mm}

\begin{figure*}[ht] % If you want the figure along the entire page width use \begin{figure*}
    \centering
	\includegraphics[width = 0.92\textwidth]{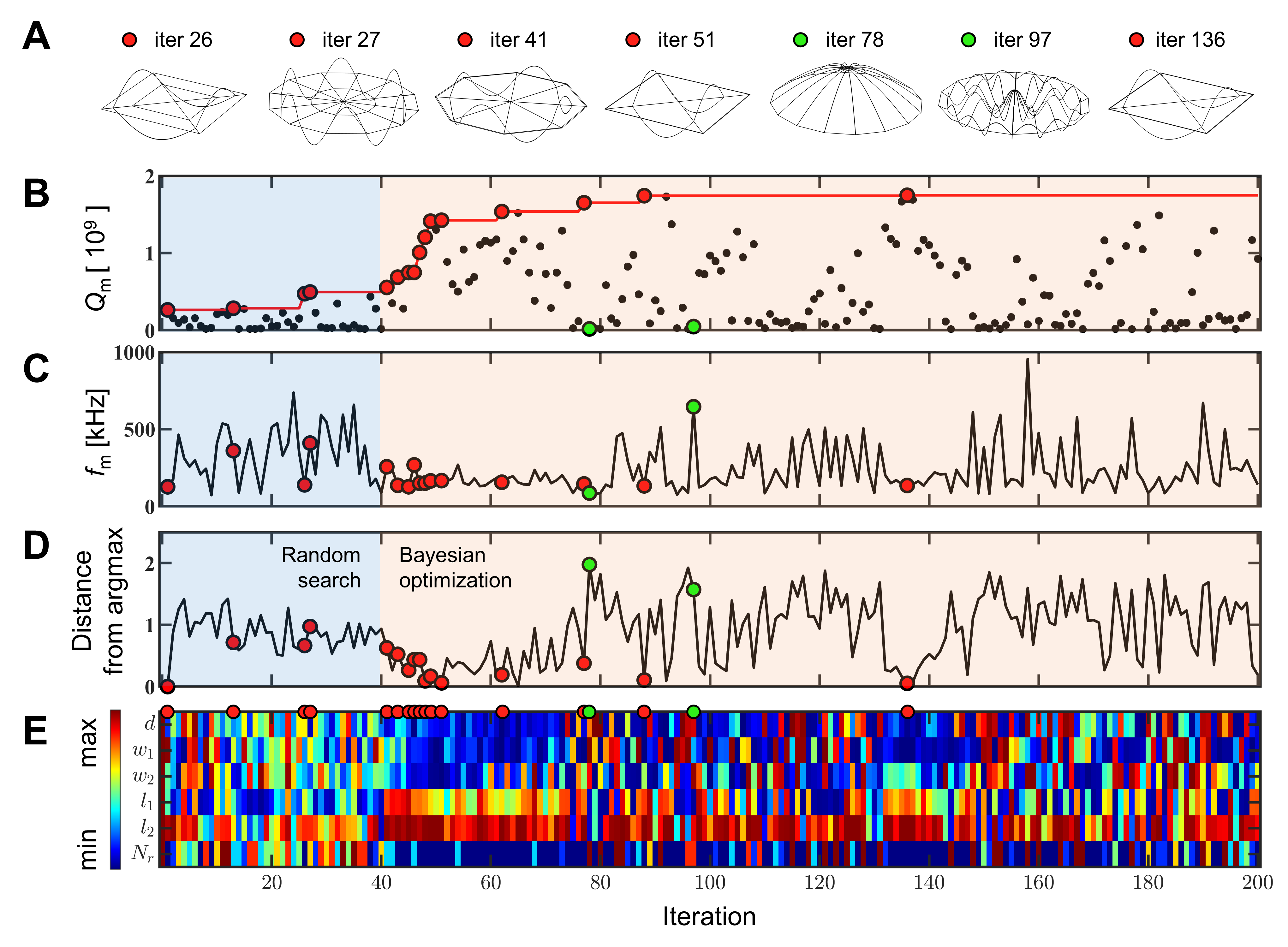} 
	\caption{Overview of the Bayesian optimization process for designing the spiderweb nanomechanical resonator: \textbf{A} designs and simulated mode shapes at the corresponding iterations highlighted by circular markers in the remaining figures (the enlarged figures are in the Supporting Information); \textbf{B} evolution of the quality factor $Q_m$ (the red markers and red line indicate the highest quality factor until that iteration); \textbf{C} the frequency $f_m$; \textbf{D} the distance from a previous optimized point to the point considered in that iteration; and \textbf{E} values of the 6 design parameters at every iteration, providing an idea of how the designs change in the optimization process (4 ${\leq}$ {$N_r$} ${\leq}$ 16, 1 {\textmu}m ${\leq}$ ({$d$}, {$w_1$}, {$w_2$}) ${\leq}$ 4 {\textmu}m, 0 mm ${<}$ {$l_1$} ${<}$ {$l_2$} ${<}$ 1.5 mm). The abscissa for \textbf{B}--\textbf{E} is the same and corresponds to the design iterations as the optimization process evolves. The blue region in \textbf{B}--\textbf{D} corresponds to the 40 initial designs that were randomly selected, i.e. before starting the Bayesian optimization process; while the red region corresponds to the Bayesian optimization iterations.}
	\label{fig_3}
\end{figure*}

% \vspace{5mm}

\textbf{Figure \ref{fig_3}} refers to the optimization history wherein the spiderweb nanomechanical resonator's quality factor is maximized. The process starts with a random search of 40 iterations to train the model (shown in light blue), followed by the Bayesian optimization phase (shown in light red). \textbf{Figure \ref{fig_3}}\textbf{B} shows the evolution of the quality factor $Q_m$, while \textbf{Figure \ref{fig_3}}\textbf{D} plots the distance from a previous optimized point to the point considered in that iteration (the distance between the normalized input vectors). As seen in \textbf{Figure \ref{fig_3}}\textbf{A}, the random search from iterations 1-40 find vibrational modes with the highest quality factors in iteration 26 and 27 which vibrate in the outer lateral ring. When the Bayesian optimization starts at iteration 41 (red markers), it begins to follow the vibrational modes concentrated on the inner lateral beams. Even though iteration 41 starts with a quality factor similar to the one found in the random search (light blue region), the algorithm continues \textit{exploiting} the optimal design region without going too far from this successful iteration, as can be seen in \textbf{Figure \ref{fig_3}}\textbf{D} between iterations 41 and 51, mostly by finely tuning design parameters as can be seen in \textbf{Figure \ref{fig_3}}\textbf{E}. In these iterations, the design improves the quality factor by more than 180\% compared to the best value obtained from the first 40 random searches and the machine learning model promotes local optimization.

% \vspace{5mm}

Note that the optimal modes found during the Bayesian optimization process correspond to vibrational modes (iterations 41, 51, 136 in \textbf{Figure \ref{fig_3}}\textbf{A}) which harbor only a slight deformation near the clamping points because the major bending elements are in the intermediate ring. Surprisingly, these lateral vibrational modes mimic actual vibrations utilized in spider webs for prey detection \cite{kawano2019detecting}. Without encoding any prior knowledge about how spiderwebs function, the machine learning algorithm was able to find how actual spiderwebs work in nature and adapt it to silicon nitride nanostructures. After iteration 51, the algorithm starts \textit{exploring} the design space more to search for a better design far from the previous optimal, which can be seen from the high values in \textbf{Figure \ref{fig_3}}\textbf{D}. This trade-off between exploration and exploitation is often responsible for the competitive advantage of Bayesian optimization when compared to other algorithms. The green markers in \textbf{Figure \ref{fig_3}} are clear examples of the Bayesian optimization \textit{exploring} far from previous optima. Note that the lateral beam's vibrational mode did not always have the highest quality factor, as optimum performance arises from the discovery of this new mode in combination with geometric parameters that promote bending and torsion in a particular way, as discussed later. By \textit{exploring} and \textit{exploiting} simultaneously to optimize the quality factor of the mechanical resonator, the algorithm reaches a maximum quality factor in iteration 136. Bayesian optimization consistently balances \textit{exploitation} and \textit{exploration}, so the solution for a large number of iterations could lead to continuous improvement. The result here considers up to 200 iterations for the optimization, considering a few days for optimization. In the case of having thousands of data sets regardless of the computational cost, the consideration of additional design parameters could be interesting for future work. As seen in \textbf{Figure \ref{fig_3}}\textbf{C}, we also found that the highest quality factors $Q_m$ occur at lower frequencies, which agrees with the simplified one-dimensional model of the beam resonator. Note that the convergence speed and the optimum result could depend on the initial searching points. The comparison study can be found in the Supporting Information ("Optimization convergence dependency on the initial random points"). Moreover, the optimization process also shows that thinner structures are not necessarily better when using micro-wide tethers -- a counter-intuitive finding given that every previous design of nanomechanical  resonators had out-of-plane mechanical modes with $Q_m$ that benefited from thinner geometries. The detailed discussion can be found in the Supporting Information ("Conversion of energy loss regarding the thickness of the spiderweb nanomechanical resonator"). 

% \vspace{5mm}

\begin{table}[b]
\centering
\caption{Optimal design parameters for the spiderweb resonator, when $d_i$, $d_o$, $w_{oh}$, and $L$ are set as 1 \textmu m, 4 \textmu m, 2 \textmu m, and 3 mm, respectively.}
  \label{tab_design_para}
  \begin{tabular}{c c c c c c c}
  \hline
 $N_r$ & $d$ & $w_1$ & $w_2$ & $l_1$ & $l_2$\\
  \hline
    4 & 1.05 \textmu m & 1 \textmu m & 2.46 \textmu m & 1.21 mm & 1.48 mm\\
  \hline
 \end{tabular}
\end{table}

% \vspace{5mm}

The optimum spiderweb nanomechanical resonator is predicted to have a quality factor $Q_m$ above 1.75 billion at 134.9kHz for a design considering a diagonal size of 3 mm and a thickness of 50 nm. \textbf{Table \ref{tab_design_para}} shows the design parameters corresponding to this design. Comparatively, what is striking about this design is that it is able to achieve a quality factor $Q_m$ above a billion without requiring any tether widths under a micron. This allows it to be readily defined using large-scale photolithography which further makes manufacturing faster and cheaper. This is extremely beneficial for real applications in that it decreases the resonators' microchip footprint. For thermal management, lower aspect ratio is very beneficial as typically Si\textsubscript{3}N\textsubscript{4} nanoresonators are interfaced with optics for high-precision sensing and quantum application. Although Si\textsubscript{3}N\textsubscript{4} is preferred for its low optical absorption \cite{steinlechner2017optical,zwickl2008high}, even small amounts of optical heating can have deleterious consequences for high precision experiments, and having a smaller aspect ratio allows for enhanced thermal conduction (to the substrate) which scales as $t/L$.  What is important to note is that previous nanomechanical resonators followed a common design paradigm wherein the maximal amplitude of the mode is in the center of the resonator and where the aim is to reduce bending losses from that center to the substrate. Here the algorithm takes an entirely different route by looking at modes that oscillate laterally like a mechanical whispering gallery mode, allowing for tiny distances between amplitude maxima and substrate, providing new insight into the nanomechanical resonator design.

% \vspace{5mm}

\begin{figure*}[t] % If you want the figure along the entire page width use \begin{figure*}
    \centering
	\includegraphics[width = 0.92\textwidth]{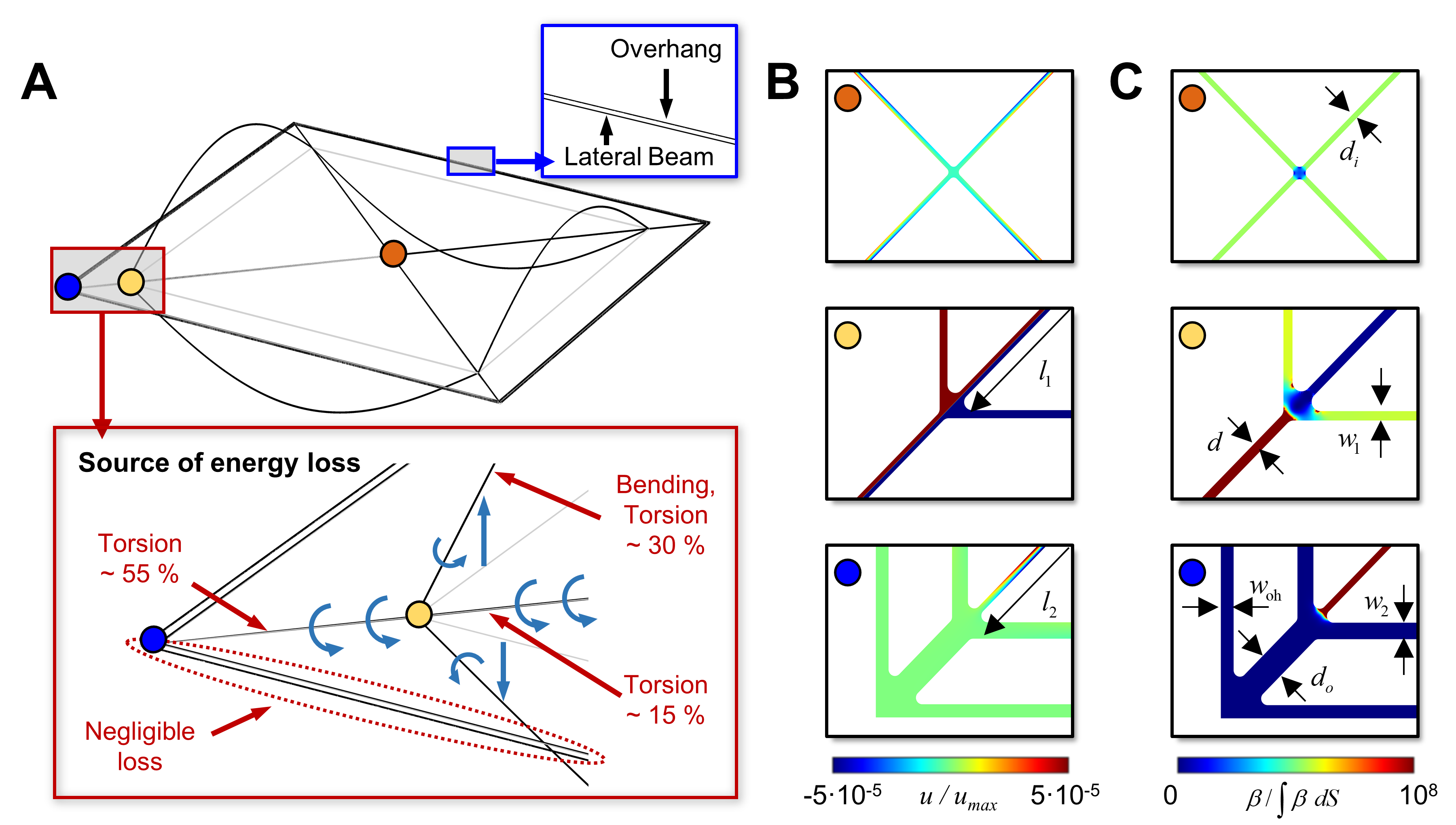} 
	\caption{Optimized spiderweb design exhibiting a soft clamping mode. \textbf{A} The full motion of the optimized vibration mode shape with a zoom around the overhang and an illustration of the portion of energy loss. The gray part in this figure shows the structure at rest (no vibration). \textbf{B} Shows the local deformations and \textbf{C} the normalized bending loss density for the three regions in \textbf{A}. From top to bottom: center (orange marker), joint of the inner lateral beam (yellow marker), and the edge (blue marker).}
	\label{fig_4}
\end{figure*}

% \vspace{5mm}

\textbf{Figure \ref{fig_4}} provides additional details about the optimum design, where \textbf{Figure \ref{fig_4}}\textbf{A} highlights the novel "torsional soft clamping" mechanism found by the data-driven strategy. This design yields an unprecedentedly high quality factor because the resonator vibrates with an out-of-plane deformation that is localized in the inner ring of lateral beams while undergoing torsional deformation of the radial beams. \textbf{Figure \ref{fig_4}}\textbf{B} and \textbf{C} also support this observation, where the displacement magnitude (\textbf{B}) and normalized bending loss (\textbf{C}) clearly demonstrate the low displacement and bending losses at the boundary (blue marker) while the joint of the inner lateral ring (yellow marker) undergoes significant deformation. As a consequence of the radial beams' torsional motion, the curvature between the bending lateral beam with the radial beam at the clamping point is not highly concentrated, thus significantly diminishing the clamping losses at these points. Although the torsional motion of the radial beams leads to 70~$\%$ of the energy dissipation, it is comparable to the bending losses in the deforming lateral beams. Note that the normalized bending loss in \textbf{Figure \ref{fig_4}}\textbf{C} indicates that the bending energy near the joint of the inner lateral beam (yellow marker region) is spread out in the region near the joint, which avoids sharp curvatures that can ultimately limit $Q_m$. Additionally, the blue marker region of \textbf{Figure \ref{fig_4}}\textbf{A} shows that the outer lateral beams near the boundary prevent deformation and bending loss from propagating toward the boundary in a subtle way by preventing torsional deformation in the radial beams from propagating to the boundary. Unlike the bending loss density of the string's bending modes \cite{schmid2011damping} (where it is highly concentrated on the clamping region), the torsional bending loss density in the spiderweb resonator does not highly concentrate where the vibration stops. For this reason, simulations without the outer lateral beams also gave a similar quality factor. Nonetheless, the optimized position of the outer ring was used to block the torsion propagation from the inner ring to the chip, enhancing the resonator's isolation from the substrate.

The optimized $N_r$ is 4, which maximizes the side beam length when all other parameters are the same. This trend shows that the optimized mode aimed to make the vibrating lateral beams as long as possible, as the quality factor of a string resonator \cite{fedorov2019generalized} increases for a longer beam. Compared with the one-dimensional approach \cite{ghadimi2017radiation, ghadimi2018elastic}, which is a subset of our structure by considering $N_r$ = 2, the web-like structure in the two-dimensional domain has the potential to achieve higher $Q_m$ by exploring novel soft clamping motions compared to the limited number of vibration modes \cite{beccari2021hierarchical} of the one-dimensional structure. The optimized $l_2$, which defines the distance between the centre and the inner radial beam, was close to the maximum limit, while the number of radial beams $N_r$ was minimized. This trend shows that the optimized mode aimed to make the vibrating lateral beams as long as possible, as the quality factor of a string resonator \cite{fedorov2019generalized} increases for a longer beam. Also, the $w_2$ was optimized at 1 \textmu m which represents the minimum limit allowed. Allowing for widths thinner than 1 micron would likely increase the quality factor. While the lateral size, $L$, was limited to 3~mm, a study of increasing $L$ (discussed in the Supporting Information) also shows that the optimized $Q_m$ scales quadratically with the size of the resonator while $f_m/Q_m$ is significantly reduced. All of these mechanisms work together to achieve a compact design with an ultra-high quality factor at a lower order vibration mode via a novel soft clamping approach that does not require the use of phononic crystals or sub-micron lithographic features. Notwithstanding, up to now this finding remains a computational prediction.

% \vspace{5mm}

\begin{figure*}[t]
  \centering
    \includegraphics[width=1\textwidth]{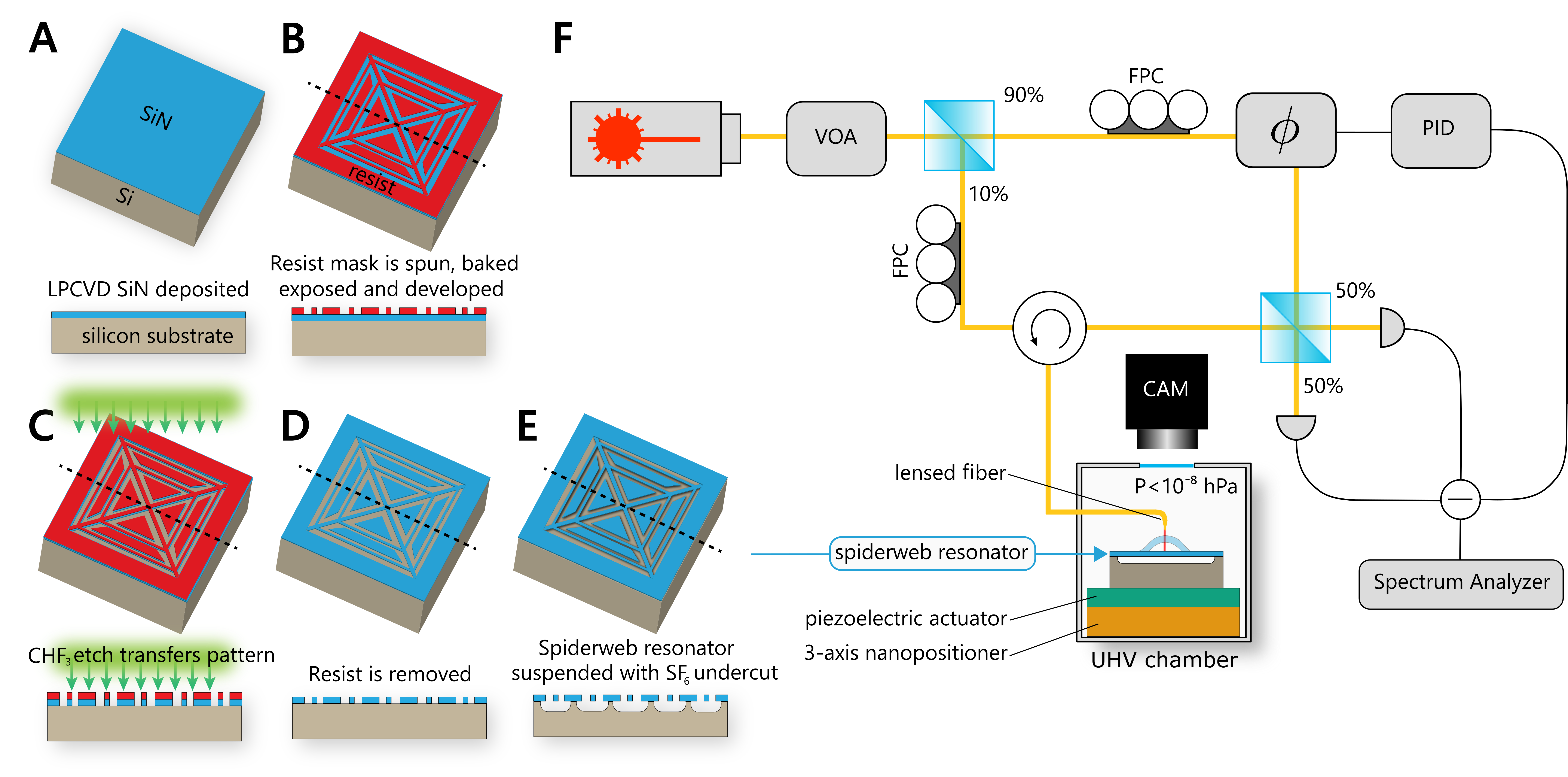}
  \caption{Schematic representation of the fabrication process flow and the mechanical characterization setup. Process steps consist of \textbf{A} deposition of Si\textsubscript{3}N\textsubscript{4} onto a silicon substrate \textbf{B} mask patterning \textbf{C} Si\textsubscript{3}N\textsubscript{4} patterning via dry etching \textbf{D} mask removing and \textbf{E} Si\textsubscript{3}N\textsubscript{4} undercut and release. \textbf{F} The spiderweb nanomechanical resonator is resonantly driven by a piezoelectric actuator and its motion is optical measured with a balanced homodyne interferometer. The resonator is placed inside an UHV chamber to reach a pressure lower than $10^{-8}$ mbar. VOA: Variable Optical Attenuator. PID: Proportional Integral Derivative controller. FPC: Fiber Polarization Controller. $\Phi$: fiber stretcher.}
  \label{fig_5}
\end{figure*}

\begin{figure*}[t] % If you want the figure along the entire page width use \begin{figure*}
    \centering
	\includegraphics[width = 0.92\textwidth]{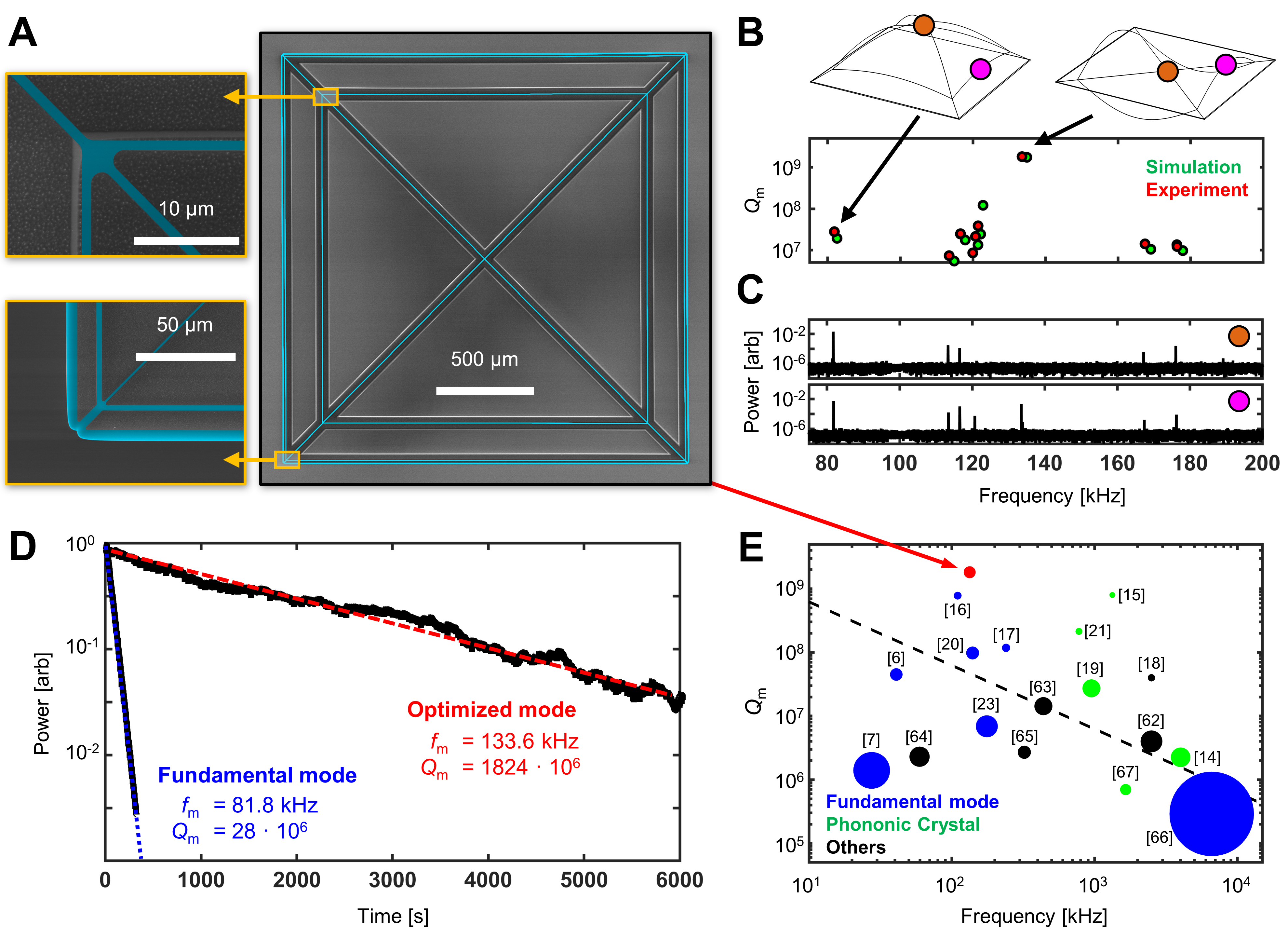} 
	\caption{Experimental characterization of the optimal spiderweb nanomechanical resonator. \textbf{A} False colored scanning electron microscope images of the optimal design. \textbf{B} $Q_m$ from the simulation and experimental results of the out-of-plane vibration modes with the figure of the fundamental and optimized mode shapes. \textbf{C} The thermomechanical noise spectra measured at the center of the resonator (orange marker) and at the center of the inner lateral beam (pink marker). The y axis are the normalized power with respect to the maximum power. \textbf{D} Ringdown measurement of the optimized 3~mm spiderweb resonator excited in its 133.6~kHz mode with an extracted quality factor higher than 1.8 billion, compared with the quality factor measured for the fundamental mode at 81.8~kHz. The y axis is the normalized power with respect to the power at time=0 for each curve. \textbf{E} Comparison of the presented experiment result with state-of-the-art experiment reports \cite{reinhardt2016ultralow,krause2012high,ghadimi2017radiation,ghadimi2018elastic,beccari2021hierarchical,hoj2021ultra,chakram2014dissipation,guo2019feedback,norte2016mechanical,tsaturyan2017ultracoherent,schmid2011damping,wilson2009cavity,serra2018silicon,usami2012optical,cole2014tensile,faust2012microwave, reetz2019analysis}. Marker area corresponds to their aspect-ratio $t/L$. The dashed line corresponds to the mechanical decoherence constraint in \textbf{Figure \ref{fig_1}}}.
	\label{fig_6}
\end{figure*}

While we have so far described the computational design process which occurred without experimental trial-and-error, the performance of the novel resonator needs to be experimentally validated. To do this we fabricate the optimal resonator and experimentally determine the quality factor of the system from a ringdown measurement in the high vacuum setup shown in \textbf{Figure \ref{fig_5}}\textbf{F} to avoid air damping. Ringdown measurements involve using piezoelectric stages to resonantly excite the motion of nanomechanical resonators, stopping the drive, and observing decay of the resonators' motion via interferometric optical readout. The rate of decay of the resonators amplitude gives its rate of energy dissipation and thus its mechanical quality factor. The spiderweb nanomechanical resonator is fabricated on high-stress Si\textsubscript{3}N\textsubscript{4} grown by low-pressure chemical vapor deposition on a silicon wafer (\textbf{Figure \ref{fig_5}}\textbf{A}). The pattern is first written on a resist mask  (\textbf{Figure \ref{fig_5}}\textbf{B}), then transferred on the Si\textsubscript{3}N\textsubscript{4} layer with a directional CHF$_3$ plasma etching (\textbf{Figure \ref{fig_5}}\textbf{C}). After that, the resist mask is removed (\textbf{Figure \ref{fig_5}}\textbf{D}) and the spiderweb nanomechanical resonator is released by a fluorine-based (SF$_6$) dry etching step (\textbf{Figure \ref{fig_5}}\textbf{E}), which does not require any mask or additional cleaning steps, making the fabrication considerably easier, higher-yield and higher-quality. Crucially, this fabrication process allows for remarkable agreement between experimental results and idealized simulations that allow us to reliably use the latter as data-points for the machine learning algorithm. A detailed explanation of the fabrication process and the mechanical characterization setup is summarized in the Methods Section.
\textbf{Figure \ref{fig_6}}\textbf{A} shows a scanning electron microscope image of the fabricated device, where the suspended Si\textsubscript{3}N\textsubscript{4} spiderweb nanomechanical resonator is highlighted in blue, surrounded by dummy Si\textsubscript{3}N\textsubscript{4} islands disconnected from the suspended structure, used to prevent overetching and overexposure (in light gray). The thermomechanical noise spectra as obtained by interferometric optical readout at the center of the spiderweb nanomechanical resonator (orange marker) and the inner lateral beam (pink marker) are plotted in \textbf{Figure \ref{fig_6}}\textbf{C}. As shown in the figure, the optimal vibration mode occurring around 133.6 kHz is visible in the spectrum of the inner ring but has no amplitude in the center of the resonator and therefore well confined, contrary to the fundamental mode at 81.8 kHz. This corroborates the presence of the novel soft clamping mode shown in \textbf{Figure \ref{fig_4}}. Furthermore the simulated out-of-plane vibration mode frequencies agrees with experiments to around 1\% as shown in \textbf{Figure \ref{fig_6}}\textbf{B}. The presence of the torsional soft clamping mode is further proved by the ringdown measurement shown in \textbf{Figure \ref{fig_6}}\textbf{D}, where the quality factor of the optimized mode at 133.6 kHz is measured to be $1.8$ billion, which is in excellent agreement with the computational predictions (1.75 billion) and the highest mechanical quality factor yet measured in this frequency range at room temperature. Compared to the fundamental mode's $Q_m$, it has more than 60 times higher value. In this research, we simulated the $Q_m$ considering the dissipation dilution but the acoustic radiation ($Q_{rad}$) \cite{norte2016mechanical,beccari2021hierarchical} and the loss from gas damping ($Q_{gas}$) \cite{schmid2008damping} can also affect the ringdown in the experiment. The excellent match between bending loss simulations and experimental results supports our hypothesis that bending loss is the dominant source of mechanical loss for the spiderweb nanomechanical resonator. While acoustic radiation loss (through the substrate) affects mechanical resonators with motion near the resonator-substrate boundary \cite{norte2016mechanical,hoj2021ultra,beccari2021hierarchical}, it is expected to be negligible in our optimal design because the resonator motion is isolated from the substrate, with the support of the outer lateral beams, as described in \textbf{Figure \ref{fig_4}}. At the same time, the gas damping effect can be ignored by performing the measurement under a sufficiently high vacuum of $4.0 \cdot 10^{-9}$ hPa. The high $Q_m$ result is especially striking when considering the short length and larger thickness of the resonator than existing solutions in the literature, making it more practical to fabricate and operate. \textbf{Figure \ref{fig_6}}\textbf{E} compares our result with the state-of-the-art nanomechanical resonator's experiment values at room temperature by plotting their $Q_m$, $f_m$ and their aspect-ratio $t/L$ via marker area when $t$ and $L$ represent the thickness and the size of each reported resonator, respectively. It supports that our spiderweb resonator has obtained a high $Q_m$ not using more challenge fabrication but considering novel vibration mode. 

\vspace{5mm}

In conclusion, a simulation based data-driven optimization approach was used to design a spiderweb nanomechanical resonator with ultralow dissipation in room temperature environments. Our approach relies on the observation that spiderwebs have evolved over millions of years through evolutionary competition to be remarkable vibration sensors \cite{cranford2012nonlinear,zaera2014uncovering}. Using silicon nitride as a base material, our machine learning algorithm hitchhikes on this natural optimization, and discovers nanomechanical designs tailored for high precision sensors. While silicon nitride is one of the most widely used thin-films for nanomechanical resonators, the design approach in this work could be extended to other materials such as diamond \cite{tensilediamond97}, gallium arsenide \cite{liu2011high, usami2012optical}, silicon carbide \cite{kermany2014microresonators, romero2020engineering}, indium gallium phosphide \cite{cole2014tensile, buckle2018stress}, fused silica glass \cite{cumming2020lowest}, silicon \cite{beccari2021strained}, phosphorus carbide \cite{Tan2017FewLayerBP,Kistanov2020PointDI}, and even superconducting films \cite{read2001tensile,nahar2017stress}. The enhancement of mechanical quality factor results from the discovery of a soft-clamping mechanism that uses a torsional motion to isolate a nanomechanical mode from ambient thermal noise. This enables high-$Q_m$ nanomechanical resonators that have smaller aspect ratios than previous state-of-the-art designs, making them significantly easier, cheaper and faster to manufacture. Our experimental validation demonstrates a new class of mechanical resonators that exhibit mechanical quality factor exceeding a billion in room temperature environments. This is achieved via a torsional soft clamping mechanism that avoids radiation losses without using phononic crystals or sub-micron lithographic features. While other state-of-the-art resonators require tethers which are hundreds of nanometers in width, our resonators (with micron-sized features) can be reliably fabricated at large scales with photolithography. While high-$Q_m$ resonators typically require $\sim 20-30$nm thicknesses, we design ours with $50$~nm thickness to simplify the fabrication. Undoubtedly, by designing our resonators tethers with sub-micron tethers and thinner geometries, we could further improve $Q_m$ at the cost of making the fabrication less accessible for general use. The low dissipation rates of the resonator, with $f_m/Q_m$ $\approx$ 75 \textmu Hz, also represent an important step towards high-precision sensing applications and room temperature quantum technologies. This includes quantum-limited force microscopy \cite{halg2021membrane}, "cavity-free" cooling scheme \cite{Pluchar:20} and quantum control of motion at room temperature \cite{rossi2018measurement}. What is fascinating is that the machine learning algorithm independently hones in on torsional vibration mechanisms which are actually used by spiderwebs in nature without any knowledge of how a spiderweb functions detect prey. Notwithstanding, we recognize that this data-driven exploration guided by machine learning is just a first step towards rational design of the next-generation of nanomechanical resonators. The demonstrated approach for realizing high-$Q_m$ resonance modes is not restricted to the specific spiderweb-like design studied in this work. The design strategy might be applied to a wide range of geometries and design problems involving low-throughput simulations or experiments (the most common scenario in engineering and science). We expect future developments in machine learning and optimization together with novel fabrication techniques to lead to unprecedented nanotechnology within the next decade.

\vspace{5mm}

\section*{Methods}

\textbf{Fabrication process:} Our nanomechanical resonators are fabricated from 58 nm thick high-stress (1.07 GPa) Si\textsubscript{3}N\textsubscript{4} deposited by low pressure chemical vapor deposition (LPCVD) on a silicon substrate (\textbf{Figure \ref{fig_5}}\textbf{A}). The disposition is carried out in-house, which allows to obtain Si\textsubscript{3}N\textsubscript{4} films of arbitrary thickness in stoichiometric form (3/4 ratio of silicon to nitrogen) leading to the uniform high tensile stress in the film. The pattern in the resonator is first written in a positive tone resist (AR-P 6200) by electron beam lithography (\textbf{Figure \ref{fig_5}}\textbf{B}) to create a mask. To do so, the resist is spin-coated on top of the Si\textsubscript{3}N\textsubscript{4}, baked at 155\textdegree{} C, exposed and developed in pentylacetate. Note that we used e-beam lithography instead of optical lithography due to the high level of control and flexibility for the device geometry which is highly beneficial during the development of a new design. However the minimum features of the resonators are designed to be 1 \textmu m to ensure a easier, large-scale fabrication with optical lithography. The pattern is then transferred into the silicon nitride thin-film layer using an inductively-coupled plasma (ICP) etching based on CHF3 plasma etch (\textbf{Figure \ref{fig_5}}\textbf{C}). Next, the resist is removed with dimethylformamide followed by two cleaning steps with hot piranha solution to remove organic residues and diluted hydrofluoric acid solution to remove surface oxides (\textbf{Figure \ref{fig_5}}\textbf{D}). Last, the Si\textsubscript{3}N\textsubscript{4} layer is released from the silicon substrate with an ICP etching with SF6 at -120\textdegree{} (\textbf{Figure \ref{fig_5}}\textbf{E}) \cite{PhysRevLett.121.030405}, performed at high pressure and low DC bias to etch isotropically the silicon substrate. This last step does not require any mask given the high selectivity of the chosen chemical against silicon nitride, avoiding any additional cleaning steps and removing any limitation posed by capillary force or stiction usually encountered in isotropic wet etchings (such as KOH or TMAH). Moreover, it is not constrained by the crystal planes of the silicon substrate, enabling the fabrication of arbitrary shapes and avoiding multiple exposures. The final thickness of the Si\textsubscript{3}N\textsubscript{4} films is expected to be 50 nm.

\textbf{Mechanical characterization setup:} All the measurements presented were performed using a custom balanced homodyne detection interferometer (\textbf{Figure \ref{fig_5}}\textbf{F}). The mechanical displacement is probed with a fiber coupled infrared laser (1550 nm). The power is divided into two arms: the local oscillator (90 $\%$) used as interference reference and the signal arm (10 $\%$) terminated with a lensed fiber. The signal arm and the device are mounted on two separate 3-axis nanopositioners placed perpendicular to each other, in order to align the device to the focal plane of the lensed fiber. In this way the signal which comes out from the lensed fiber is focused on the device and its reflection collected back inside the fiber. A piezoelectric plate is connected to the sample holder to actuate the devices mechanically. To reduce the effect of gas damping on the measurements, the lensed fiber and sample stage are placed inside a vacuum chamber. With the aid of a turbomolecular and a diaphragm pump, the system can reach a pressure of $4.0$$\cdot$$10^{-9}$ hPa. The sensitivity of the measured signal to phase oscillations is maximal in the linear region of the interference signal. To this end, the phase of the local oscillator signal is controlled with a fiber stretcher driven by a proportional-integral-derivative (PID) controller implemented with a FPGA board (RedPitaya 125-14) in order to stabilize the interferometer's low-frequency fluctuations using the signal measured from the balanced photodetector as an error signal for a feedback loop.
Thermomechanical noise spectra were acquired with an electronic spectrum analyzer without mechanical excitation applied to the piezoelectric plate. On the contrary, for the ringdown measurements the device was first actuated close to the mechanical resonance frequency of interest with the piezoelectric plate until it reached and excited steady-state. Second, the mechanical actuation was turned off and the decay in time of the measured displacement signal was measured with an electronic spectrum analyzer by setting a resolution bandwidth larger than 5 Hz. The bandwidth needs to be larger than the expected linewidth of the resonator, but small enough to increase the signal-to-noise ratio.

\vspace{5mm}

% Acknowledgements
\medskip
\textbf{Acknowledgements} \par %delete if not applicable))
The research leading to these results has received funding from the European Union’s Horizon 2020 research and innovation programme under Grant Agreement Nos. 785219 and 881603 Graphene Flagship. This work has received funding from the EMPIR programme co-financed by the Participating States and from the European Union’s Horizon 2020 research and innovation programme (No. 17FUN05 PhotoQuant). This publication is part of the project, Probing the physics of exotic superconductors with microchip Casimir experiments (740.018.020) of the research programme NWO Start-up which is partly financed by the Dutch Research Council (NWO).  A.C and R. N. acknowledge valuable support from the Kavli Nanolab Delft, in particular from C. de Boer, and from the Technical Support Staff at PME 3mE Delft, in particular from Gideon Emmaneel and Patrick van Holst. A.C and R. N. would like to thank Minxing Xu and Martin Lee for stimulating discussions and early assistance with fabrication and experiments. R. N. would also like to thank Simon Groeblacher for initial support. D.S., M.A.B and R.N. would like to acknowledge the TU Delft's 3mE Faculty Cohesion grant that enabled to start this project.

\vspace{5mm}

\medskip
\textbf{Conﬂict of Interest} \par
The authors declare no conﬂict of interest.

\vspace{5mm}

\medskip
\textbf{Author Contributions} \par
D.S., M.A.B. and R.N. designed the research; D.S. and M.A.B. conducted the data-driven computational design; A.C. fabricated the spiderweb resonators with support from R.N., A.C. and R.N led the experiment with contributions from D.S. and M.J.; D.S. and A.C. analyzed the data; D.S., A.C., M.J., P.S., M.A.B., and R.N. wrote the paper. D.S. and A.C. contributed equally to this work, as well as M.A.B and R.N.

\vspace{5mm}

% References
\medskip

\bibliography{references_web_resonator}

\begin{thebibliography}{81}%
\makeatletter
\providecommand \@ifxundefined [1]{%
 \@ifx{#1\undefined}
}%
\providecommand \@ifnum [1]{%
 \ifnum #1\expandafter \@firstoftwo
 \else \expandafter \@secondoftwo
 \fi
}%
\providecommand \@ifx [1]{%
 \ifx #1\expandafter \@firstoftwo
 \else \expandafter \@secondoftwo
 \fi
}%
\providecommand \natexlab [1]{#1}%
\providecommand \enquote  [1]{``#1''}%
\providecommand \bibnamefont  [1]{#1}%
\providecommand \bibfnamefont [1]{#1}%
\providecommand \citenamefont [1]{#1}%
\providecommand \href@noop [0]{\@secondoftwo}%
\providecommand \href [0]{\begingroup \@sanitize@url \@href}%
\providecommand \@href[1]{\@@startlink{#1}\@@href}%
\providecommand \@@href[1]{\endgroup#1\@@endlink}%
\providecommand \@sanitize@url [0]{\catcode `\\12\catcode `\$12\catcode
  `\&12\catcode `\#12\catcode `\^12\catcode `\_12\catcode `\%12\relax}%
\providecommand \@@startlink[1]{}%
\providecommand \@@endlink[0]{}%
\providecommand \url  [0]{\begingroup\@sanitize@url \@url }%
\providecommand \@url [1]{\endgroup\@href {#1}{\urlprefix }}%
\providecommand \urlprefix  [0]{URL }%
\providecommand \Eprint [0]{\href }%
\providecommand \doibase [0]{https://doi.org/}%
\providecommand \selectlanguage [0]{\@gobble}%
\providecommand \bibinfo  [0]{\@secondoftwo}%
\providecommand \bibfield  [0]{\@secondoftwo}%
\providecommand \translation [1]{[#1]}%
\providecommand \BibitemOpen [0]{}%
\providecommand \bibitemStop [0]{}%
\providecommand \bibitemNoStop [0]{.\EOS\space}%
\providecommand \EOS [0]{\spacefactor3000\relax}%
\providecommand \BibitemShut  [1]{\csname bibitem#1\endcsname}%
\let\auto@bib@innerbib\@empty
%</preamble>
\bibitem [{\citenamefont {Marquardt}\ \emph {et~al.}(2007)\citenamefont
  {Marquardt}, \citenamefont {Chen}, \citenamefont {Clerk},\ and\ \citenamefont
  {Girvin}}]{marquardt2007quantum}%
  \BibitemOpen
  \bibfield  {author} {\bibinfo {author} {\bibfnamefont {F.}~\bibnamefont
  {Marquardt}}, \bibinfo {author} {\bibfnamefont {J.~P.}\ \bibnamefont {Chen}},
  \bibinfo {author} {\bibfnamefont {A.~A.}\ \bibnamefont {Clerk}},\ and\
  \bibinfo {author} {\bibfnamefont {S.}~\bibnamefont {Girvin}},\ }\bibfield
  {title} {\bibinfo {title} {Quantum theory of cavity-assisted sideband cooling
  of mechanical motion},\ }\href@noop {} {\bibfield  {journal} {\bibinfo
  {journal} {Physical Review Letters}\ }\textbf {\bibinfo {volume} {99}},\
  \bibinfo {pages} {093902} (\bibinfo {year} {2007})}\BibitemShut {NoStop}%
\bibitem [{\citenamefont {H{\"a}lg}\ \emph {et~al.}(2021)\citenamefont
  {H{\"a}lg}, \citenamefont {Gisler}, \citenamefont {Tsaturyan}, \citenamefont
  {Catalini}, \citenamefont {Grob}, \citenamefont {Krass}, \citenamefont
  {H'eritier}, \citenamefont {Mattiat}, \citenamefont {Thamm}, \citenamefont
  {Schirhagl}, \citenamefont {Langman}, \citenamefont {Schliesser},
  \citenamefont {Degen},\ and\ \citenamefont {Eichler}}]{halg2021membrane}%
  \BibitemOpen
  \bibfield  {author} {\bibinfo {author} {\bibfnamefont {D.}~\bibnamefont
  {H{\"a}lg}}, \bibinfo {author} {\bibfnamefont {T.}~\bibnamefont {Gisler}},
  \bibinfo {author} {\bibfnamefont {Y.}~\bibnamefont {Tsaturyan}}, \bibinfo
  {author} {\bibfnamefont {L.}~\bibnamefont {Catalini}}, \bibinfo {author}
  {\bibfnamefont {U.}~\bibnamefont {Grob}}, \bibinfo {author} {\bibfnamefont
  {M.}~\bibnamefont {Krass}}, \bibinfo {author} {\bibfnamefont
  {M.}~\bibnamefont {H'eritier}}, \bibinfo {author} {\bibfnamefont
  {H.}~\bibnamefont {Mattiat}}, \bibinfo {author} {\bibfnamefont
  {A.}~\bibnamefont {Thamm}}, \bibinfo {author} {\bibfnamefont
  {R.}~\bibnamefont {Schirhagl}}, \bibinfo {author} {\bibfnamefont
  {E.}~\bibnamefont {Langman}}, \bibinfo {author} {\bibfnamefont
  {A.}~\bibnamefont {Schliesser}}, \bibinfo {author} {\bibfnamefont
  {C.}~\bibnamefont {Degen}},\ and\ \bibinfo {author} {\bibfnamefont
  {A.}~\bibnamefont {Eichler}},\ }\bibfield  {title} {\bibinfo {title}
  {Membrane-based scanning force microscopy},\ }\href@noop {} {\bibfield
  {journal} {\bibinfo  {journal} {Physical Review Applied}\ }\textbf {\bibinfo
  {volume} {15}},\ \bibinfo {pages} {L021001} (\bibinfo {year}
  {2021})}\BibitemShut {NoStop}%
\bibitem [{\citenamefont {Metcalfe}(2014)}]{metcalfe2014applications}%
  \BibitemOpen
  \bibfield  {author} {\bibinfo {author} {\bibfnamefont {M.}~\bibnamefont
  {Metcalfe}},\ }\bibfield  {title} {\bibinfo {title} {Applications of cavity
  optomechanics},\ }\href@noop {} {\bibfield  {journal} {\bibinfo  {journal}
  {Applied Physics Reviews}\ }\textbf {\bibinfo {volume} {1}},\ \bibinfo
  {pages} {031105} (\bibinfo {year} {2014})}\BibitemShut {NoStop}%
\bibitem [{\citenamefont {Page}\ \emph {et~al.}(2021)\citenamefont {Page},
  \citenamefont {Goryachev}, \citenamefont {Miao}, \citenamefont {Chen},
  \citenamefont {Ma}, \citenamefont {Mason}, \citenamefont {Rossi},
  \citenamefont {Blair}, \citenamefont {Ju}, \citenamefont {Blair},
  \citenamefont {Schliesser}, \citenamefont {Tobar},\ and\ \citenamefont
  {Zhao}}]{page2021gravitational}%
  \BibitemOpen
  \bibfield  {author} {\bibinfo {author} {\bibfnamefont {M.~A.}\ \bibnamefont
  {Page}}, \bibinfo {author} {\bibfnamefont {M.}~\bibnamefont {Goryachev}},
  \bibinfo {author} {\bibfnamefont {H.}~\bibnamefont {Miao}}, \bibinfo {author}
  {\bibfnamefont {Y.}~\bibnamefont {Chen}}, \bibinfo {author} {\bibfnamefont
  {Y.}~\bibnamefont {Ma}}, \bibinfo {author} {\bibfnamefont {D.}~\bibnamefont
  {Mason}}, \bibinfo {author} {\bibfnamefont {M.}~\bibnamefont {Rossi}},
  \bibinfo {author} {\bibfnamefont {C.~D.}\ \bibnamefont {Blair}}, \bibinfo
  {author} {\bibfnamefont {L.}~\bibnamefont {Ju}}, \bibinfo {author}
  {\bibfnamefont {D.~G.}\ \bibnamefont {Blair}}, \bibinfo {author}
  {\bibfnamefont {A.}~\bibnamefont {Schliesser}}, \bibinfo {author}
  {\bibfnamefont {M.~E.}\ \bibnamefont {Tobar}},\ and\ \bibinfo {author}
  {\bibfnamefont {C.}~\bibnamefont {Zhao}},\ }\bibfield  {title} {\bibinfo
  {title} {Gravitational wave detectors with broadband high frequency
  sensitivity},\ }\href@noop {} {\bibfield  {journal} {\bibinfo  {journal}
  {Communications Physics}\ }\textbf {\bibinfo {volume} {4}},\ \bibinfo {pages}
  {1} (\bibinfo {year} {2021})}\BibitemShut {NoStop}%
\bibitem [{\citenamefont {Safavi-Naeini}\ and\ \citenamefont
  {Painter}(2011)}]{safavi2011proposal}%
  \BibitemOpen
  \bibfield  {author} {\bibinfo {author} {\bibfnamefont {A.~H.}\ \bibnamefont
  {Safavi-Naeini}}\ and\ \bibinfo {author} {\bibfnamefont {O.}~\bibnamefont
  {Painter}},\ }\bibfield  {title} {\bibinfo {title} {Proposal for an
  optomechanical traveling wave phonon--photon translator},\ }\href@noop {}
  {\bibfield  {journal} {\bibinfo  {journal} {New Journal of Physics}\ }\textbf
  {\bibinfo {volume} {13}},\ \bibinfo {pages} {013017} (\bibinfo {year}
  {2011})}\BibitemShut {NoStop}%
\bibitem [{\citenamefont {Reinhardt}\ \emph {et~al.}(2016)\citenamefont
  {Reinhardt}, \citenamefont {M{\"u}ller}, \citenamefont {Bourassa},\ and\
  \citenamefont {Sankey}}]{reinhardt2016ultralow}%
  \BibitemOpen
  \bibfield  {author} {\bibinfo {author} {\bibfnamefont {C.}~\bibnamefont
  {Reinhardt}}, \bibinfo {author} {\bibfnamefont {T.}~\bibnamefont
  {M{\"u}ller}}, \bibinfo {author} {\bibfnamefont {A.}~\bibnamefont
  {Bourassa}},\ and\ \bibinfo {author} {\bibfnamefont {J.~C.}\ \bibnamefont
  {Sankey}},\ }\bibfield  {title} {\bibinfo {title} {Ultralow-noise sin
  trampoline resonators for sensing and optomechanics},\ }\href@noop {}
  {\bibfield  {journal} {\bibinfo  {journal} {Physical Review X}\ }\textbf
  {\bibinfo {volume} {6}},\ \bibinfo {pages} {021001} (\bibinfo {year}
  {2016})}\BibitemShut {NoStop}%
\bibitem [{\citenamefont {Krause}\ \emph {et~al.}(2012)\citenamefont {Krause},
  \citenamefont {Winger}, \citenamefont {Blasius}, \citenamefont {Lin},\ and\
  \citenamefont {Painter}}]{krause2012high}%
  \BibitemOpen
  \bibfield  {author} {\bibinfo {author} {\bibfnamefont {A.~G.}\ \bibnamefont
  {Krause}}, \bibinfo {author} {\bibfnamefont {M.}~\bibnamefont {Winger}},
  \bibinfo {author} {\bibfnamefont {T.~D.}\ \bibnamefont {Blasius}}, \bibinfo
  {author} {\bibfnamefont {Q.}~\bibnamefont {Lin}},\ and\ \bibinfo {author}
  {\bibfnamefont {O.}~\bibnamefont {Painter}},\ }\bibfield  {title} {\bibinfo
  {title} {A high-resolution microchip optomechanical accelerometer},\
  }\href@noop {} {\bibfield  {journal} {\bibinfo  {journal} {Nature Photonics}\
  }\textbf {\bibinfo {volume} {6}},\ \bibinfo {pages} {768} (\bibinfo {year}
  {2012})}\BibitemShut {NoStop}%
\bibitem [{\citenamefont {Carney}\ \emph {et~al.}(2021)\citenamefont {Carney},
  \citenamefont {Krnjaic}, \citenamefont {Moore}, \citenamefont {Regal},
  \citenamefont {Afek}, \citenamefont {Bhave}, \citenamefont {Brubaker},
  \citenamefont {Corbitt}, \citenamefont {Cripe}, \citenamefont {Crisosto},
  \citenamefont {Geraci}, \citenamefont {Ghosh}, \citenamefont {Harris},
  \citenamefont {Hook}, \citenamefont {Kolb}, \citenamefont {Kunjummen},
  \citenamefont {Lang}, \citenamefont {Li}, \citenamefont {Lin}, \citenamefont
  {Liu}, \citenamefont {Lykken}, \citenamefont {Magrini}, \citenamefont
  {Manley}, \citenamefont {Matsumoto}, \citenamefont {Monte}, \citenamefont
  {Monteiro}, \citenamefont {Purdy}, \citenamefont {Riedel}, \citenamefont
  {Singh}, \citenamefont {Singh}, \citenamefont {Sinha}, \citenamefont
  {Taylor}, \citenamefont {Qin}, \citenamefont {Wilson},\ and\ \citenamefont
  {Zhao}}]{carney2021mechanical}%
  \BibitemOpen
  \bibfield  {author} {\bibinfo {author} {\bibfnamefont {D.}~\bibnamefont
  {Carney}}, \bibinfo {author} {\bibfnamefont {G.}~\bibnamefont {Krnjaic}},
  \bibinfo {author} {\bibfnamefont {D.}~\bibnamefont {Moore}}, \bibinfo
  {author} {\bibfnamefont {C.}~\bibnamefont {Regal}}, \bibinfo {author}
  {\bibfnamefont {G.}~\bibnamefont {Afek}}, \bibinfo {author} {\bibfnamefont
  {S.}~\bibnamefont {Bhave}}, \bibinfo {author} {\bibfnamefont
  {B.}~\bibnamefont {Brubaker}}, \bibinfo {author} {\bibfnamefont
  {T.}~\bibnamefont {Corbitt}}, \bibinfo {author} {\bibfnamefont
  {J.}~\bibnamefont {Cripe}}, \bibinfo {author} {\bibfnamefont
  {N.}~\bibnamefont {Crisosto}}, \bibinfo {author} {\bibfnamefont
  {A.}~\bibnamefont {Geraci}}, \bibinfo {author} {\bibfnamefont
  {S.}~\bibnamefont {Ghosh}}, \bibinfo {author} {\bibfnamefont {J.~G.~E.}\
  \bibnamefont {Harris}}, \bibinfo {author} {\bibfnamefont {A.}~\bibnamefont
  {Hook}}, \bibinfo {author} {\bibfnamefont {E.}~\bibnamefont {Kolb}}, \bibinfo
  {author} {\bibfnamefont {J.}~\bibnamefont {Kunjummen}}, \bibinfo {author}
  {\bibfnamefont {R.}~\bibnamefont {Lang}}, \bibinfo {author} {\bibfnamefont
  {T.}~\bibnamefont {Li}}, \bibinfo {author} {\bibfnamefont {T.}~\bibnamefont
  {Lin}}, \bibinfo {author} {\bibfnamefont {Z.}~\bibnamefont {Liu}}, \bibinfo
  {author} {\bibfnamefont {J.}~\bibnamefont {Lykken}}, \bibinfo {author}
  {\bibfnamefont {L.}~\bibnamefont {Magrini}}, \bibinfo {author} {\bibfnamefont
  {J.}~\bibnamefont {Manley}}, \bibinfo {author} {\bibfnamefont
  {N.}~\bibnamefont {Matsumoto}}, \bibinfo {author} {\bibfnamefont
  {A.}~\bibnamefont {Monte}}, \bibinfo {author} {\bibfnamefont
  {F.}~\bibnamefont {Monteiro}}, \bibinfo {author} {\bibfnamefont
  {T.}~\bibnamefont {Purdy}}, \bibinfo {author} {\bibfnamefont {C.~J.}\
  \bibnamefont {Riedel}}, \bibinfo {author} {\bibfnamefont {R.}~\bibnamefont
  {Singh}}, \bibinfo {author} {\bibfnamefont {S.}~\bibnamefont {Singh}},
  \bibinfo {author} {\bibfnamefont {K.}~\bibnamefont {Sinha}}, \bibinfo
  {author} {\bibfnamefont {J.~M.}\ \bibnamefont {Taylor}}, \bibinfo {author}
  {\bibfnamefont {J.}~\bibnamefont {Qin}}, \bibinfo {author} {\bibfnamefont
  {D.~J.}\ \bibnamefont {Wilson}},\ and\ \bibinfo {author} {\bibfnamefont
  {Y.}~\bibnamefont {Zhao}},\ }\bibfield  {title} {\bibinfo {title} {Mechanical
  quantum sensing in the search for dark matter},\ }\href@noop {} {\bibfield
  {journal} {\bibinfo  {journal} {Quantum Science and Technology}\ }\textbf
  {\bibinfo {volume} {6}},\ \bibinfo {pages} {024002} (\bibinfo {year}
  {2021})}\BibitemShut {NoStop}%
\bibitem [{\citenamefont {Manley}\ \emph {et~al.}(2021)\citenamefont {Manley},
  \citenamefont {Chowdhury}, \citenamefont {Grin}, \citenamefont {Singh},\ and\
  \citenamefont {Wilson}}]{manley2021searching}%
  \BibitemOpen
  \bibfield  {author} {\bibinfo {author} {\bibfnamefont {J.}~\bibnamefont
  {Manley}}, \bibinfo {author} {\bibfnamefont {M.~D.}\ \bibnamefont
  {Chowdhury}}, \bibinfo {author} {\bibfnamefont {D.}~\bibnamefont {Grin}},
  \bibinfo {author} {\bibfnamefont {S.}~\bibnamefont {Singh}},\ and\ \bibinfo
  {author} {\bibfnamefont {D.~J.}\ \bibnamefont {Wilson}},\ }\bibfield  {title}
  {\bibinfo {title} {Searching for vector dark matter with an optomechanical
  accelerometer},\ }\href@noop {} {\bibfield  {journal} {\bibinfo  {journal}
  {Physical Review Letters}\ }\textbf {\bibinfo {volume} {126}},\ \bibinfo
  {pages} {061301} (\bibinfo {year} {2021})}\BibitemShut {NoStop}%
\bibitem [{\citenamefont {Schm{\"o}le}\ \emph {et~al.}(2016)\citenamefont
  {Schm{\"o}le}, \citenamefont {Dragosits}, \citenamefont {Hepach},\ and\
  \citenamefont {Aspelmeyer}}]{schmole2016micromechanical}%
  \BibitemOpen
  \bibfield  {author} {\bibinfo {author} {\bibfnamefont {J.}~\bibnamefont
  {Schm{\"o}le}}, \bibinfo {author} {\bibfnamefont {M.}~\bibnamefont
  {Dragosits}}, \bibinfo {author} {\bibfnamefont {H.}~\bibnamefont {Hepach}},\
  and\ \bibinfo {author} {\bibfnamefont {M.}~\bibnamefont {Aspelmeyer}},\
  }\bibfield  {title} {\bibinfo {title} {A micromechanical proof-of-principle
  experiment for measuring the gravitational force of milligram masses},\
  }\href@noop {} {\bibfield  {journal} {\bibinfo  {journal} {Classical and
  Quantum Gravity}\ }\textbf {\bibinfo {volume} {33}},\ \bibinfo {pages}
  {125031} (\bibinfo {year} {2016})}\BibitemShut {NoStop}%
\bibitem [{\citenamefont {Miao}\ \emph {et~al.}(2020)\citenamefont {Miao},
  \citenamefont {Martynov}, \citenamefont {Yang},\ and\ \citenamefont
  {Datta}}]{miao2020quantum}%
  \BibitemOpen
  \bibfield  {author} {\bibinfo {author} {\bibfnamefont {H.}~\bibnamefont
  {Miao}}, \bibinfo {author} {\bibfnamefont {D.}~\bibnamefont {Martynov}},
  \bibinfo {author} {\bibfnamefont {H.}~\bibnamefont {Yang}},\ and\ \bibinfo
  {author} {\bibfnamefont {A.}~\bibnamefont {Datta}},\ }\bibfield  {title}
  {\bibinfo {title} {Quantum correlations of light mediated by gravity},\
  }\href@noop {} {\bibfield  {journal} {\bibinfo  {journal} {Physical Review
  A}\ }\textbf {\bibinfo {volume} {101}},\ \bibinfo {pages} {063804} (\bibinfo
  {year} {2020})}\BibitemShut {NoStop}%
\bibitem [{\citenamefont {Chan}\ \emph {et~al.}(2011)\citenamefont {Chan},
  \citenamefont {Alegre}, \citenamefont {Safavi-Naeini}, \citenamefont {Hill},
  \citenamefont {Krause}, \citenamefont {Gr{\"o}blacher}, \citenamefont
  {Aspelmeyer},\ and\ \citenamefont {Painter}}]{chan2011laser}%
  \BibitemOpen
  \bibfield  {author} {\bibinfo {author} {\bibfnamefont {J.}~\bibnamefont
  {Chan}}, \bibinfo {author} {\bibfnamefont {T.~M.}\ \bibnamefont {Alegre}},
  \bibinfo {author} {\bibfnamefont {A.~H.}\ \bibnamefont {Safavi-Naeini}},
  \bibinfo {author} {\bibfnamefont {J.~T.}\ \bibnamefont {Hill}}, \bibinfo
  {author} {\bibfnamefont {A.}~\bibnamefont {Krause}}, \bibinfo {author}
  {\bibfnamefont {S.}~\bibnamefont {Gr{\"o}blacher}}, \bibinfo {author}
  {\bibfnamefont {M.}~\bibnamefont {Aspelmeyer}},\ and\ \bibinfo {author}
  {\bibfnamefont {O.}~\bibnamefont {Painter}},\ }\bibfield  {title} {\bibinfo
  {title} {Laser cooling of a nanomechanical oscillator into its quantum ground
  state},\ }\href@noop {} {\bibfield  {journal} {\bibinfo  {journal} {Nature}\
  }\textbf {\bibinfo {volume} {478}},\ \bibinfo {pages} {89} (\bibinfo {year}
  {2011})}\BibitemShut {NoStop}%
\bibitem [{\citenamefont {Deli{\'c}}\ \emph {et~al.}(2020)\citenamefont
  {Deli{\'c}}, \citenamefont {Reisenbauer}, \citenamefont {Dare}, \citenamefont
  {Grass}, \citenamefont {Vuleti{\'c}}, \citenamefont {Kiesel},\ and\
  \citenamefont {Aspelmeyer}}]{Delic892}%
  \BibitemOpen
  \bibfield  {author} {\bibinfo {author} {\bibfnamefont {U.}~\bibnamefont
  {Deli{\'c}}}, \bibinfo {author} {\bibfnamefont {M.}~\bibnamefont
  {Reisenbauer}}, \bibinfo {author} {\bibfnamefont {K.}~\bibnamefont {Dare}},
  \bibinfo {author} {\bibfnamefont {D.}~\bibnamefont {Grass}}, \bibinfo
  {author} {\bibfnamefont {V.}~\bibnamefont {Vuleti{\'c}}}, \bibinfo {author}
  {\bibfnamefont {N.}~\bibnamefont {Kiesel}},\ and\ \bibinfo {author}
  {\bibfnamefont {M.}~\bibnamefont {Aspelmeyer}},\ }\bibfield  {title}
  {\bibinfo {title} {Cooling of a levitated nanoparticle to the motional
  quantum ground state},\ }\href {https://doi.org/10.1126/science.aba3993}
  {\bibfield  {journal} {\bibinfo  {journal} {Science}\ }\textbf {\bibinfo
  {volume} {367}},\ \bibinfo {pages} {892} (\bibinfo {year} {2020})},\ \Eprint
  {https://arxiv.org/abs/https://science.sciencemag.org/content/367/6480/892.full.pdf}
  {https://science.sciencemag.org/content/367/6480/892.full.pdf} \BibitemShut
  {NoStop}%
\bibitem [{\citenamefont {Ghadimi}\ \emph {et~al.}(2017)\citenamefont
  {Ghadimi}, \citenamefont {Wilson},\ and\ \citenamefont
  {Kippenberg}}]{ghadimi2017radiation}%
  \BibitemOpen
  \bibfield  {author} {\bibinfo {author} {\bibfnamefont {A.~H.}\ \bibnamefont
  {Ghadimi}}, \bibinfo {author} {\bibfnamefont {D.~J.}\ \bibnamefont
  {Wilson}},\ and\ \bibinfo {author} {\bibfnamefont {T.~J.}\ \bibnamefont
  {Kippenberg}},\ }\bibfield  {title} {\bibinfo {title} {Radiation and internal
  loss engineering of high-stress silicon nitride nanobeams},\ }\href@noop {}
  {\bibfield  {journal} {\bibinfo  {journal} {Nano letters}\ }\textbf {\bibinfo
  {volume} {17}},\ \bibinfo {pages} {3501} (\bibinfo {year}
  {2017})}\BibitemShut {NoStop}%
\bibitem [{\citenamefont {Ghadimi}\ \emph {et~al.}(2018)\citenamefont
  {Ghadimi}, \citenamefont {Fedorov}, \citenamefont {Engelsen}, \citenamefont
  {Bereyhi}, \citenamefont {Schilling}, \citenamefont {Wilson},\ and\
  \citenamefont {Kippenberg}}]{ghadimi2018elastic}%
  \BibitemOpen
  \bibfield  {author} {\bibinfo {author} {\bibfnamefont {A.~H.}\ \bibnamefont
  {Ghadimi}}, \bibinfo {author} {\bibfnamefont {S.~A.}\ \bibnamefont
  {Fedorov}}, \bibinfo {author} {\bibfnamefont {N.~J.}\ \bibnamefont
  {Engelsen}}, \bibinfo {author} {\bibfnamefont {M.~J.}\ \bibnamefont
  {Bereyhi}}, \bibinfo {author} {\bibfnamefont {R.}~\bibnamefont {Schilling}},
  \bibinfo {author} {\bibfnamefont {D.~J.}\ \bibnamefont {Wilson}},\ and\
  \bibinfo {author} {\bibfnamefont {T.~J.}\ \bibnamefont {Kippenberg}},\
  }\bibfield  {title} {\bibinfo {title} {Elastic strain engineering for
  ultralow mechanical dissipation},\ }\href@noop {} {\bibfield  {journal}
  {\bibinfo  {journal} {Science}\ }\textbf {\bibinfo {volume} {360}},\ \bibinfo
  {pages} {764} (\bibinfo {year} {2018})}\BibitemShut {NoStop}%
\bibitem [{\citenamefont {Beccari}\ \emph
  {et~al.}(2021{\natexlab{a}})\citenamefont {Beccari}, \citenamefont {Bereyhi},
  \citenamefont {Groth}, \citenamefont {Fedorov}, \citenamefont {Arabmoheghi},
  \citenamefont {Engelsen},\ and\ \citenamefont
  {Kippenberg}}]{beccari2021hierarchical}%
  \BibitemOpen
  \bibfield  {author} {\bibinfo {author} {\bibfnamefont {A.}~\bibnamefont
  {Beccari}}, \bibinfo {author} {\bibfnamefont {M.~J.}\ \bibnamefont
  {Bereyhi}}, \bibinfo {author} {\bibfnamefont {R.}~\bibnamefont {Groth}},
  \bibinfo {author} {\bibfnamefont {S.~A.}\ \bibnamefont {Fedorov}}, \bibinfo
  {author} {\bibfnamefont {A.}~\bibnamefont {Arabmoheghi}}, \bibinfo {author}
  {\bibfnamefont {N.~J.}\ \bibnamefont {Engelsen}},\ and\ \bibinfo {author}
  {\bibfnamefont {T.~J.}\ \bibnamefont {Kippenberg}},\ }\bibfield  {title}
  {\bibinfo {title} {Hierarchical tensile structures with ultralow mechanical
  dissipation},\ }\href@noop {} {\bibfield  {journal} {\bibinfo  {journal}
  {arXiv preprint arXiv:2103.09785}\ } (\bibinfo {year}
  {2021}{\natexlab{a}})}\BibitemShut {NoStop}%
\bibitem [{\citenamefont {H{\o}j}\ \emph {et~al.}()\citenamefont {H{\o}j},
  \citenamefont {Wang}, \citenamefont {Gao}, \citenamefont {Hoff},
  \citenamefont {Sigmund},\ and\ \citenamefont {Andersen}}]{hoj2021ultra}%
  \BibitemOpen
  \bibfield  {author} {\bibinfo {author} {\bibfnamefont {D.}~\bibnamefont
  {H{\o}j}}, \bibinfo {author} {\bibfnamefont {F.}~\bibnamefont {Wang}},
  \bibinfo {author} {\bibfnamefont {W.}~\bibnamefont {Gao}}, \bibinfo {author}
  {\bibfnamefont {U.~B.}\ \bibnamefont {Hoff}}, \bibinfo {author}
  {\bibfnamefont {O.}~\bibnamefont {Sigmund}},\ and\ \bibinfo {author}
  {\bibfnamefont {U.~L.}\ \bibnamefont {Andersen}},\ }\bibfield  {title}
  {\bibinfo {title} {Ultra-coherent nanomechanical resonators based on inverse
  design},\ }\href@noop {} {\bibfield  {journal} {\bibinfo  {journal} {Nature
  communications}\ }\textbf {\bibinfo {volume} {12 1}},\ \bibinfo {pages}
  {5766}}\BibitemShut {NoStop}%
\bibitem [{\citenamefont {Chakram}\ \emph {et~al.}(2014)\citenamefont
  {Chakram}, \citenamefont {Patil}, \citenamefont {Chang},\ and\ \citenamefont
  {Vengalattore}}]{chakram2014dissipation}%
  \BibitemOpen
  \bibfield  {author} {\bibinfo {author} {\bibfnamefont {S.}~\bibnamefont
  {Chakram}}, \bibinfo {author} {\bibfnamefont {Y.}~\bibnamefont {Patil}},
  \bibinfo {author} {\bibfnamefont {L.}~\bibnamefont {Chang}},\ and\ \bibinfo
  {author} {\bibfnamefont {M.}~\bibnamefont {Vengalattore}},\ }\bibfield
  {title} {\bibinfo {title} {Dissipation in ultrahigh quality factor sin
  membrane resonators},\ }\href@noop {} {\bibfield  {journal} {\bibinfo
  {journal} {Physical Review Letters}\ }\textbf {\bibinfo {volume} {112}},\
  \bibinfo {pages} {127201} (\bibinfo {year} {2014})}\BibitemShut {NoStop}%
\bibitem [{\citenamefont {Guo}\ \emph {et~al.}(2019)\citenamefont {Guo},
  \citenamefont {Norte},\ and\ \citenamefont
  {Gr{\"o}blacher}}]{guo2019feedback}%
  \BibitemOpen
  \bibfield  {author} {\bibinfo {author} {\bibfnamefont {J.}~\bibnamefont
  {Guo}}, \bibinfo {author} {\bibfnamefont {R.}~\bibnamefont {Norte}},\ and\
  \bibinfo {author} {\bibfnamefont {S.}~\bibnamefont {Gr{\"o}blacher}},\
  }\bibfield  {title} {\bibinfo {title} {Feedback cooling of a room temperature
  mechanical oscillator close to its motional ground state},\ }\href@noop {}
  {\bibfield  {journal} {\bibinfo  {journal} {Physical Review Letters}\
  }\textbf {\bibinfo {volume} {123}},\ \bibinfo {pages} {223602} (\bibinfo
  {year} {2019})}\BibitemShut {NoStop}%
\bibitem [{\citenamefont {Norte}\ \emph {et~al.}(2016)\citenamefont {Norte},
  \citenamefont {Moura},\ and\ \citenamefont
  {Gr{\"o}blacher}}]{norte2016mechanical}%
  \BibitemOpen
  \bibfield  {author} {\bibinfo {author} {\bibfnamefont {R.~A.}\ \bibnamefont
  {Norte}}, \bibinfo {author} {\bibfnamefont {J.~P.}\ \bibnamefont {Moura}},\
  and\ \bibinfo {author} {\bibfnamefont {S.}~\bibnamefont {Gr{\"o}blacher}},\
  }\bibfield  {title} {\bibinfo {title} {Mechanical resonators for quantum
  optomechanics experiments at room temperature},\ }\href@noop {} {\bibfield
  {journal} {\bibinfo  {journal} {Physical Review Letters}\ }\textbf {\bibinfo
  {volume} {116}},\ \bibinfo {pages} {147202} (\bibinfo {year}
  {2016})}\BibitemShut {NoStop}%
\bibitem [{\citenamefont {Tsaturyan}\ \emph {et~al.}(2017)\citenamefont
  {Tsaturyan}, \citenamefont {Barg}, \citenamefont {Polzik},\ and\
  \citenamefont {Schliesser}}]{tsaturyan2017ultracoherent}%
  \BibitemOpen
  \bibfield  {author} {\bibinfo {author} {\bibfnamefont {Y.}~\bibnamefont
  {Tsaturyan}}, \bibinfo {author} {\bibfnamefont {A.}~\bibnamefont {Barg}},
  \bibinfo {author} {\bibfnamefont {E.~S.}\ \bibnamefont {Polzik}},\ and\
  \bibinfo {author} {\bibfnamefont {A.}~\bibnamefont {Schliesser}},\ }\bibfield
   {title} {\bibinfo {title} {Ultracoherent nanomechanical resonators via soft
  clamping and dissipation dilution},\ }\href@noop {} {\bibfield  {journal}
  {\bibinfo  {journal} {Nature nanotechnology}\ }\textbf {\bibinfo {volume}
  {12}},\ \bibinfo {pages} {776} (\bibinfo {year} {2017})}\BibitemShut
  {NoStop}%
\bibitem [{\citenamefont {Fedorov}\ \emph {et~al.}(2020)\citenamefont
  {Fedorov}, \citenamefont {Beccari}, \citenamefont {Engelsen},\ and\
  \citenamefont {Kippenberg}}]{fedorov2020fractal}%
  \BibitemOpen
  \bibfield  {author} {\bibinfo {author} {\bibfnamefont {S.~A.}\ \bibnamefont
  {Fedorov}}, \bibinfo {author} {\bibfnamefont {A.}~\bibnamefont {Beccari}},
  \bibinfo {author} {\bibfnamefont {N.~J.}\ \bibnamefont {Engelsen}},\ and\
  \bibinfo {author} {\bibfnamefont {T.~J.}\ \bibnamefont {Kippenberg}},\
  }\bibfield  {title} {\bibinfo {title} {Fractal-like mechanical resonators
  with a soft-clamped fundamental mode},\ }\href@noop {} {\bibfield  {journal}
  {\bibinfo  {journal} {Physical Review Letters}\ }\textbf {\bibinfo {volume}
  {124}},\ \bibinfo {pages} {025502} (\bibinfo {year} {2020})}\BibitemShut
  {NoStop}%
\bibitem [{\citenamefont {Schmid}\ \emph {et~al.}(2011)\citenamefont {Schmid},
  \citenamefont {Jensen}, \citenamefont {Nielsen},\ and\ \citenamefont
  {Boisen}}]{schmid2011damping}%
  \BibitemOpen
  \bibfield  {author} {\bibinfo {author} {\bibfnamefont {S.}~\bibnamefont
  {Schmid}}, \bibinfo {author} {\bibfnamefont {K.}~\bibnamefont {Jensen}},
  \bibinfo {author} {\bibfnamefont {K.}~\bibnamefont {Nielsen}},\ and\ \bibinfo
  {author} {\bibfnamefont {A.}~\bibnamefont {Boisen}},\ }\bibfield  {title}
  {\bibinfo {title} {Damping mechanisms in high-q micro and nanomechanical
  string resonators},\ }\href@noop {} {\bibfield  {journal} {\bibinfo
  {journal} {Physical Review B}\ }\textbf {\bibinfo {volume} {84}},\ \bibinfo
  {pages} {165307} (\bibinfo {year} {2011})}\BibitemShut {NoStop}%
\bibitem [{\citenamefont {Verbridge}\ \emph {et~al.}(2006)\citenamefont
  {Verbridge}, \citenamefont {Parpia}, \citenamefont {Reichenbach},
  \citenamefont {Bellan},\ and\ \citenamefont {Craighead}}]{verbridge2006high}%
  \BibitemOpen
  \bibfield  {author} {\bibinfo {author} {\bibfnamefont {S.~S.}\ \bibnamefont
  {Verbridge}}, \bibinfo {author} {\bibfnamefont {J.~M.}\ \bibnamefont
  {Parpia}}, \bibinfo {author} {\bibfnamefont {R.~B.}\ \bibnamefont
  {Reichenbach}}, \bibinfo {author} {\bibfnamefont {L.~M.}\ \bibnamefont
  {Bellan}},\ and\ \bibinfo {author} {\bibfnamefont {H.~G.}\ \bibnamefont
  {Craighead}},\ }\bibfield  {title} {\bibinfo {title} {High quality factor
  resonance at room temperature with nanostrings under high tensile stress},\
  }\href@noop {} {\bibfield  {journal} {\bibinfo  {journal} {Journal of Applied
  Physics}\ }\textbf {\bibinfo {volume} {99}},\ \bibinfo {pages} {124304}
  (\bibinfo {year} {2006})}\BibitemShut {NoStop}%
\bibitem [{\citenamefont {Schmid}\ and\ \citenamefont
  {Hierold}(2008)}]{schmid2008damping}%
  \BibitemOpen
  \bibfield  {author} {\bibinfo {author} {\bibfnamefont {S.}~\bibnamefont
  {Schmid}}\ and\ \bibinfo {author} {\bibfnamefont {C.}~\bibnamefont
  {Hierold}},\ }\bibfield  {title} {\bibinfo {title} {Damping mechanisms of
  single-clamped and prestressed double-clamped resonant polymer microbeams},\
  }\href@noop {} {\bibfield  {journal} {\bibinfo  {journal} {Journal of Applied
  Physics}\ }\textbf {\bibinfo {volume} {104}},\ \bibinfo {pages} {093516}
  (\bibinfo {year} {2008})}\BibitemShut {NoStop}%
\bibitem [{\citenamefont {Zwickl}\ \emph {et~al.}(2008)\citenamefont {Zwickl},
  \citenamefont {Shanks}, \citenamefont {Jayich}, \citenamefont {Yang},
  \citenamefont {Bleszynski~Jayich}, \citenamefont {Thompson},\ and\
  \citenamefont {Harris}}]{zwickl2008high}%
  \BibitemOpen
  \bibfield  {author} {\bibinfo {author} {\bibfnamefont {B.}~\bibnamefont
  {Zwickl}}, \bibinfo {author} {\bibfnamefont {W.}~\bibnamefont {Shanks}},
  \bibinfo {author} {\bibfnamefont {A.}~\bibnamefont {Jayich}}, \bibinfo
  {author} {\bibfnamefont {C.}~\bibnamefont {Yang}}, \bibinfo {author}
  {\bibfnamefont {A.}~\bibnamefont {Bleszynski~Jayich}}, \bibinfo {author}
  {\bibfnamefont {J.}~\bibnamefont {Thompson}},\ and\ \bibinfo {author}
  {\bibfnamefont {J.}~\bibnamefont {Harris}},\ }\bibfield  {title} {\bibinfo
  {title} {High quality mechanical and optical properties of commercial silicon
  nitride membranes},\ }\href@noop {} {\bibfield  {journal} {\bibinfo
  {journal} {Applied Physics Letters}\ }\textbf {\bibinfo {volume} {92}},\
  \bibinfo {pages} {103125} (\bibinfo {year} {2008})}\BibitemShut {NoStop}%
\bibitem [{\citenamefont {Fedorov}\ \emph {et~al.}(2019)\citenamefont
  {Fedorov}, \citenamefont {Engelsen}, \citenamefont {Ghadimi}, \citenamefont
  {Bereyhi}, \citenamefont {Schilling}, \citenamefont {Wilson},\ and\
  \citenamefont {Kippenberg}}]{fedorov2019generalized}%
  \BibitemOpen
  \bibfield  {author} {\bibinfo {author} {\bibfnamefont {S.~A.}\ \bibnamefont
  {Fedorov}}, \bibinfo {author} {\bibfnamefont {N.~J.}\ \bibnamefont
  {Engelsen}}, \bibinfo {author} {\bibfnamefont {A.~H.}\ \bibnamefont
  {Ghadimi}}, \bibinfo {author} {\bibfnamefont {M.~J.}\ \bibnamefont
  {Bereyhi}}, \bibinfo {author} {\bibfnamefont {R.}~\bibnamefont {Schilling}},
  \bibinfo {author} {\bibfnamefont {D.~J.}\ \bibnamefont {Wilson}},\ and\
  \bibinfo {author} {\bibfnamefont {T.~J.}\ \bibnamefont {Kippenberg}},\
  }\bibfield  {title} {\bibinfo {title} {Generalized dissipation dilution in
  strained mechanical resonators},\ }\href@noop {} {\bibfield  {journal}
  {\bibinfo  {journal} {Physical Review B}\ }\textbf {\bibinfo {volume} {99}},\
  \bibinfo {pages} {054107} (\bibinfo {year} {2019})}\BibitemShut {NoStop}%
\bibitem [{\citenamefont {Gao}\ \emph {et~al.}(2020)\citenamefont {Gao},
  \citenamefont {Wang},\ and\ \citenamefont {Sigmund}}]{gao2020systematic}%
  \BibitemOpen
  \bibfield  {author} {\bibinfo {author} {\bibfnamefont {W.}~\bibnamefont
  {Gao}}, \bibinfo {author} {\bibfnamefont {F.}~\bibnamefont {Wang}},\ and\
  \bibinfo {author} {\bibfnamefont {O.}~\bibnamefont {Sigmund}},\ }\bibfield
  {title} {\bibinfo {title} {Systematic design of high-q prestressed micro
  membrane resonators},\ }\href@noop {} {\bibfield  {journal} {\bibinfo
  {journal} {Computer Methods in Applied Mechanics and Engineering}\ }\textbf
  {\bibinfo {volume} {361}},\ \bibinfo {pages} {112692} (\bibinfo {year}
  {2020})}\BibitemShut {NoStop}%
\bibitem [{\citenamefont {Cranford}\ \emph {et~al.}(2012)\citenamefont
  {Cranford}, \citenamefont {Tarakanova}, \citenamefont {Pugno},\ and\
  \citenamefont {Buehler}}]{cranford2012nonlinear}%
  \BibitemOpen
  \bibfield  {author} {\bibinfo {author} {\bibfnamefont {S.~W.}\ \bibnamefont
  {Cranford}}, \bibinfo {author} {\bibfnamefont {A.}~\bibnamefont
  {Tarakanova}}, \bibinfo {author} {\bibfnamefont {N.~M.}\ \bibnamefont
  {Pugno}},\ and\ \bibinfo {author} {\bibfnamefont {M.~J.}\ \bibnamefont
  {Buehler}},\ }\bibfield  {title} {\bibinfo {title} {Nonlinear material
  behaviour of spider silk yields robust webs},\ }\href@noop {} {\bibfield
  {journal} {\bibinfo  {journal} {Nature}\ }\textbf {\bibinfo {volume} {482}},\
  \bibinfo {pages} {72} (\bibinfo {year} {2012})}\BibitemShut {NoStop}%
\bibitem [{\citenamefont {Zaera}\ \emph {et~al.}(2014)\citenamefont {Zaera},
  \citenamefont {Soler},\ and\ \citenamefont {Teus}}]{zaera2014uncovering}%
  \BibitemOpen
  \bibfield  {author} {\bibinfo {author} {\bibfnamefont {R.}~\bibnamefont
  {Zaera}}, \bibinfo {author} {\bibfnamefont {A.}~\bibnamefont {Soler}},\ and\
  \bibinfo {author} {\bibfnamefont {J.}~\bibnamefont {Teus}},\ }\bibfield
  {title} {\bibinfo {title} {Uncovering changes in spider orb-web topology
  owing to aerodynamic effects},\ }\href@noop {} {\bibfield  {journal}
  {\bibinfo  {journal} {Journal of the Royal Society Interface}\ }\textbf
  {\bibinfo {volume} {11}},\ \bibinfo {pages} {20140484} (\bibinfo {year}
  {2014})}\BibitemShut {NoStop}%
\bibitem [{\citenamefont {Jyoti}\ \emph {et~al.}(2019)\citenamefont {Jyoti},
  \citenamefont {Kumar}, \citenamefont {Lakhani}, \citenamefont {Kumar},\ and\
  \citenamefont {Bhushan}}]{jyoti2019structural}%
  \BibitemOpen
  \bibfield  {author} {\bibinfo {author} {\bibfnamefont {J.}~\bibnamefont
  {Jyoti}}, \bibinfo {author} {\bibfnamefont {A.}~\bibnamefont {Kumar}},
  \bibinfo {author} {\bibfnamefont {P.}~\bibnamefont {Lakhani}}, \bibinfo
  {author} {\bibfnamefont {N.}~\bibnamefont {Kumar}},\ and\ \bibinfo {author}
  {\bibfnamefont {B.}~\bibnamefont {Bhushan}},\ }\bibfield  {title} {\bibinfo
  {title} {Structural properties and their influence on the prey retention in
  the spider web},\ }\href@noop {} {\bibfield  {journal} {\bibinfo  {journal}
  {Philosophical Transactions of the Royal Society A}\ }\textbf {\bibinfo
  {volume} {377}},\ \bibinfo {pages} {20180271} (\bibinfo {year}
  {2019})}\BibitemShut {NoStop}%
\bibitem [{\citenamefont {Vollrath}(1992)}]{vollrath1992spider}%
  \BibitemOpen
  \bibfield  {author} {\bibinfo {author} {\bibfnamefont {F.}~\bibnamefont
  {Vollrath}},\ }\bibfield  {title} {\bibinfo {title} {Spider webs and silks},\
  }\href@noop {} {\bibfield  {journal} {\bibinfo  {journal} {Scientific
  American}\ }\textbf {\bibinfo {volume} {266}},\ \bibinfo {pages} {70}
  (\bibinfo {year} {1992})}\BibitemShut {NoStop}%
\bibitem [{\citenamefont {Gosline}\ \emph {et~al.}(1999)\citenamefont
  {Gosline}, \citenamefont {Guerette}, \citenamefont {Ortlepp},\ and\
  \citenamefont {Savage}}]{gosline1999mechanical}%
  \BibitemOpen
  \bibfield  {author} {\bibinfo {author} {\bibfnamefont {J.}~\bibnamefont
  {Gosline}}, \bibinfo {author} {\bibfnamefont {P.}~\bibnamefont {Guerette}},
  \bibinfo {author} {\bibfnamefont {C.}~\bibnamefont {Ortlepp}},\ and\ \bibinfo
  {author} {\bibfnamefont {K.}~\bibnamefont {Savage}},\ }\bibfield  {title}
  {\bibinfo {title} {The mechanical design of spider silks: from fibroin
  sequence to mechanical function},\ }\href@noop {} {\bibfield  {journal}
  {\bibinfo  {journal} {Journal of Experimental Biology}\ }\textbf {\bibinfo
  {volume} {202}},\ \bibinfo {pages} {3295} (\bibinfo {year}
  {1999})}\BibitemShut {NoStop}%
\bibitem [{\citenamefont {Boutry}\ and\ \citenamefont
  {Blackledge}(2009)}]{boutry2009biomechanical}%
  \BibitemOpen
  \bibfield  {author} {\bibinfo {author} {\bibfnamefont {C.}~\bibnamefont
  {Boutry}}\ and\ \bibinfo {author} {\bibfnamefont {T.~A.}\ \bibnamefont
  {Blackledge}},\ }\bibfield  {title} {\bibinfo {title} {Biomechanical
  variation of silk links spinning plasticity to spider web function},\
  }\href@noop {} {\bibfield  {journal} {\bibinfo  {journal} {Zoology}\ }\textbf
  {\bibinfo {volume} {112}},\ \bibinfo {pages} {451} (\bibinfo {year}
  {2009})}\BibitemShut {NoStop}%
\bibitem [{\citenamefont {Meyer}\ \emph {et~al.}(2014)\citenamefont {Meyer},
  \citenamefont {Pugno},\ and\ \citenamefont {Cranford}}]{meyer2014compliant}%
  \BibitemOpen
  \bibfield  {author} {\bibinfo {author} {\bibfnamefont {A.}~\bibnamefont
  {Meyer}}, \bibinfo {author} {\bibfnamefont {N.~M.}\ \bibnamefont {Pugno}},\
  and\ \bibinfo {author} {\bibfnamefont {S.~W.}\ \bibnamefont {Cranford}},\
  }\bibfield  {title} {\bibinfo {title} {Compliant threads maximize spider silk
  connection strength and toughness},\ }\href@noop {} {\bibfield  {journal}
  {\bibinfo  {journal} {Journal of The Royal Society Interface}\ }\textbf
  {\bibinfo {volume} {11}},\ \bibinfo {pages} {20140561} (\bibinfo {year}
  {2014})}\BibitemShut {NoStop}%
\bibitem [{\citenamefont {Masters}(1984)}]{masters1984vibrations}%
  \BibitemOpen
  \bibfield  {author} {\bibinfo {author} {\bibfnamefont {W.~M.}\ \bibnamefont
  {Masters}},\ }\bibfield  {title} {\bibinfo {title} {Vibrations in the orbwebs
  of nuctenea sclopetaria (araneidae)},\ }\href@noop {} {\bibfield  {journal}
  {\bibinfo  {journal} {Behavioral Ecology and Sociobiology}\ }\textbf
  {\bibinfo {volume} {15}},\ \bibinfo {pages} {207} (\bibinfo {year}
  {1984})}\BibitemShut {NoStop}%
\bibitem [{\citenamefont {Du}\ \emph {et~al.}(2011)\citenamefont {Du},
  \citenamefont {Yang}, \citenamefont {Liu}, \citenamefont {Li},\ and\
  \citenamefont {Xu}}]{du2011structural}%
  \BibitemOpen
  \bibfield  {author} {\bibinfo {author} {\bibfnamefont {N.}~\bibnamefont
  {Du}}, \bibinfo {author} {\bibfnamefont {Z.}~\bibnamefont {Yang}}, \bibinfo
  {author} {\bibfnamefont {X.~Y.}\ \bibnamefont {Liu}}, \bibinfo {author}
  {\bibfnamefont {Y.}~\bibnamefont {Li}},\ and\ \bibinfo {author}
  {\bibfnamefont {H.~Y.}\ \bibnamefont {Xu}},\ }\bibfield  {title} {\bibinfo
  {title} {Structural origin of the strain-hardening of spider silk},\
  }\href@noop {} {\bibfield  {journal} {\bibinfo  {journal} {Advanced
  Functional Materials}\ }\textbf {\bibinfo {volume} {21}},\ \bibinfo {pages}
  {772} (\bibinfo {year} {2011})}\BibitemShut {NoStop}%
\bibitem [{\citenamefont {Barrows}(1915)}]{barrows1915reactions}%
  \BibitemOpen
  \bibfield  {author} {\bibinfo {author} {\bibfnamefont {W.~M.}\ \bibnamefont
  {Barrows}},\ }\bibfield  {title} {\bibinfo {title} {The reactions of an
  orb-weaving spider, epeira sclopetaria clerck, to rhythmic vibrations of its
  web},\ }\href@noop {} {\bibfield  {journal} {\bibinfo  {journal} {The
  Biological Bulletin}\ }\textbf {\bibinfo {volume} {29}},\ \bibinfo {pages}
  {316} (\bibinfo {year} {1915})}\BibitemShut {NoStop}%
\bibitem [{\citenamefont {Mortimer}\ \emph {et~al.}(2015)\citenamefont
  {Mortimer}, \citenamefont {Holland}, \citenamefont {Windmill},\ and\
  \citenamefont {Vollrath}}]{mortimer2015unpicking}%
  \BibitemOpen
  \bibfield  {author} {\bibinfo {author} {\bibfnamefont {B.}~\bibnamefont
  {Mortimer}}, \bibinfo {author} {\bibfnamefont {C.}~\bibnamefont {Holland}},
  \bibinfo {author} {\bibfnamefont {J.~F.}\ \bibnamefont {Windmill}},\ and\
  \bibinfo {author} {\bibfnamefont {F.}~\bibnamefont {Vollrath}},\ }\bibfield
  {title} {\bibinfo {title} {Unpicking the signal thread of the sector web
  spider zygiella x-notata},\ }\href@noop {} {\bibfield  {journal} {\bibinfo
  {journal} {Journal of The Royal Society Interface}\ }\textbf {\bibinfo
  {volume} {12}},\ \bibinfo {pages} {20150633} (\bibinfo {year}
  {2015})}\BibitemShut {NoStop}%
\bibitem [{\citenamefont {Gao}\ \emph {et~al.}(2003)\citenamefont {Gao},
  \citenamefont {Ji}, \citenamefont {J{\"a}ger}, \citenamefont {Arzt},\ and\
  \citenamefont {Fratzl}}]{gao2003materials}%
  \BibitemOpen
  \bibfield  {author} {\bibinfo {author} {\bibfnamefont {H.}~\bibnamefont
  {Gao}}, \bibinfo {author} {\bibfnamefont {B.}~\bibnamefont {Ji}}, \bibinfo
  {author} {\bibfnamefont {I.~L.}\ \bibnamefont {J{\"a}ger}}, \bibinfo {author}
  {\bibfnamefont {E.}~\bibnamefont {Arzt}},\ and\ \bibinfo {author}
  {\bibfnamefont {P.}~\bibnamefont {Fratzl}},\ }\bibfield  {title} {\bibinfo
  {title} {Materials become insensitive to flaws at nanoscale: lessons from
  nature},\ }\href@noop {} {\bibfield  {journal} {\bibinfo  {journal}
  {Proceedings of the national Academy of Sciences}\ }\textbf {\bibinfo
  {volume} {100}},\ \bibinfo {pages} {5597} (\bibinfo {year}
  {2003})}\BibitemShut {NoStop}%
\bibitem [{\citenamefont {Aizenberg}\ \emph {et~al.}(2005)\citenamefont
  {Aizenberg}, \citenamefont {Weaver}, \citenamefont {Thanawala}, \citenamefont
  {Sundar}, \citenamefont {Morse},\ and\ \citenamefont
  {Fratzl}}]{aizenberg2005skeleton}%
  \BibitemOpen
  \bibfield  {author} {\bibinfo {author} {\bibfnamefont {J.}~\bibnamefont
  {Aizenberg}}, \bibinfo {author} {\bibfnamefont {J.~C.}\ \bibnamefont
  {Weaver}}, \bibinfo {author} {\bibfnamefont {M.~S.}\ \bibnamefont
  {Thanawala}}, \bibinfo {author} {\bibfnamefont {V.~C.}\ \bibnamefont
  {Sundar}}, \bibinfo {author} {\bibfnamefont {D.~E.}\ \bibnamefont {Morse}},\
  and\ \bibinfo {author} {\bibfnamefont {P.}~\bibnamefont {Fratzl}},\
  }\bibfield  {title} {\bibinfo {title} {Skeleton of euplectella sp.:
  structural hierarchy from the nanoscale to the macroscale},\ }\href@noop {}
  {\bibfield  {journal} {\bibinfo  {journal} {Science}\ }\textbf {\bibinfo
  {volume} {309}},\ \bibinfo {pages} {275} (\bibinfo {year}
  {2005})}\BibitemShut {NoStop}%
\bibitem [{\citenamefont {Kamat}\ \emph {et~al.}(2000)\citenamefont {Kamat},
  \citenamefont {Su}, \citenamefont {Ballarini},\ and\ \citenamefont
  {Heuer}}]{kamat2000structural}%
  \BibitemOpen
  \bibfield  {author} {\bibinfo {author} {\bibfnamefont {S.}~\bibnamefont
  {Kamat}}, \bibinfo {author} {\bibfnamefont {X.}~\bibnamefont {Su}}, \bibinfo
  {author} {\bibfnamefont {R.}~\bibnamefont {Ballarini}},\ and\ \bibinfo
  {author} {\bibfnamefont {A.}~\bibnamefont {Heuer}},\ }\bibfield  {title}
  {\bibinfo {title} {Structural basis for the fracture toughness of the shell
  of the conch strombus gigas},\ }\href@noop {} {\bibfield  {journal} {\bibinfo
   {journal} {Nature}\ }\textbf {\bibinfo {volume} {405}},\ \bibinfo {pages}
  {1036} (\bibinfo {year} {2000})}\BibitemShut {NoStop}%
\bibitem [{\citenamefont {Miniaci}\ \emph {et~al.}(2016)\citenamefont
  {Miniaci}, \citenamefont {Krushynska}, \citenamefont {Movchan}, \citenamefont
  {Bosia},\ and\ \citenamefont {Pugno}}]{miniaci2016spider}%
  \BibitemOpen
  \bibfield  {author} {\bibinfo {author} {\bibfnamefont {M.}~\bibnamefont
  {Miniaci}}, \bibinfo {author} {\bibfnamefont {A.}~\bibnamefont {Krushynska}},
  \bibinfo {author} {\bibfnamefont {A.~B.}\ \bibnamefont {Movchan}}, \bibinfo
  {author} {\bibfnamefont {F.}~\bibnamefont {Bosia}},\ and\ \bibinfo {author}
  {\bibfnamefont {N.~M.}\ \bibnamefont {Pugno}},\ }\bibfield  {title} {\bibinfo
  {title} {Spider web-inspired acoustic metamaterials},\ }\href@noop {}
  {\bibfield  {journal} {\bibinfo  {journal} {Applied Physics Letters}\
  }\textbf {\bibinfo {volume} {109}},\ \bibinfo {pages} {071905} (\bibinfo
  {year} {2016})}\BibitemShut {NoStop}%
\bibitem [{\citenamefont {Krushynska}\ \emph {et~al.}(2017)\citenamefont
  {Krushynska}, \citenamefont {Bosia}, \citenamefont {Miniaci},\ and\
  \citenamefont {Pugno}}]{krushynska2017spider}%
  \BibitemOpen
  \bibfield  {author} {\bibinfo {author} {\bibfnamefont {A.}~\bibnamefont
  {Krushynska}}, \bibinfo {author} {\bibfnamefont {F.}~\bibnamefont {Bosia}},
  \bibinfo {author} {\bibfnamefont {M.}~\bibnamefont {Miniaci}},\ and\ \bibinfo
  {author} {\bibfnamefont {N.}~\bibnamefont {Pugno}},\ }\bibfield  {title}
  {\bibinfo {title} {Spider web-structured labyrinthine acoustic metamaterials
  for low-frequency sound control},\ }\href@noop {} {\bibfield  {journal}
  {\bibinfo  {journal} {New Journal of Physics}\ }\textbf {\bibinfo {volume}
  {19}},\ \bibinfo {pages} {105001} (\bibinfo {year} {2017})}\BibitemShut
  {NoStop}%
\bibitem [{\citenamefont {Liu}\ \emph {et~al.}(2018)\citenamefont {Liu},
  \citenamefont {Liu}, \citenamefont {hun Lee}, \citenamefont {Zheng},
  \citenamefont {Du}, \citenamefont {Zhang}, \citenamefont {Xu}, \citenamefont
  {Wang}, \citenamefont {Wu}, \citenamefont {Shen}, \citenamefont {Cui},
  \citenamefont {Mai},\ and\ \citenamefont {Kim}}]{liu2018spider}%
  \BibitemOpen
  \bibfield  {author} {\bibinfo {author} {\bibfnamefont {X.}~\bibnamefont
  {Liu}}, \bibinfo {author} {\bibfnamefont {D.}~\bibnamefont {Liu}}, \bibinfo
  {author} {\bibfnamefont {J.}~\bibnamefont {hun Lee}}, \bibinfo {author}
  {\bibfnamefont {Q.}~\bibnamefont {Zheng}}, \bibinfo {author} {\bibfnamefont
  {X.}~\bibnamefont {Du}}, \bibinfo {author} {\bibfnamefont {X.}~\bibnamefont
  {Zhang}}, \bibinfo {author} {\bibfnamefont {H.}~\bibnamefont {Xu}}, \bibinfo
  {author} {\bibfnamefont {Z.}~\bibnamefont {Wang}}, \bibinfo {author}
  {\bibfnamefont {Y.}~\bibnamefont {Wu}}, \bibinfo {author} {\bibfnamefont
  {X.}~\bibnamefont {Shen}}, \bibinfo {author} {\bibfnamefont {J.}~\bibnamefont
  {Cui}}, \bibinfo {author} {\bibfnamefont {Y.}~\bibnamefont {Mai}},\ and\
  \bibinfo {author} {\bibfnamefont {J.-K.}\ \bibnamefont {Kim}},\ }\bibfield
  {title} {\bibinfo {title} {Spider-web-inspired stretchable graphene woven
  fabric for highly sensitive, transparent, wearable strain sensors},\
  }\href@noop {} {\bibfield  {journal} {\bibinfo  {journal} {ACS applied
  materials \& interfaces}\ }\textbf {\bibinfo {volume} {11}},\ \bibinfo
  {pages} {2282} (\bibinfo {year} {2018})}\BibitemShut {NoStop}%
\bibitem [{\citenamefont {Yu}\ \emph {et~al.}(2012)\citenamefont {Yu},
  \citenamefont {Purdy},\ and\ \citenamefont {Regal}}]{yu2012control}%
  \BibitemOpen
  \bibfield  {author} {\bibinfo {author} {\bibfnamefont {P.-L.}\ \bibnamefont
  {Yu}}, \bibinfo {author} {\bibfnamefont {T.}~\bibnamefont {Purdy}},\ and\
  \bibinfo {author} {\bibfnamefont {C.}~\bibnamefont {Regal}},\ }\bibfield
  {title} {\bibinfo {title} {Control of material damping in high-q membrane
  microresonators},\ }\href@noop {} {\bibfield  {journal} {\bibinfo  {journal}
  {Physical Review Letters}\ }\textbf {\bibinfo {volume} {108}},\ \bibinfo
  {pages} {083603} (\bibinfo {year} {2012})}\BibitemShut {NoStop}%
\bibitem [{\citenamefont {Villanueva}\ and\ \citenamefont
  {Schmid}(2014)}]{villanueva2014evidence}%
  \BibitemOpen
  \bibfield  {author} {\bibinfo {author} {\bibfnamefont {L.~G.}\ \bibnamefont
  {Villanueva}}\ and\ \bibinfo {author} {\bibfnamefont {S.}~\bibnamefont
  {Schmid}},\ }\bibfield  {title} {\bibinfo {title} {Evidence of surface loss
  as ubiquitous limiting damping mechanism in sin micro-and nanomechanical
  resonators},\ }\href@noop {} {\bibfield  {journal} {\bibinfo  {journal}
  {Physical Review Letters}\ }\textbf {\bibinfo {volume} {113}},\ \bibinfo
  {pages} {227201} (\bibinfo {year} {2014})}\BibitemShut {NoStop}%
\bibitem [{\citenamefont {Gu}\ \emph {et~al.}(2018)\citenamefont {Gu},
  \citenamefont {Chen}, \citenamefont {Richmond},\ and\ \citenamefont
  {Buehler}}]{gu2018bioinspired}%
  \BibitemOpen
  \bibfield  {author} {\bibinfo {author} {\bibfnamefont {G.~X.}\ \bibnamefont
  {Gu}}, \bibinfo {author} {\bibfnamefont {C.-T.}\ \bibnamefont {Chen}},
  \bibinfo {author} {\bibfnamefont {D.~J.}\ \bibnamefont {Richmond}},\ and\
  \bibinfo {author} {\bibfnamefont {M.~J.}\ \bibnamefont {Buehler}},\
  }\bibfield  {title} {\bibinfo {title} {Bioinspired hierarchical composite
  design using machine learning: simulation, additive manufacturing, and
  experiment},\ }\href@noop {} {\bibfield  {journal} {\bibinfo  {journal}
  {Materials Horizons}\ }\textbf {\bibinfo {volume} {5}},\ \bibinfo {pages}
  {939} (\bibinfo {year} {2018})}\BibitemShut {NoStop}%
\bibitem [{\citenamefont {Carrasquilla}\ and\ \citenamefont
  {Melko}(2017)}]{carrasquilla2017machine}%
  \BibitemOpen
  \bibfield  {author} {\bibinfo {author} {\bibfnamefont {J.}~\bibnamefont
  {Carrasquilla}}\ and\ \bibinfo {author} {\bibfnamefont {R.~G.}\ \bibnamefont
  {Melko}},\ }\bibfield  {title} {\bibinfo {title} {Machine learning phases of
  matter},\ }\href@noop {} {\bibfield  {journal} {\bibinfo  {journal} {Nature
  Physics}\ }\textbf {\bibinfo {volume} {13}},\ \bibinfo {pages} {431}
  (\bibinfo {year} {2017})}\BibitemShut {NoStop}%
\bibitem [{\citenamefont {Jiao}\ and\ \citenamefont
  {Alavi}(2020)}]{jiao2020evolutionary}%
  \BibitemOpen
  \bibfield  {author} {\bibinfo {author} {\bibfnamefont {P.}~\bibnamefont
  {Jiao}}\ and\ \bibinfo {author} {\bibfnamefont {A.~H.}\ \bibnamefont
  {Alavi}},\ }\bibfield  {title} {\bibinfo {title} {Evolutionary computation
  for design and characterization of nanoscale metastructures},\ }\href@noop {}
  {\bibfield  {journal} {\bibinfo  {journal} {Applied Materials Today}\
  }\textbf {\bibinfo {volume} {21}},\ \bibinfo {pages} {100816} (\bibinfo
  {year} {2020})}\BibitemShut {NoStop}%
\bibitem [{\citenamefont {Ding}\ \emph {et~al.}(2018)\citenamefont {Ding},
  \citenamefont {Kim}, \citenamefont {Kuindersma},\ and\ \citenamefont
  {Walsh}}]{ding2018human}%
  \BibitemOpen
  \bibfield  {author} {\bibinfo {author} {\bibfnamefont {Y.}~\bibnamefont
  {Ding}}, \bibinfo {author} {\bibfnamefont {M.}~\bibnamefont {Kim}}, \bibinfo
  {author} {\bibfnamefont {S.}~\bibnamefont {Kuindersma}},\ and\ \bibinfo
  {author} {\bibfnamefont {C.~J.}\ \bibnamefont {Walsh}},\ }\bibfield  {title}
  {\bibinfo {title} {Human-in-the-loop optimization of hip assistance with a
  soft exosuit during walking},\ }\href@noop {} {\bibfield  {journal} {\bibinfo
   {journal} {Science Robotics}\ }\textbf {\bibinfo {volume} {3}} (\bibinfo
  {year} {2018})}\BibitemShut {NoStop}%
\bibitem [{\citenamefont {Shalloo}\ \emph {et~al.}(2020)\citenamefont
  {Shalloo}, \citenamefont {Dann}, \citenamefont {Gruse}, \citenamefont
  {Underwood}, \citenamefont {Antoine}, \citenamefont {Arran}, \citenamefont
  {Backhouse}, \citenamefont {Baird}, \citenamefont {Balcazar}, \citenamefont
  {Bourgeois}, \citenamefont {Cardarelli}, \citenamefont {Hatfield},
  \citenamefont {Kang}, \citenamefont {Krushelnick}, \citenamefont {Mangles},
  \citenamefont {Murphy}, \citenamefont {Lu}, \citenamefont {Osterhoff},
  \citenamefont {P{\~o}der}, \citenamefont {Rajeev}, \citenamefont {Ridgers},
  \citenamefont {Rozario}, \citenamefont {Selwood}, \citenamefont {Shahani},
  \citenamefont {Symes}, \citenamefont {Thomas}, \citenamefont {Thornton},
  \citenamefont {Najmudin},\ and\ \citenamefont
  {Streeter}}]{shalloo2020automation}%
  \BibitemOpen
  \bibfield  {author} {\bibinfo {author} {\bibfnamefont {R.}~\bibnamefont
  {Shalloo}}, \bibinfo {author} {\bibfnamefont {S.}~\bibnamefont {Dann}},
  \bibinfo {author} {\bibfnamefont {J.-N.}\ \bibnamefont {Gruse}}, \bibinfo
  {author} {\bibfnamefont {C.}~\bibnamefont {Underwood}}, \bibinfo {author}
  {\bibfnamefont {A.}~\bibnamefont {Antoine}}, \bibinfo {author} {\bibfnamefont
  {C.}~\bibnamefont {Arran}}, \bibinfo {author} {\bibfnamefont
  {M.}~\bibnamefont {Backhouse}}, \bibinfo {author} {\bibfnamefont
  {C.}~\bibnamefont {Baird}}, \bibinfo {author} {\bibfnamefont
  {M.}~\bibnamefont {Balcazar}}, \bibinfo {author} {\bibfnamefont
  {N.}~\bibnamefont {Bourgeois}}, \bibinfo {author} {\bibfnamefont {J.~A.}\
  \bibnamefont {Cardarelli}}, \bibinfo {author} {\bibfnamefont
  {P.}~\bibnamefont {Hatfield}}, \bibinfo {author} {\bibfnamefont
  {J.}~\bibnamefont {Kang}}, \bibinfo {author} {\bibfnamefont {K.}~\bibnamefont
  {Krushelnick}}, \bibinfo {author} {\bibfnamefont {S.}~\bibnamefont
  {Mangles}}, \bibinfo {author} {\bibfnamefont {C.}~\bibnamefont {Murphy}},
  \bibinfo {author} {\bibfnamefont {N.}~\bibnamefont {Lu}}, \bibinfo {author}
  {\bibfnamefont {J.}~\bibnamefont {Osterhoff}}, \bibinfo {author}
  {\bibfnamefont {K.}~\bibnamefont {P{\~o}der}}, \bibinfo {author}
  {\bibfnamefont {P.}~\bibnamefont {Rajeev}}, \bibinfo {author} {\bibfnamefont
  {C.}~\bibnamefont {Ridgers}}, \bibinfo {author} {\bibfnamefont
  {S.}~\bibnamefont {Rozario}}, \bibinfo {author} {\bibfnamefont
  {M.}~\bibnamefont {Selwood}}, \bibinfo {author} {\bibfnamefont
  {A.}~\bibnamefont {Shahani}}, \bibinfo {author} {\bibfnamefont
  {D.}~\bibnamefont {Symes}}, \bibinfo {author} {\bibfnamefont
  {A.}~\bibnamefont {Thomas}}, \bibinfo {author} {\bibfnamefont
  {C.}~\bibnamefont {Thornton}}, \bibinfo {author} {\bibfnamefont
  {Z.}~\bibnamefont {Najmudin}},\ and\ \bibinfo {author} {\bibfnamefont
  {M.}~\bibnamefont {Streeter}},\ }\bibfield  {title} {\bibinfo {title}
  {Automation and control of laser wakefield accelerators using bayesian
  optimization},\ }\href@noop {} {\bibfield  {journal} {\bibinfo  {journal}
  {Nature communications}\ }\textbf {\bibinfo {volume} {11}},\ \bibinfo {pages}
  {1} (\bibinfo {year} {2020})}\BibitemShut {NoStop}%
\bibitem [{\citenamefont {Shields}\ \emph {et~al.}(2021)\citenamefont
  {Shields}, \citenamefont {Stevens}, \citenamefont {Li}, \citenamefont
  {Parasram}, \citenamefont {Damani}, \citenamefont {Alvarado}, \citenamefont
  {Janey}, \citenamefont {Adams},\ and\ \citenamefont
  {Doyle}}]{shields2021bayesian}%
  \BibitemOpen
  \bibfield  {author} {\bibinfo {author} {\bibfnamefont {B.~J.}\ \bibnamefont
  {Shields}}, \bibinfo {author} {\bibfnamefont {J.}~\bibnamefont {Stevens}},
  \bibinfo {author} {\bibfnamefont {J.}~\bibnamefont {Li}}, \bibinfo {author}
  {\bibfnamefont {M.}~\bibnamefont {Parasram}}, \bibinfo {author}
  {\bibfnamefont {F.}~\bibnamefont {Damani}}, \bibinfo {author} {\bibfnamefont
  {J.~I.~M.}\ \bibnamefont {Alvarado}}, \bibinfo {author} {\bibfnamefont
  {J.~M.}\ \bibnamefont {Janey}}, \bibinfo {author} {\bibfnamefont {R.~P.}\
  \bibnamefont {Adams}},\ and\ \bibinfo {author} {\bibfnamefont {A.~G.}\
  \bibnamefont {Doyle}},\ }\bibfield  {title} {\bibinfo {title} {Bayesian
  reaction optimization as a tool for chemical synthesis},\ }\href@noop {}
  {\bibfield  {journal} {\bibinfo  {journal} {Nature}\ }\textbf {\bibinfo
  {volume} {590}},\ \bibinfo {pages} {89} (\bibinfo {year} {2021})}\BibitemShut
  {NoStop}%
\bibitem [{\citenamefont {Pelikan}\ \emph {et~al.}(1999)\citenamefont
  {Pelikan}, \citenamefont {Goldberg},\ and\ \citenamefont
  {Cant{\'u}-Paz}}]{pelikan1999boa}%
  \BibitemOpen
  \bibfield  {author} {\bibinfo {author} {\bibfnamefont {M.}~\bibnamefont
  {Pelikan}}, \bibinfo {author} {\bibfnamefont {D.~E.}\ \bibnamefont
  {Goldberg}},\ and\ \bibinfo {author} {\bibfnamefont {E.}~\bibnamefont
  {Cant{\'u}-Paz}},\ }\bibfield  {title} {\bibinfo {title} {Boa: The bayesian
  optimization algorithm},\ }in\ \href@noop {} {\emph {\bibinfo {booktitle}
  {Proceedings of the genetic and evolutionary computation conference
  GECCO-99}}},\ Vol.~\bibinfo {volume} {1}\ (\bibinfo {organization}
  {Citeseer},\ \bibinfo {year} {1999})\ pp.\ \bibinfo {pages}
  {525--532}\BibitemShut {NoStop}%
\bibitem [{\citenamefont {Bessa}\ \emph {et~al.}(2019)\citenamefont {Bessa},
  \citenamefont {Glowacki},\ and\ \citenamefont {Houlder}}]{bessa2019bayesian}%
  \BibitemOpen
  \bibfield  {author} {\bibinfo {author} {\bibfnamefont {M.~A.}\ \bibnamefont
  {Bessa}}, \bibinfo {author} {\bibfnamefont {P.}~\bibnamefont {Glowacki}},\
  and\ \bibinfo {author} {\bibfnamefont {M.}~\bibnamefont {Houlder}},\
  }\bibfield  {title} {\bibinfo {title} {Bayesian machine learning in
  metamaterial design: Fragile becomes supercompressible},\ }\href@noop {}
  {\bibfield  {journal} {\bibinfo  {journal} {Advanced Materials}\ }\textbf
  {\bibinfo {volume} {31}},\ \bibinfo {pages} {1904845} (\bibinfo {year}
  {2019})}\BibitemShut {NoStop}%
\bibitem [{\citenamefont {Frazier}\ and\ \citenamefont
  {Wang}(2016)}]{Frazier2016}%
  \BibitemOpen
  \bibfield  {author} {\bibinfo {author} {\bibfnamefont {P.~I.}\ \bibnamefont
  {Frazier}}\ and\ \bibinfo {author} {\bibfnamefont {J.}~\bibnamefont {Wang}},\
  }\bibfield  {title} {\bibinfo {title} {Bayesian optimization for materials
  design},\ }in\ \href@noop {} {\emph {\bibinfo {booktitle} {Information
  science for materials discovery and design}}}\ (\bibinfo  {publisher}
  {Springer},\ \bibinfo {year} {2016})\ pp.\ \bibinfo {pages}
  {45--75}\BibitemShut {NoStop}%
\bibitem [{\citenamefont {Shahriari}\ \emph {et~al.}(2015)\citenamefont
  {Shahriari}, \citenamefont {Swersky}, \citenamefont {Wang}, \citenamefont
  {Adams},\ and\ \citenamefont {De~Freitas}}]{shahriari2015taking}%
  \BibitemOpen
  \bibfield  {author} {\bibinfo {author} {\bibfnamefont {B.}~\bibnamefont
  {Shahriari}}, \bibinfo {author} {\bibfnamefont {K.}~\bibnamefont {Swersky}},
  \bibinfo {author} {\bibfnamefont {Z.}~\bibnamefont {Wang}}, \bibinfo {author}
  {\bibfnamefont {R.~P.}\ \bibnamefont {Adams}},\ and\ \bibinfo {author}
  {\bibfnamefont {N.}~\bibnamefont {De~Freitas}},\ }\bibfield  {title}
  {\bibinfo {title} {Taking the human out of the loop: A review of bayesian
  optimization},\ }\href@noop {} {\bibfield  {journal} {\bibinfo  {journal}
  {Proceedings of the IEEE}\ }\textbf {\bibinfo {volume} {104}},\ \bibinfo
  {pages} {148} (\bibinfo {year} {2015})}\BibitemShut {NoStop}%
\bibitem [{\citenamefont {The-GPyOpt-authors}(2016)}]{gpyopt2016}%
  \BibitemOpen
  \bibfield  {author} {\bibinfo {author} {\bibnamefont {The-GPyOpt-authors}},\
  }\href@noop {} {\bibinfo {title} {Gpyopt: A bayesian optimization framework
  in python}},\ \bibinfo {howpublished}
  {\url{http://github.com/SheffieldML/GPyOpt}} (\bibinfo {year}
  {2016})\BibitemShut {NoStop}%
\bibitem [{\citenamefont {Inc.}(2018)}]{comsol}%
  \BibitemOpen
  \bibfield  {author} {\bibinfo {author} {\bibfnamefont {C.}~\bibnamefont
  {Inc.}},\ }\href {http://www.comsol.com/products/multiphysics/} {\bibinfo
  {title} {Comsol}} (\bibinfo {year} {2018})\BibitemShut {NoStop}%
\bibitem [{\citenamefont {Kawano}\ and\ \citenamefont
  {Morassi}(2019)}]{kawano2019detecting}%
  \BibitemOpen
  \bibfield  {author} {\bibinfo {author} {\bibfnamefont {A.}~\bibnamefont
  {Kawano}}\ and\ \bibinfo {author} {\bibfnamefont {A.}~\bibnamefont
  {Morassi}},\ }\bibfield  {title} {\bibinfo {title} {Detecting a prey in a
  spider orb web},\ }\href@noop {} {\bibfield  {journal} {\bibinfo  {journal}
  {SIAM Journal on Applied Mathematics}\ }\textbf {\bibinfo {volume} {79}},\
  \bibinfo {pages} {2506} (\bibinfo {year} {2019})}\BibitemShut {NoStop}%
\bibitem [{\citenamefont {Steinlechner}\ \emph {et~al.}(2017)\citenamefont
  {Steinlechner}, \citenamefont {Kr{\"u}ger}, \citenamefont {Martin},
  \citenamefont {Bell}, \citenamefont {Hough}, \citenamefont {Kaufer},
  \citenamefont {Rowan}, \citenamefont {Schnabel},\ and\ \citenamefont
  {Steinlechner}}]{steinlechner2017optical}%
  \BibitemOpen
  \bibfield  {author} {\bibinfo {author} {\bibfnamefont {J.}~\bibnamefont
  {Steinlechner}}, \bibinfo {author} {\bibfnamefont {C.}~\bibnamefont
  {Kr{\"u}ger}}, \bibinfo {author} {\bibfnamefont {I.~W.}\ \bibnamefont
  {Martin}}, \bibinfo {author} {\bibfnamefont {A.}~\bibnamefont {Bell}},
  \bibinfo {author} {\bibfnamefont {J.}~\bibnamefont {Hough}}, \bibinfo
  {author} {\bibfnamefont {H.}~\bibnamefont {Kaufer}}, \bibinfo {author}
  {\bibfnamefont {S.}~\bibnamefont {Rowan}}, \bibinfo {author} {\bibfnamefont
  {R.}~\bibnamefont {Schnabel}},\ and\ \bibinfo {author} {\bibfnamefont
  {S.}~\bibnamefont {Steinlechner}},\ }\bibfield  {title} {\bibinfo {title}
  {Optical absorption of silicon nitride membranes at 1064 nm and at 1550 nm},\
  }\href@noop {} {\bibfield  {journal} {\bibinfo  {journal} {Physical Review
  D}\ }\textbf {\bibinfo {volume} {96}},\ \bibinfo {pages} {022007} (\bibinfo
  {year} {2017})}\BibitemShut {NoStop}%
\bibitem [{\citenamefont {Wilson}\ \emph {et~al.}(2009)\citenamefont {Wilson},
  \citenamefont {Regal}, \citenamefont {Papp},\ and\ \citenamefont
  {Kimble}}]{wilson2009cavity}%
  \BibitemOpen
  \bibfield  {author} {\bibinfo {author} {\bibfnamefont {D.~J.}\ \bibnamefont
  {Wilson}}, \bibinfo {author} {\bibfnamefont {C.~A.}\ \bibnamefont {Regal}},
  \bibinfo {author} {\bibfnamefont {S.~B.}\ \bibnamefont {Papp}},\ and\
  \bibinfo {author} {\bibfnamefont {H.}~\bibnamefont {Kimble}},\ }\bibfield
  {title} {\bibinfo {title} {Cavity optomechanics with stoichiometric sin
  films},\ }\href@noop {} {\bibfield  {journal} {\bibinfo  {journal} {Physical
  Review Letters}\ }\textbf {\bibinfo {volume} {103}},\ \bibinfo {pages}
  {207204} (\bibinfo {year} {2009})}\BibitemShut {NoStop}%
\bibitem [{\citenamefont {Serra}\ \emph {et~al.}(2018)\citenamefont {Serra},
  \citenamefont {Morana}, \citenamefont {Borrielli}, \citenamefont {Marin},
  \citenamefont {Pandraud}, \citenamefont {Pontin}, \citenamefont {Prodi},
  \citenamefont {Sarro},\ and\ \citenamefont {Bonaldi}}]{serra2018silicon}%
  \BibitemOpen
  \bibfield  {author} {\bibinfo {author} {\bibfnamefont {E.}~\bibnamefont
  {Serra}}, \bibinfo {author} {\bibfnamefont {B.}~\bibnamefont {Morana}},
  \bibinfo {author} {\bibfnamefont {A.}~\bibnamefont {Borrielli}}, \bibinfo
  {author} {\bibfnamefont {F.}~\bibnamefont {Marin}}, \bibinfo {author}
  {\bibfnamefont {G.}~\bibnamefont {Pandraud}}, \bibinfo {author}
  {\bibfnamefont {A.}~\bibnamefont {Pontin}}, \bibinfo {author} {\bibfnamefont
  {G.~A.}\ \bibnamefont {Prodi}}, \bibinfo {author} {\bibfnamefont {P.~M.}\
  \bibnamefont {Sarro}},\ and\ \bibinfo {author} {\bibfnamefont
  {M.}~\bibnamefont {Bonaldi}},\ }\bibfield  {title} {\bibinfo {title} {Silicon
  nitride moms oscillator for room temperature quantum optomechanics},\
  }\href@noop {} {\bibfield  {journal} {\bibinfo  {journal} {Journal of
  Microelectromechanical Systems}\ }\textbf {\bibinfo {volume} {27}},\ \bibinfo
  {pages} {1193} (\bibinfo {year} {2018})}\BibitemShut {NoStop}%
\bibitem [{\citenamefont {Usami}\ \emph {et~al.}(2012)\citenamefont {Usami},
  \citenamefont {Naesby}, \citenamefont {Bagci}, \citenamefont {Nielsen},
  \citenamefont {Liu}, \citenamefont {Stobbe}, \citenamefont {Lodahl},\ and\
  \citenamefont {Polzik}}]{usami2012optical}%
  \BibitemOpen
  \bibfield  {author} {\bibinfo {author} {\bibfnamefont {K.}~\bibnamefont
  {Usami}}, \bibinfo {author} {\bibfnamefont {A.}~\bibnamefont {Naesby}},
  \bibinfo {author} {\bibfnamefont {T.}~\bibnamefont {Bagci}}, \bibinfo
  {author} {\bibfnamefont {B.~M.}\ \bibnamefont {Nielsen}}, \bibinfo {author}
  {\bibfnamefont {J.}~\bibnamefont {Liu}}, \bibinfo {author} {\bibfnamefont
  {S.}~\bibnamefont {Stobbe}}, \bibinfo {author} {\bibfnamefont
  {P.}~\bibnamefont {Lodahl}},\ and\ \bibinfo {author} {\bibfnamefont {E.~S.}\
  \bibnamefont {Polzik}},\ }\bibfield  {title} {\bibinfo {title} {Optical
  cavity cooling of mechanical modes of a semiconductor nanomembrane},\
  }\href@noop {} {\bibfield  {journal} {\bibinfo  {journal} {Nature Physics}\
  }\textbf {\bibinfo {volume} {8}},\ \bibinfo {pages} {168} (\bibinfo {year}
  {2012})}\BibitemShut {NoStop}%
\bibitem [{\citenamefont {Cole}\ \emph {et~al.}(2014)\citenamefont {Cole},
  \citenamefont {Yu}, \citenamefont {G{\"a}rtner}, \citenamefont {Siquans},
  \citenamefont {Moghadas~Nia}, \citenamefont {Schm{\"o}le}, \citenamefont
  {Hoelscher-Obermaier}, \citenamefont {Purdy}, \citenamefont {Wieczorek},
  \citenamefont {Regal},\ and\ \citenamefont {Aspelmeyer}}]{cole2014tensile}%
  \BibitemOpen
  \bibfield  {author} {\bibinfo {author} {\bibfnamefont {G.~D.}\ \bibnamefont
  {Cole}}, \bibinfo {author} {\bibfnamefont {P.-L.}\ \bibnamefont {Yu}},
  \bibinfo {author} {\bibfnamefont {C.}~\bibnamefont {G{\"a}rtner}}, \bibinfo
  {author} {\bibfnamefont {K.}~\bibnamefont {Siquans}}, \bibinfo {author}
  {\bibfnamefont {R.}~\bibnamefont {Moghadas~Nia}}, \bibinfo {author}
  {\bibfnamefont {J.}~\bibnamefont {Schm{\"o}le}}, \bibinfo {author}
  {\bibfnamefont {J.}~\bibnamefont {Hoelscher-Obermaier}}, \bibinfo {author}
  {\bibfnamefont {T.~P.}\ \bibnamefont {Purdy}}, \bibinfo {author}
  {\bibfnamefont {W.}~\bibnamefont {Wieczorek}}, \bibinfo {author}
  {\bibfnamefont {C.~A.}\ \bibnamefont {Regal}},\ and\ \bibinfo {author}
  {\bibfnamefont {M.}~\bibnamefont {Aspelmeyer}},\ }\bibfield  {title}
  {\bibinfo {title} {Tensile-strained inxga1- xp membranes for cavity
  optomechanics},\ }\href@noop {} {\bibfield  {journal} {\bibinfo  {journal}
  {Applied Physics Letters}\ }\textbf {\bibinfo {volume} {104}},\ \bibinfo
  {pages} {201908} (\bibinfo {year} {2014})}\BibitemShut {NoStop}%
\bibitem [{\citenamefont {Faust}\ \emph {et~al.}(2012)\citenamefont {Faust},
  \citenamefont {Krenn}, \citenamefont {Manus}, \citenamefont {Kotthaus},\ and\
  \citenamefont {Weig}}]{faust2012microwave}%
  \BibitemOpen
  \bibfield  {author} {\bibinfo {author} {\bibfnamefont {T.}~\bibnamefont
  {Faust}}, \bibinfo {author} {\bibfnamefont {P.}~\bibnamefont {Krenn}},
  \bibinfo {author} {\bibfnamefont {S.}~\bibnamefont {Manus}}, \bibinfo
  {author} {\bibfnamefont {J.~P.}\ \bibnamefont {Kotthaus}},\ and\ \bibinfo
  {author} {\bibfnamefont {E.~M.}\ \bibnamefont {Weig}},\ }\bibfield  {title}
  {\bibinfo {title} {Microwave cavity-enhanced transduction for plug and play
  nanomechanics at room temperature},\ }\href@noop {} {\bibfield  {journal}
  {\bibinfo  {journal} {Nature communications}\ }\textbf {\bibinfo {volume}
  {3}},\ \bibinfo {pages} {1} (\bibinfo {year} {2012})}\BibitemShut {NoStop}%
\bibitem [{\citenamefont {Reetz}\ \emph {et~al.}(2019)\citenamefont {Reetz},
  \citenamefont {Fischer}, \citenamefont {Assumpcao}, \citenamefont {McNally},
  \citenamefont {Burns}, \citenamefont {Sankey},\ and\ \citenamefont
  {Regal}}]{reetz2019analysis}%
  \BibitemOpen
  \bibfield  {author} {\bibinfo {author} {\bibfnamefont {C.}~\bibnamefont
  {Reetz}}, \bibinfo {author} {\bibfnamefont {R.}~\bibnamefont {Fischer}},
  \bibinfo {author} {\bibfnamefont {G.~G.}\ \bibnamefont {Assumpcao}}, \bibinfo
  {author} {\bibfnamefont {D.~P.}\ \bibnamefont {McNally}}, \bibinfo {author}
  {\bibfnamefont {P.~S.}\ \bibnamefont {Burns}}, \bibinfo {author}
  {\bibfnamefont {J.~C.}\ \bibnamefont {Sankey}},\ and\ \bibinfo {author}
  {\bibfnamefont {C.~A.}\ \bibnamefont {Regal}},\ }\bibfield  {title} {\bibinfo
  {title} {Analysis of membrane phononic crystals with wide band gaps and
  low-mass defects},\ }\href@noop {} {\bibfield  {journal} {\bibinfo  {journal}
  {Physical Review Applied}\ }\textbf {\bibinfo {volume} {12}},\ \bibinfo
  {pages} {044027} (\bibinfo {year} {2019})}\BibitemShut {NoStop}%
\bibitem [{\citenamefont {Nakamura}\ \emph {et~al.}(1997)\citenamefont
  {Nakamura}, \citenamefont {Sakagami}, \citenamefont {Amamoto},\ and\
  \citenamefont {Watanabe}}]{tensilediamond97}%
  \BibitemOpen
  \bibfield  {author} {\bibinfo {author} {\bibfnamefont {Y.}~\bibnamefont
  {Nakamura}}, \bibinfo {author} {\bibfnamefont {S.}~\bibnamefont {Sakagami}},
  \bibinfo {author} {\bibfnamefont {Y.}~\bibnamefont {Amamoto}},\ and\ \bibinfo
  {author} {\bibfnamefont {Y.}~\bibnamefont {Watanabe}},\ }\bibfield  {title}
  {\bibinfo {title} {Measurement of internal stresses in cvd diamond films},\
  }\href@noop {} {\bibfield  {journal} {\bibinfo  {journal} {Thin Solid Films}\
  }\textbf {\bibinfo {volume} {308}},\ \bibinfo {pages} {249} (\bibinfo {year}
  {1997})}\BibitemShut {NoStop}%
\bibitem [{\citenamefont {Liu}\ \emph {et~al.}(2011)\citenamefont {Liu},
  \citenamefont {Usami}, \citenamefont {Naesby}, \citenamefont {Bagci},
  \citenamefont {Polzik}, \citenamefont {Lodahl},\ and\ \citenamefont
  {Stobbe}}]{liu2011high}%
  \BibitemOpen
  \bibfield  {author} {\bibinfo {author} {\bibfnamefont {J.}~\bibnamefont
  {Liu}}, \bibinfo {author} {\bibfnamefont {K.}~\bibnamefont {Usami}}, \bibinfo
  {author} {\bibfnamefont {A.}~\bibnamefont {Naesby}}, \bibinfo {author}
  {\bibfnamefont {T.}~\bibnamefont {Bagci}}, \bibinfo {author} {\bibfnamefont
  {E.~S.}\ \bibnamefont {Polzik}}, \bibinfo {author} {\bibfnamefont
  {P.}~\bibnamefont {Lodahl}},\ and\ \bibinfo {author} {\bibfnamefont
  {S.}~\bibnamefont {Stobbe}},\ }\bibfield  {title} {\bibinfo {title} {High-q
  optomechanical gaas nanomembranes},\ }\href@noop {} {\bibfield  {journal}
  {\bibinfo  {journal} {Applied Physics Letters}\ }\textbf {\bibinfo {volume}
  {99}},\ \bibinfo {pages} {243102} (\bibinfo {year} {2011})}\BibitemShut
  {NoStop}%
\bibitem [{\citenamefont {Kermany}\ \emph {et~al.}(2014)\citenamefont
  {Kermany}, \citenamefont {Brawley}, \citenamefont {Mishra}, \citenamefont
  {Sheridan}, \citenamefont {Bowen},\ and\ \citenamefont
  {Iacopi}}]{kermany2014microresonators}%
  \BibitemOpen
  \bibfield  {author} {\bibinfo {author} {\bibfnamefont {A.~R.}\ \bibnamefont
  {Kermany}}, \bibinfo {author} {\bibfnamefont {G.}~\bibnamefont {Brawley}},
  \bibinfo {author} {\bibfnamefont {N.}~\bibnamefont {Mishra}}, \bibinfo
  {author} {\bibfnamefont {E.}~\bibnamefont {Sheridan}}, \bibinfo {author}
  {\bibfnamefont {W.~P.}\ \bibnamefont {Bowen}},\ and\ \bibinfo {author}
  {\bibfnamefont {F.}~\bibnamefont {Iacopi}},\ }\bibfield  {title} {\bibinfo
  {title} {Microresonators with q-factors over a million from highly stressed
  epitaxial silicon carbide on silicon},\ }\href@noop {} {\bibfield  {journal}
  {\bibinfo  {journal} {Applied Physics Letters}\ }\textbf {\bibinfo {volume}
  {104}},\ \bibinfo {pages} {081901} (\bibinfo {year} {2014})}\BibitemShut
  {NoStop}%
\bibitem [{\citenamefont {Romero}\ \emph {et~al.}(2020)\citenamefont {Romero},
  \citenamefont {Valenzuela}, \citenamefont {Kermany}, \citenamefont
  {Sementilli}, \citenamefont {Iacopi},\ and\ \citenamefont
  {Bowen}}]{romero2020engineering}%
  \BibitemOpen
  \bibfield  {author} {\bibinfo {author} {\bibfnamefont {E.}~\bibnamefont
  {Romero}}, \bibinfo {author} {\bibfnamefont {V.~M.}\ \bibnamefont
  {Valenzuela}}, \bibinfo {author} {\bibfnamefont {A.~R.}\ \bibnamefont
  {Kermany}}, \bibinfo {author} {\bibfnamefont {L.}~\bibnamefont {Sementilli}},
  \bibinfo {author} {\bibfnamefont {F.}~\bibnamefont {Iacopi}},\ and\ \bibinfo
  {author} {\bibfnamefont {W.~P.}\ \bibnamefont {Bowen}},\ }\bibfield  {title}
  {\bibinfo {title} {Engineering the dissipation of crystalline micromechanical
  resonators},\ }\href@noop {} {\bibfield  {journal} {\bibinfo  {journal}
  {Physical Review Applied}\ }\textbf {\bibinfo {volume} {13}},\ \bibinfo
  {pages} {044007} (\bibinfo {year} {2020})}\BibitemShut {NoStop}%
\bibitem [{\citenamefont {B{\"u}ckle}\ \emph {et~al.}(2018)\citenamefont
  {B{\"u}ckle}, \citenamefont {Hauber}, \citenamefont {Cole}, \citenamefont
  {G{\"a}rtner}, \citenamefont {Zeimer}, \citenamefont {Grenzer},\ and\
  \citenamefont {Weig}}]{buckle2018stress}%
  \BibitemOpen
  \bibfield  {author} {\bibinfo {author} {\bibfnamefont {M.}~\bibnamefont
  {B{\"u}ckle}}, \bibinfo {author} {\bibfnamefont {V.~C.}\ \bibnamefont
  {Hauber}}, \bibinfo {author} {\bibfnamefont {G.~D.}\ \bibnamefont {Cole}},
  \bibinfo {author} {\bibfnamefont {C.}~\bibnamefont {G{\"a}rtner}}, \bibinfo
  {author} {\bibfnamefont {U.}~\bibnamefont {Zeimer}}, \bibinfo {author}
  {\bibfnamefont {J.}~\bibnamefont {Grenzer}},\ and\ \bibinfo {author}
  {\bibfnamefont {E.~M.}\ \bibnamefont {Weig}},\ }\bibfield  {title} {\bibinfo
  {title} {Stress control of tensile-strained in1- x ga x p nanomechanical
  string resonators},\ }\href@noop {} {\bibfield  {journal} {\bibinfo
  {journal} {Applied Physics Letters}\ }\textbf {\bibinfo {volume} {113}},\
  \bibinfo {pages} {201903} (\bibinfo {year} {2018})}\BibitemShut {NoStop}%
\bibitem [{\citenamefont {Cumming}\ \emph {et~al.}(2020)\citenamefont
  {Cumming}, \citenamefont {Sorazu}, \citenamefont {Daw}, \citenamefont
  {Hammond}, \citenamefont {Hough}, \citenamefont {Jones}, \citenamefont
  {Martin}, \citenamefont {Rowan}, \citenamefont {Strain},\ and\ \citenamefont
  {Williams}}]{cumming2020lowest}%
  \BibitemOpen
  \bibfield  {author} {\bibinfo {author} {\bibfnamefont {A.}~\bibnamefont
  {Cumming}}, \bibinfo {author} {\bibfnamefont {B.}~\bibnamefont {Sorazu}},
  \bibinfo {author} {\bibfnamefont {E.}~\bibnamefont {Daw}}, \bibinfo {author}
  {\bibfnamefont {G.}~\bibnamefont {Hammond}}, \bibinfo {author} {\bibfnamefont
  {J.}~\bibnamefont {Hough}}, \bibinfo {author} {\bibfnamefont
  {R.}~\bibnamefont {Jones}}, \bibinfo {author} {\bibfnamefont
  {I.}~\bibnamefont {Martin}}, \bibinfo {author} {\bibfnamefont
  {S.}~\bibnamefont {Rowan}}, \bibinfo {author} {\bibfnamefont
  {K.}~\bibnamefont {Strain}},\ and\ \bibinfo {author} {\bibfnamefont
  {D.}~\bibnamefont {Williams}},\ }\bibfield  {title} {\bibinfo {title} {Lowest
  observed surface and weld losses in fused silica fibres for gravitational
  wave detectors},\ }\href@noop {} {\bibfield  {journal} {\bibinfo  {journal}
  {Classical and Quantum Gravity}\ }\textbf {\bibinfo {volume} {37}},\ \bibinfo
  {pages} {195019} (\bibinfo {year} {2020})}\BibitemShut {NoStop}%
\bibitem [{\citenamefont {Beccari}\ \emph
  {et~al.}(2021{\natexlab{b}})\citenamefont {Beccari}, \citenamefont {Visani},
  \citenamefont {Fedorov}, \citenamefont {Bereyhi}, \citenamefont {Boureau},
  \citenamefont {Engelsen},\ and\ \citenamefont
  {Kippenberg}}]{beccari2021strained}%
  \BibitemOpen
  \bibfield  {author} {\bibinfo {author} {\bibfnamefont {A.}~\bibnamefont
  {Beccari}}, \bibinfo {author} {\bibfnamefont {D.~A.}\ \bibnamefont {Visani}},
  \bibinfo {author} {\bibfnamefont {S.~A.}\ \bibnamefont {Fedorov}}, \bibinfo
  {author} {\bibfnamefont {M.~J.}\ \bibnamefont {Bereyhi}}, \bibinfo {author}
  {\bibfnamefont {V.}~\bibnamefont {Boureau}}, \bibinfo {author} {\bibfnamefont
  {N.~J.}\ \bibnamefont {Engelsen}},\ and\ \bibinfo {author} {\bibfnamefont
  {T.~J.}\ \bibnamefont {Kippenberg}},\ }\bibfield  {title} {\bibinfo {title}
  {Strained crystalline nanomechanical resonators with ultralow dissipation},\
  }\href@noop {} {\bibfield  {journal} {\bibinfo  {journal} {arXiv preprint
  arXiv:2107.02124}\ } (\bibinfo {year} {2021}{\natexlab{b}})}\BibitemShut
  {NoStop}%
\bibitem [{\citenamefont {Tan}\ \emph {et~al.}(2017)\citenamefont {Tan},
  \citenamefont {Cai}, \citenamefont {Ng}, \citenamefont {Huang}, \citenamefont
  {Feng}, \citenamefont {Zhang}, \citenamefont {Zhang}, \citenamefont
  {Nijhuis}, \citenamefont {Liu},\ and\ \citenamefont
  {Ang}}]{Tan2017FewLayerBP}%
  \BibitemOpen
  \bibfield  {author} {\bibinfo {author} {\bibfnamefont {W.~C.}\ \bibnamefont
  {Tan}}, \bibinfo {author} {\bibfnamefont {Y.}~\bibnamefont {Cai}}, \bibinfo
  {author} {\bibfnamefont {R.~J.}\ \bibnamefont {Ng}}, \bibinfo {author}
  {\bibfnamefont {L.}~\bibnamefont {Huang}}, \bibinfo {author} {\bibfnamefont
  {X.}~\bibnamefont {Feng}}, \bibinfo {author} {\bibfnamefont {G.}~\bibnamefont
  {Zhang}}, \bibinfo {author} {\bibfnamefont {Y.-W.}\ \bibnamefont {Zhang}},
  \bibinfo {author} {\bibfnamefont {C.~A.}\ \bibnamefont {Nijhuis}}, \bibinfo
  {author} {\bibfnamefont {X.}~\bibnamefont {Liu}},\ and\ \bibinfo {author}
  {\bibfnamefont {K.}~\bibnamefont {Ang}},\ }\bibfield  {title} {\bibinfo
  {title} {Few-layer black phosphorus carbide field-effect transistor via
  carbon doping.},\ }\href@noop {} {\bibfield  {journal} {\bibinfo  {journal}
  {Advanced materials}\ }\textbf {\bibinfo {volume} {29}},\ \bibinfo {pages}
  {1700503} (\bibinfo {year} {2017})}\BibitemShut {NoStop}%
\bibitem [{\citenamefont {Kistanov}\ \emph {et~al.}(2021)\citenamefont
  {Kistanov}, \citenamefont {Nikitenko},\ and\ \citenamefont
  {Prezhdo}}]{Kistanov2020PointDI}%
  \BibitemOpen
  \bibfield  {author} {\bibinfo {author} {\bibfnamefont {A.~A.}\ \bibnamefont
  {Kistanov}}, \bibinfo {author} {\bibfnamefont {V.~R.}\ \bibnamefont
  {Nikitenko}},\ and\ \bibinfo {author} {\bibfnamefont {O.~V.}\ \bibnamefont
  {Prezhdo}},\ }\bibfield  {title} {\bibinfo {title} {Point defects in
  two-dimensional $\gamma$-phosphorus carbide.},\ }\href
  {https://doi.org/10.1021/acs.jpclett.0c03608} {\bibfield  {journal} {\bibinfo
   {journal} {The Journal of Physical Chemistry Letters}\ }\textbf {\bibinfo
  {volume} {12}},\ \bibinfo {pages} {620} (\bibinfo {year} {2021})}\BibitemShut
  {NoStop}%
\bibitem [{\citenamefont {Read}\ \emph {et~al.}(2001)\citenamefont {Read},
  \citenamefont {Cheng}, \citenamefont {Keller},\ and\ \citenamefont
  {McColskey}}]{read2001tensile}%
  \BibitemOpen
  \bibfield  {author} {\bibinfo {author} {\bibfnamefont {D.~T.}\ \bibnamefont
  {Read}}, \bibinfo {author} {\bibfnamefont {Y.-W.}\ \bibnamefont {Cheng}},
  \bibinfo {author} {\bibfnamefont {R.~R.}\ \bibnamefont {Keller}},\ and\
  \bibinfo {author} {\bibfnamefont {J.~D.}\ \bibnamefont {McColskey}},\
  }\bibfield  {title} {\bibinfo {title} {Tensile properties of free-standing
  aluminum thin films},\ }\href@noop {} {\bibfield  {journal} {\bibinfo
  {journal} {Scripta Materialia}\ }\textbf {\bibinfo {volume} {45}},\ \bibinfo
  {pages} {583} (\bibinfo {year} {2001})}\BibitemShut {NoStop}%
\bibitem [{\citenamefont {Nahar}\ \emph {et~al.}(2017)\citenamefont {Nahar},
  \citenamefont {Rocklein}, \citenamefont {Andreas}, \citenamefont {Funston},\
  and\ \citenamefont {Goodner}}]{nahar2017stress}%
  \BibitemOpen
  \bibfield  {author} {\bibinfo {author} {\bibfnamefont {M.}~\bibnamefont
  {Nahar}}, \bibinfo {author} {\bibfnamefont {N.}~\bibnamefont {Rocklein}},
  \bibinfo {author} {\bibfnamefont {M.}~\bibnamefont {Andreas}}, \bibinfo
  {author} {\bibfnamefont {G.}~\bibnamefont {Funston}},\ and\ \bibinfo {author}
  {\bibfnamefont {D.}~\bibnamefont {Goodner}},\ }\bibfield  {title} {\bibinfo
  {title} {Stress modulation of titanium nitride thin films deposited using
  atomic layer deposition},\ }\href@noop {} {\bibfield  {journal} {\bibinfo
  {journal} {Journal of Vacuum Science \& Technology A: Vacuum, Surfaces, and
  Films}\ }\textbf {\bibinfo {volume} {35}},\ \bibinfo {pages} {01B144}
  (\bibinfo {year} {2017})}\BibitemShut {NoStop}%
\bibitem [{\citenamefont {Pluchar}\ \emph {et~al.}(2020)\citenamefont
  {Pluchar}, \citenamefont {Agrawal}, \citenamefont {Schenk},\ and\
  \citenamefont {Wilson}}]{Pluchar:20}%
  \BibitemOpen
  \bibfield  {author} {\bibinfo {author} {\bibfnamefont {C.~M.}\ \bibnamefont
  {Pluchar}}, \bibinfo {author} {\bibfnamefont {A.~R.}\ \bibnamefont
  {Agrawal}}, \bibinfo {author} {\bibfnamefont {E.}~\bibnamefont {Schenk}},\
  and\ \bibinfo {author} {\bibfnamefont {D.~J.}\ \bibnamefont {Wilson}},\
  }\bibfield  {title} {\bibinfo {title} {Towards cavity-free ground-state
  cooling of an acoustic-frequency silicon nitride membrane},\ }\href
  {https://doi.org/10.1364/AO.394388} {\bibfield  {journal} {\bibinfo
  {journal} {Applied Optics}\ }\textbf {\bibinfo {volume} {59}},\ \bibinfo
  {pages} {G107} (\bibinfo {year} {2020})}\BibitemShut {NoStop}%
\bibitem [{\citenamefont {Rossi}\ \emph {et~al.}(2018)\citenamefont {Rossi},
  \citenamefont {Mason}, \citenamefont {Chen}, \citenamefont {Tsaturyan},\ and\
  \citenamefont {Schliesser}}]{rossi2018measurement}%
  \BibitemOpen
  \bibfield  {author} {\bibinfo {author} {\bibfnamefont {M.}~\bibnamefont
  {Rossi}}, \bibinfo {author} {\bibfnamefont {D.}~\bibnamefont {Mason}},
  \bibinfo {author} {\bibfnamefont {J.}~\bibnamefont {Chen}}, \bibinfo {author}
  {\bibfnamefont {Y.}~\bibnamefont {Tsaturyan}},\ and\ \bibinfo {author}
  {\bibfnamefont {A.}~\bibnamefont {Schliesser}},\ }\bibfield  {title}
  {\bibinfo {title} {Measurement-based quantum control of mechanical motion},\
  }\href@noop {} {\bibfield  {journal} {\bibinfo  {journal} {Nature}\ }\textbf
  {\bibinfo {volume} {563}},\ \bibinfo {pages} {53} (\bibinfo {year}
  {2018})}\BibitemShut {NoStop}%
\bibitem [{\citenamefont {Norte}\ \emph {et~al.}(2018)\citenamefont {Norte},
  \citenamefont {Forsch}, \citenamefont {Wallucks}, \citenamefont
  {Marinkovi\ifmmode~\acute{c}\else \'{c}\fi{}},\ and\ \citenamefont
  {Gr\"oblacher}}]{PhysRevLett.121.030405}%
  \BibitemOpen
  \bibfield  {author} {\bibinfo {author} {\bibfnamefont {R.~A.}\ \bibnamefont
  {Norte}}, \bibinfo {author} {\bibfnamefont {M.}~\bibnamefont {Forsch}},
  \bibinfo {author} {\bibfnamefont {A.}~\bibnamefont {Wallucks}}, \bibinfo
  {author} {\bibfnamefont {I.}~\bibnamefont {Marinkovi\ifmmode~\acute{c}\else
  \'{c}\fi{}}},\ and\ \bibinfo {author} {\bibfnamefont {S.}~\bibnamefont
  {Gr\"oblacher}},\ }\bibfield  {title} {\bibinfo {title} {Platform for
  measurements of the casimir force between two superconductors},\ }\href
  {https://doi.org/10.1103/PhysRevLett.121.030405} {\bibfield  {journal}
  {\bibinfo  {journal} {Physical Review Letters}\ }\textbf {\bibinfo {volume}
  {121}},\ \bibinfo {pages} {030405} (\bibinfo {year} {2018})}\BibitemShut
  {NoStop}%
\end{thebibliography}%

\setcounter{figure}{0}
\renewcommand{\thefigure}{S\arabic{figure}}
\setcounter{equation}{0}
\renewcommand{\theequation}{S\arabic{equation}}
\setcounter{table}{0}
\renewcommand{\thetable}{S\arabic{table}}

\newpage

\clearpage

\begin{widetext} 

\section*{Supporting information}

\section*{Lessons from string resonators}

Generally, room temperature resonators with state-of-the-art quality factors have converged towards shared similarities in their geometry, material properties and mechanical modes of interest. To date, they all exhibit some of the highest aspect-ratios in microchip technology with free-standing structures nanometers thick but laterally suspended over several millimeters. As discussed in the main text, increasing the lateral size ($L$) of the resonator with respect to its thickness in the direction of motion ($t$) improves $Q_\mathrm{m}$ for beam and membrane nanomechanical resonators since generally $Q_\mathrm{m} \propto L/t$. Another common attribute of all high $Q_\mathrm{m}$ resonators is the use of highly stretched silicon nitride (Si\textsubscript{3}N\textsubscript{4}) to fabricate these high-aspect-ratio nanostructures. A resonators' $Q_\mathrm{m}$ is defined as $2\pi W/\Delta W$ where $W$ is the total stored energy of the resonator, and $\Delta W$ corresponds to the energy dissipated for each vibration cycle. Fabricating resonators from tensile materials enables extremely low mechanical dissipation ($\Delta W$) while increasing the resonator's stored mechanical energy $W$ as a consequence of geometric strain engineering  \cite{fedorov2019generalized}. The large initial stress before the vibration increases the potential energy dominantly, whereas the energy loss for each cycle is determined by the bending of the resonator's vibration mode. It means that the critical design parameters for the nanomechanical resonator's $Q_\mathrm{m}$ are the strain (stress) distribution after releasing the initial pre-stress and the vibration mode shape. The analytical formulation for a one-dimensional beam's bending $f_\mathrm{m}$ and the corresponding $Q_\mathrm{m}$ follows the equations below \cite{schmid2011damping}.

\vspace{5mm}

\begin{equation} \label{eq_1D_beam_f} \tag{S1}
f_\mathrm{m}=\frac{n}{2L}\sqrt{\frac{\sigma}{\rho}}\sqrt{1+n^2\pi^2\lambda^2}
\end{equation}

\vspace{5mm}

\begin{equation} \label{eq_1D_beam_Q} \tag{S2}
Q_\mathrm{m} \approx (2\lambda+\pi^2n^2\lambda^2)^{-1} Q_0 
\end{equation}

\vspace{5mm}

\noindent $Q_0$ is the intrinsic quality factor defined as $Q_0^{-1}=Q_\mathrm{volume}^{-1}+Q_\mathrm{surface}^{-1}$, where $Q_\mathrm{volume}$ is the bulk material loss of Si\textsubscript{3}N\textsubscript{4} and $Q_\mathrm{surface}$ is the surface loss, linear to the resonator's thickness. For thin resonators operating at room temperature, we assume $Q_0$ $\approx$ 6900 $t$/100 nm \cite{villanueva2014evidence}. $n$, $\sigma$, $\rho$ are bending mode number, initial stress and density, respectively. $\lambda$ is $\sqrt{E/12\sigma} t/L$ where $E$ is the Young's modulus. Note that the dominant factor $2\lambda$ in eq. \ref{eq_1D_beam_Q} happens because of the sharp curvature change near the boundaries around a narrow length $L_c=\sqrt{2}L\lambda$ \cite{schmid2011damping}, which reduces the quality factor dominantly. A similar formulation for the membrane structure could be found in \cite{yu2012control}. \textbf{Figure 1} in the main text shows the $Q_\mathrm{m}$ versus vibration frequency ($f_\mathrm{m}$) graph for a 50 nm thick 3 mm Si\textsubscript{3}N\textsubscript{4} beam considering 1.07 GPa initial stress, when we vary the mode number. As it can be seen in it, the pre-stressed double clamped beam shows a significant improvement of $Q_\mathrm{m}$ compared to the unstressed beam's intrinsic quality factor $Q_0$. Also, the perfect soft clamping assumption, which excludes the clamping loss part in eq. \ref{eq_1D_beam_Q}, emphasizes the major loss of the quality factor from the sharp curvature change in the boundary, especially at lower order modes. Finally, all of these Si\textsubscript{3}N\textsubscript{4} resonators have relied on mechanical modes which have a highest amplitude of oscillation in the resonator center and where most of the induced losses come from how the mechanical mode connects to the microchip substrate where bending is highest. Given that the dominant channel of mechanical resonator loss comes from bending losses, until now most of the design strategies have typically consisted of engineering the decay of the resonator's mechanical mode to reduce bending in the Si\textsubscript{3}N\textsubscript{4} resonators near the clamping points via hierarchical \cite{fedorov2020fractal,beccari2021hierarchical} structures or phononic crystal designs \cite{ghadimi2017radiation,ghadimi2018elastic}.

\section*{Derivation of the quality factor for two-dimensional structures}

The formulation here is the generalized form of the uniform membrane equations from Yu and co-workers \cite{yu2012control}. Considering the out-of-plane deformation as the displacement field corresponding to the vibration mode shape, the strain can be derived from the generalized Euler-Bernoulli equation. Note that we calculate the strain assuming plane stress, constant deformation at the center of the plate ${u}_z$, as well as constant thickness. The oscillation of the plate $u_z(x,y)e^{i2{\pi}{f_\mathrm{m}}t}$ induces strain in the plate due to bending and elongation during the vibration.

\begin{equation} \label{eq_strain_plane_stress} \tag{S3}
 \begin{Bmatrix} 
 \tilde{\epsilon}_{xx} \\ 
 \tilde{\epsilon}_{yy} \\ 
 2\tilde{\epsilon}_{xy} 
 \end{Bmatrix}
 =
 -
 \underbrace{
 {ze^{2i\pi{f_\mathrm{m}}t}}
 \begin{Bmatrix} 
 u_{z,xx} \\ 
 u_{z,yy} \\ 
 2u_{z,xy} 
 \end{Bmatrix}}
 _{bending}
+ 
\underbrace{
 \frac{e^{4i\pi{f_\mathrm{m}}t}}{2}
 \begin{Bmatrix} 
 u_{z,x}^2 \\ 
 u_{z,y}^2 \\ 
 2u_{z,x}u_{z,y} 
 \end{Bmatrix}}
 _{elongation}
\end{equation}

Then, the corresponding variation of stress is derived from generalized Hooke's law with the complex Young's modulus $\tilde{E}$=$E_1$+$iE_2$.

\begin{equation} \label{eq_stress_plane_stress} \tag{S4}
 \begin{Bmatrix} 
 \tilde{\sigma}_{xx} \\ 
 \tilde{\sigma}_{yy} \\ 
 \tilde{\sigma}_{xy} 
 \end{Bmatrix}
 =
 \frac{\tilde{E}}{1-\nu^2}
  \begin{bmatrix}
  1   & \nu & 0 \\
  \nu & 1   & 0 \\
  0   & 0   & (1-\nu)/{2} \\
  \end{bmatrix}
\begin{Bmatrix} 
\tilde{\epsilon}_{xx} \\ 
\tilde{\epsilon}_{yy} \\ 
2\tilde{\epsilon}_{xy} 
\end{Bmatrix}   
\end{equation}

The mechanical work done per oscillation, which corresponds to the energy loss, can be derived as

\begin{equation} \label{energy_loss} \tag{S5}
 \Delta U=\int_{Volume}
 \int_0^{1/{f_\mathrm{m}}} 
   \Re(\tilde{\sigma}_{xx})\Re(\tilde{\epsilon}_{xx,t})
  +\Re(\tilde{\sigma}_{yy})\Re(\tilde{\epsilon}_{yy,t})
 +2\Re(\tilde{\sigma}_{xy})\Re(\tilde{\epsilon}_{xy,t})dt dV
\end{equation}

Inserting eq. \ref{eq_strain_plane_stress} and eq. \ref{eq_stress_plane_stress} into eq. \ref{energy_loss} and by ignoring the elongation term which is a few orders of magnitude lower than the other terms for the small deformation regime, eq. \ref{energy_loss} can be summarized as

\begin{equation} \label{energy_loss_modified} \tag{S6}
 \begin{split}
 \Delta U &
 =\iiint
 \frac{\pi E_2}{1-\nu^2}((u_{z,xx}+u_{z,yy})^2z^2-2(1-\nu)(u_{z,xx}u_{z,yy}-u_{z,xy}^2)z^2) dV\\
 &
 =\frac{\pi E_2 t^3}{12(1-\nu^2)}
 \iint (u_{z,xx}^2+u_{z,yy}^2+2\nu u_{z,xx}u_{z,yy}+2(1-\nu)u_{z,xy}^2) dS
 \end{split}
\end{equation}

The system's stored energy can be obtained by calculating the maximum kinetic energy or the maximum elastic energy. Between the two, in this research, we considered the maximum elastic energy of the stretched structure due to bending and elongation. The work done due to elongation happens to be a few orders of magnitude higher than the bending energy, so we ignore the bending energy here. Therefore,

\begin{equation} \label{energy_stored} \tag{S7}
 \begin{split}
U &
 = \iiint 
 ({\frac{1}{2}} u^2_{z,x} \sigma_{xx}+ \frac{1}{2}u^2_{z,y}{\sigma}_{yy}+u_{z,x}u_{z,y}{\sigma}_{xy}) dV \\
&
 =\frac{t}{2} \iint 
 ({\sigma}_{xx} u^2_{z,x}+{\sigma}_{yy} u^2_{z,y}+2{\sigma}_{xy} u_{z,x}u_{z,y}) dS
 \end{split}
\end{equation}

From this result we obtain $2\pi U/\Delta U$:

\begin{equation} \label{Q_m_general} \tag{S8}
Q_\mathrm{m} =
\frac{12(1-\nu^2)}{E_2 t^2}
\frac
{\iint ({\sigma}_{xx} u^2_{z,x}+{\sigma}_{yy} u^2_{z,y}+2{\sigma}_{xy} u_{z,x}u_{z,y}) dS}
{\iint (u_{z,xx}^2+u_{z,yy}^2+2\nu u_{z,xx}u_{z,yy}+2(1-\nu)u_{z,xy}^2) dS} 
\end{equation}

Leading to eqs. (1)-(3) in the main text, when $Q_{0}$=$E/E_2$ considering $E_1 \gg E_2$.

\section*{Bayesian optimization}

\begin{figure}[htb]
    \centering
    \includegraphics[width=0.95\linewidth]{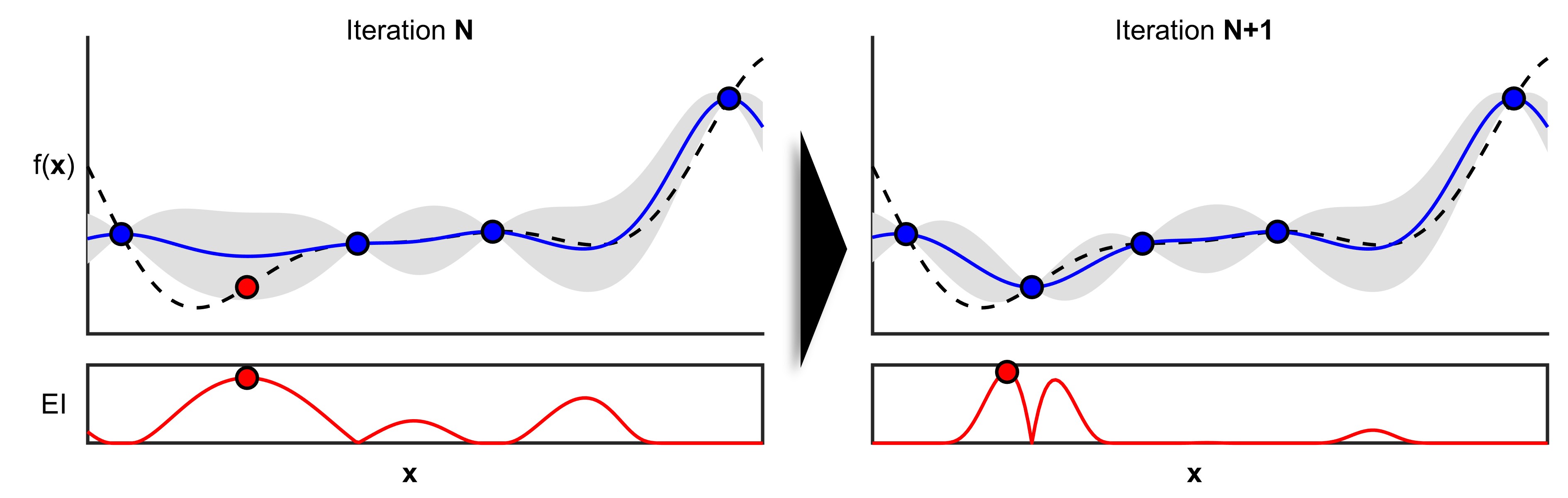}
    \caption{An overview of Bayesian optimization with a one-dimensional minimization example. A Gaussian process regression model is fitted to the observed data (blue markers) and its acquisition function is maximized to select the design parameter (red markers) for the next iteration. The dotted black line corresponds to the unknown black-box function and the solid blue line corresponds to the regression result based on Gaussian process regression. The regression model is plotted as a posterior mean with shaded areas representing 95 percent confidence interval.}
    \label{fig_S1}
\end{figure}

Handling simulation-based optimization of spider web resonators requires the optimizer to use simulations with high computational cost. Bayesian optimization \cite{shahriari2015taking,pelikan1999boa} is a global optimizer that is expected to avoid many local solutions associated to different modes of vibration. Using an online machine learning approach, this method performs optimization while updating limited information in the design space. If the response surface of $Q_\mathrm{m}$ is unknown, we can train the model with scarce information, starting with a few initial observations and add additional feature evaluations sequentially. The goal is to optimize and track optimal design parameters using several evaluations of the finite element model of the spider web resonator. As shown in \textbf{Figure \ref{fig_S1}}, Bayesian optimization uses Gaussian process regression (solid blue line) to approximate the unknown response of the function (dashed line) at each iteration. With the obtained data and the probability distribution taking into account the variance of the unmeasured region, the optimizer uses an acquisition function that determines the design variables for the next iteration. Here we used Expected Improvement (EI) as the most commonly used acquisition function. It estimates where the most considerable improvement over the current best results will be so that both exploitation for local area search as well as the exploration for global optimization could be performed simultaneously. Compared to other optimization algorithms, additional computational resources are required to determine the next optimization point. However, when considering expensive function calls and small or medium design dimensions as in our study, they are negligible compared to the finite element simulation time. In this study, the Python library GPyOpt \cite{gpyopt2016} was used. A more detailed information about the method can be found at \cite{shahriari2015taking}.

\section*{Optimization convergence dependency on the initial random points}

\begin{figure}[h]
  \centering
    \includegraphics[width=0.9\textwidth]{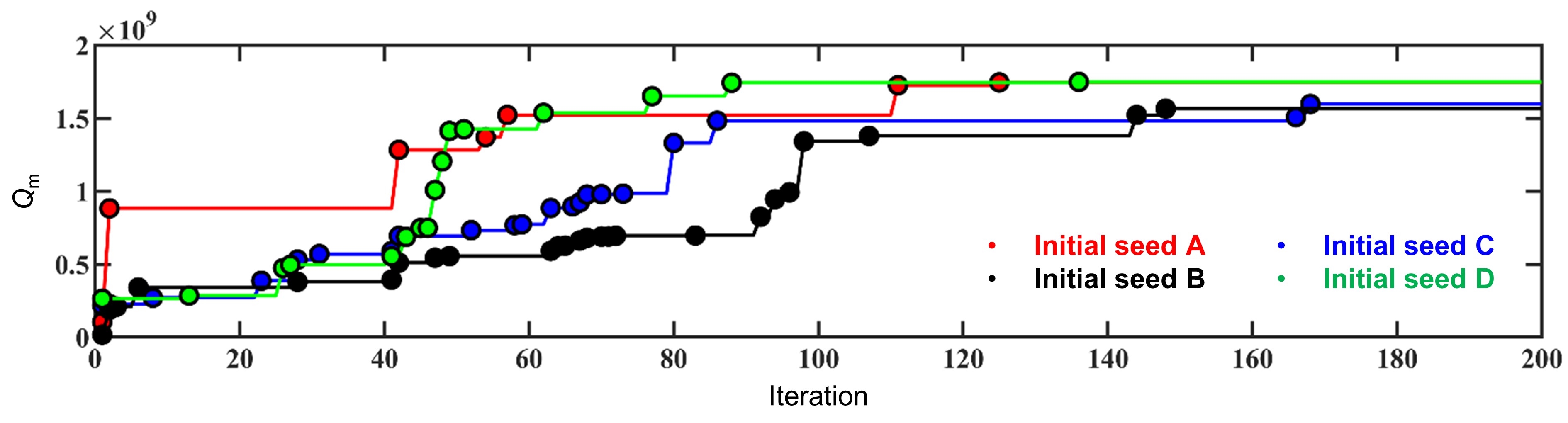}
  \caption{Iteration history of the web-like nanomechanical resonator for optimizing $Q_\mathrm{m}$ with four different randomly selected initial points. The initial seed D is the result discussed in the main text.}
  \label{fig_S2}
\end{figure}

The initial design of experiments affects the convergence to the optimum solution. Since our problem is dealing with global optimization, the curse of dimensionality is a crucial issue to be addressed. Even though we use Bayesian optimization to reduce the total number of design evaluations, the number of initial points ($=$40) is small compared to the design space (six design parameters). The result in \textbf{Figure \ref{fig_S2}} shows the optimization history considering four different sets of randomly selected initial points. The figure shows that changing the initial points affects the convergence speed, as expected. However, the optimized design obtained for all the four cases exhibits the same vibration mechanism described in the main text with a similar range of quality factors. 

\begin{table}[h]
\centering
 \caption{Optimal design parameters and the corresponding $f_\mathrm{m}$, $Q_\mathrm{m}$, and $S_F$ for different initial random points.}
  \label{tab_design_para_initial}
  \begin{tabular}{c c c c c c c c c c}
  \hline
  Design & $N_r$ & $d$(\textmu m) & $w_1$(\textmu m) & $w_2$(\textmu m) & $l_1$(mm) & $l_2$(mm) & $f_\mathrm{m}$(kHz) & $Q_\mathrm{m}$  & $S_F$(aN/$\sqrt{\text{Hz}}$)\\
  \hline
    Initial seed A & 4 & 1.01 & 1.00  & 2.00  & 1.14  & 1.48 & 141   & 1.75 $\cdot 10^9$ & 2.1   \\
    Initial seed B & 4 & 1.20 & 1.03  & 3.59  & 1.15  & 1.45 & 142  & 1.60 $\cdot 10^9$ & 2.2    \\
    Initial seed C & 4 & 1.16 & 1.11  & 3.30  & 1.09  & 1.46 & 145  & 1.56 $\cdot 10^9$ & 2.3    \\    
    Initial seed D & 4 & 1.05 & 1.00  & 2.46  & 1.21  & 1.48 & 135  & 1.75 $\cdot 10^9$ & 2.1    \\        
  \hline
 \end{tabular}
\end{table}

\section*{Conversion of energy loss regarding the thickness of the web-like resonator}

\begin{figure}[h]
  \centering
    \includegraphics[width=0.9\textwidth]{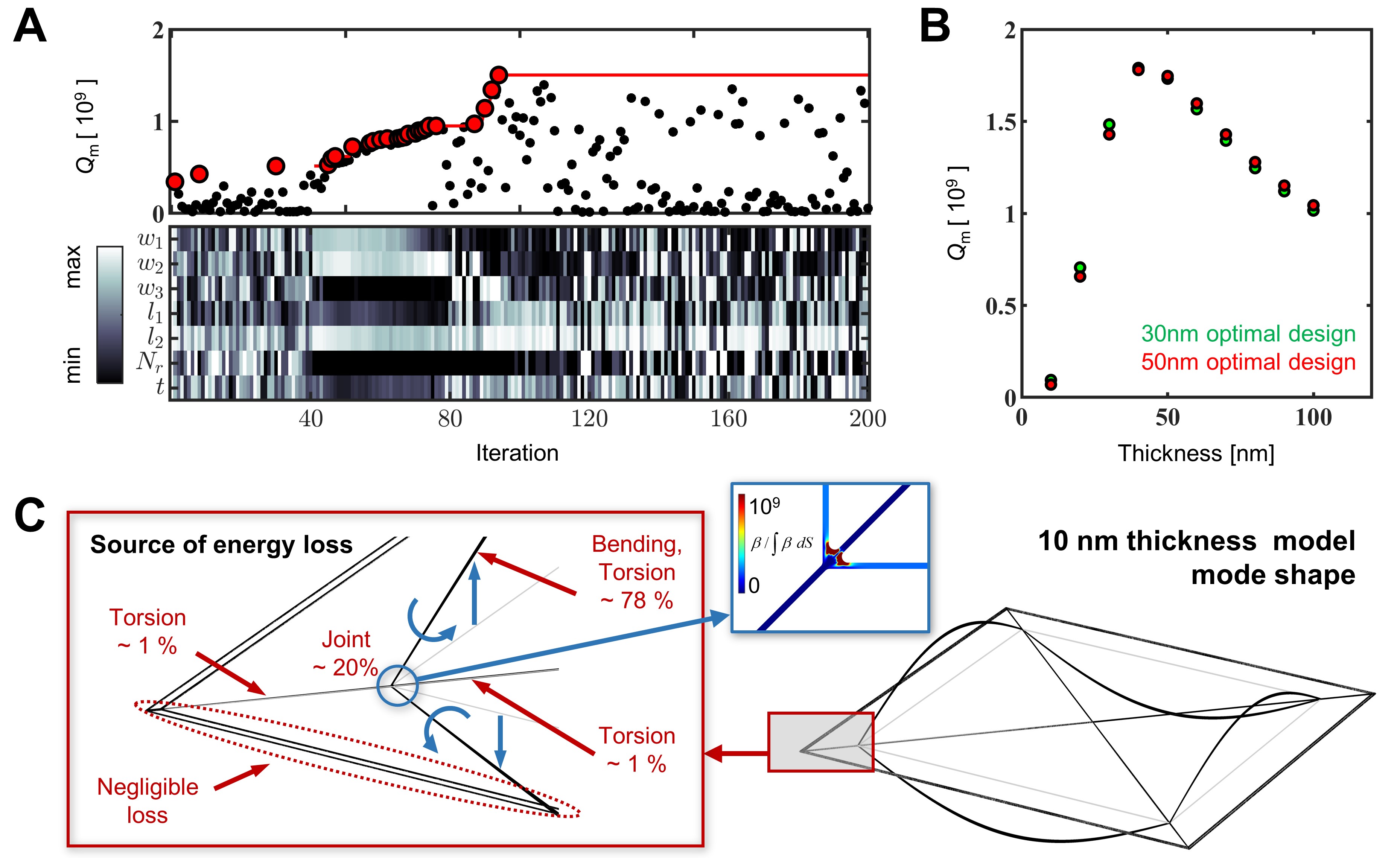}
  \caption{Overview of the thickness study for designing the spiderweb nanomechanical resonator. \textbf{A} Evolution of the quality factor $Q_\mathrm{m}$ and values of the seven design parameters at every iteration, including the thickness of the resonator. \textbf{B} Sweep of the thickness for two spiderweb nanomechanical resonators which are optimized for 30 nm and 50 nm, respectively. 50 nm design corresponds to the optimal result in \textbf{Figure 4} in the main text. \textbf{C} The full motion of the vibration mode shape and an illustration of the portion of energy loss of a 10 nm thickness spiderweb nanomechanical resonator. Other design parameters refer to \textbf{Table 1} in the main text. }
  \label{fig_S3}
\end{figure}

The optimization considering thickness as a design parameter has shown that minimizing the thickness was not giving the best quality factor, contrary to what is generally observed for straight strings (eq. \ref{eq_1D_beam_Q}), or phononic crystal resonators \cite{ghadimi2018elastic}. \textbf{Figure \ref{fig_S3}}\textbf{A} shows the optimization iteration history with the same conditions of the main text except considering the thickness as an additional design parameter. The optimized result converged at around 60 nm (local optimum), which differs from the general trend of minimizing the resonator thickness. To study more in details this effect, we performed two optimizations constraining the resonator thickness to 30 nm and 50 nm, respectively. Afterward we swept the thickness of the two obtained designs from 10 nm to 100 nm. \textbf{Figure \ref{fig_S3}}\textbf{B} shows the trend. As usual, the quality factor increases for thinner resonators when the thickness is larger than 50 nm. However, below 40 nm, it starts to decrease sharply. The reason of this effect can be understood from \textbf{Figure \ref{fig_S3}}\textbf{C}, which shows the same design in the main text \textbf{Figure 4}, but modifying the thickness to 10 nm for comparison. As shown in the figure, most of the energy loss starts to focus on the lateral vibration beam, especially because the torsional motion of the lateral beam becomes significant. Furthermore, unlike the thicker designs, the joint region of the inner lateral beams starts to have a concentration of bending energy which starts to keep a large portion of energy loss because the sharp curvature change occurs, even though the soft clamping mechanism isolates the resonator. Considering this motion, it is expected to improve the quality factor for thinner resonators when considering more challenging fabrication with a width below 1 \textmu m. Note that the research in the main text designed a 50 nm thickness spiderweb nanomechanical resonator considering our fabrication process.

\section*{Mechanical quality factor of the web-like resonator regarding the resonator size}

\begin{figure}[h]
  \centering
    \includegraphics[width=0.45\textwidth]{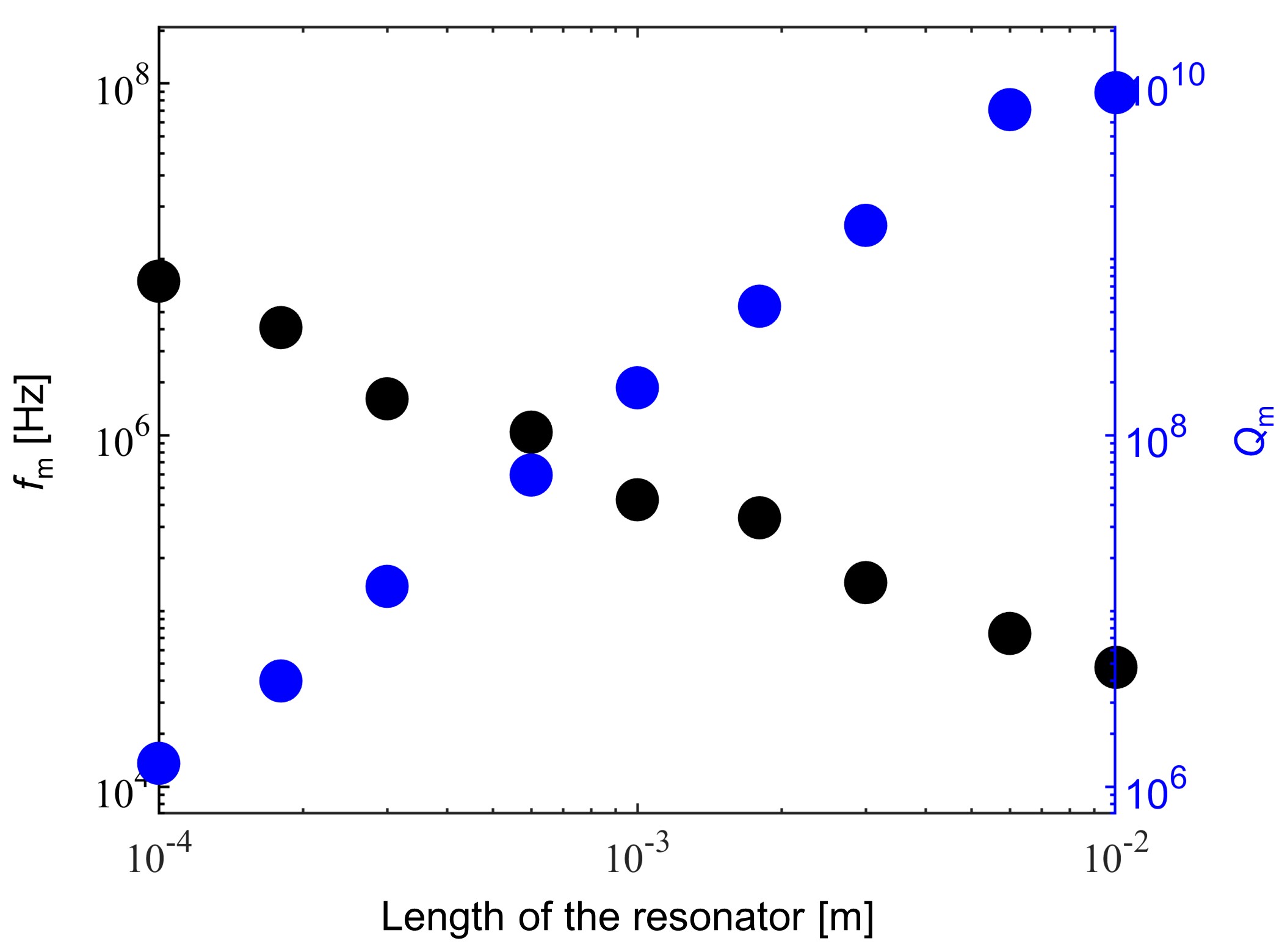}
  \caption{The mechanical quality factor and the vibration frequency for various sizes of optimized nanomechanical resonators.}
  \label{fig_S4}
\end{figure}

The length of the mechanical resonator has been discussed as an essential factor for the $f_\mathrm{m}$ and $Q_\mathrm{m}$. Following the trend of straight beams (eqs. \ref{eq_1D_beam_f} and \ref{eq_1D_beam_Q}), $Q_\mathrm{m}$ and $1/f_\mathrm{m}$ is linear to $L$ when $L$ $\gg$ $t$, and $Q_\mathrm{m}$ becomes quadratic to $L$ if we assume perfect soft clamping. To measure the length effects on the mechanical quality factor and the frequency of the spiderweb nanomechanical resonator, we optimized the quality factor of each design starting from 100 \textmu m to 10 mm length of the resonator. Other design parameters and constraints were considered the same as in the main text. As shown in \textbf{Figure \ref{fig_S4}}, increasing the resonator's length increased the quality factor quadratically, while the frequency was proportional to the inverse of the length. This result also supports that our new resonator follows the expected trend of the effect of soft clamping. Since soft clamping becomes more crucial for longer resonators, we expect to achieve a $Q_\mathrm{m}$ around 10 billion with a 1 cm web-like resonator.

\section*{Fundamental and center defect mode shape}

\begin{figure}[h]
    \centering
    \includegraphics[width=0.9\linewidth]{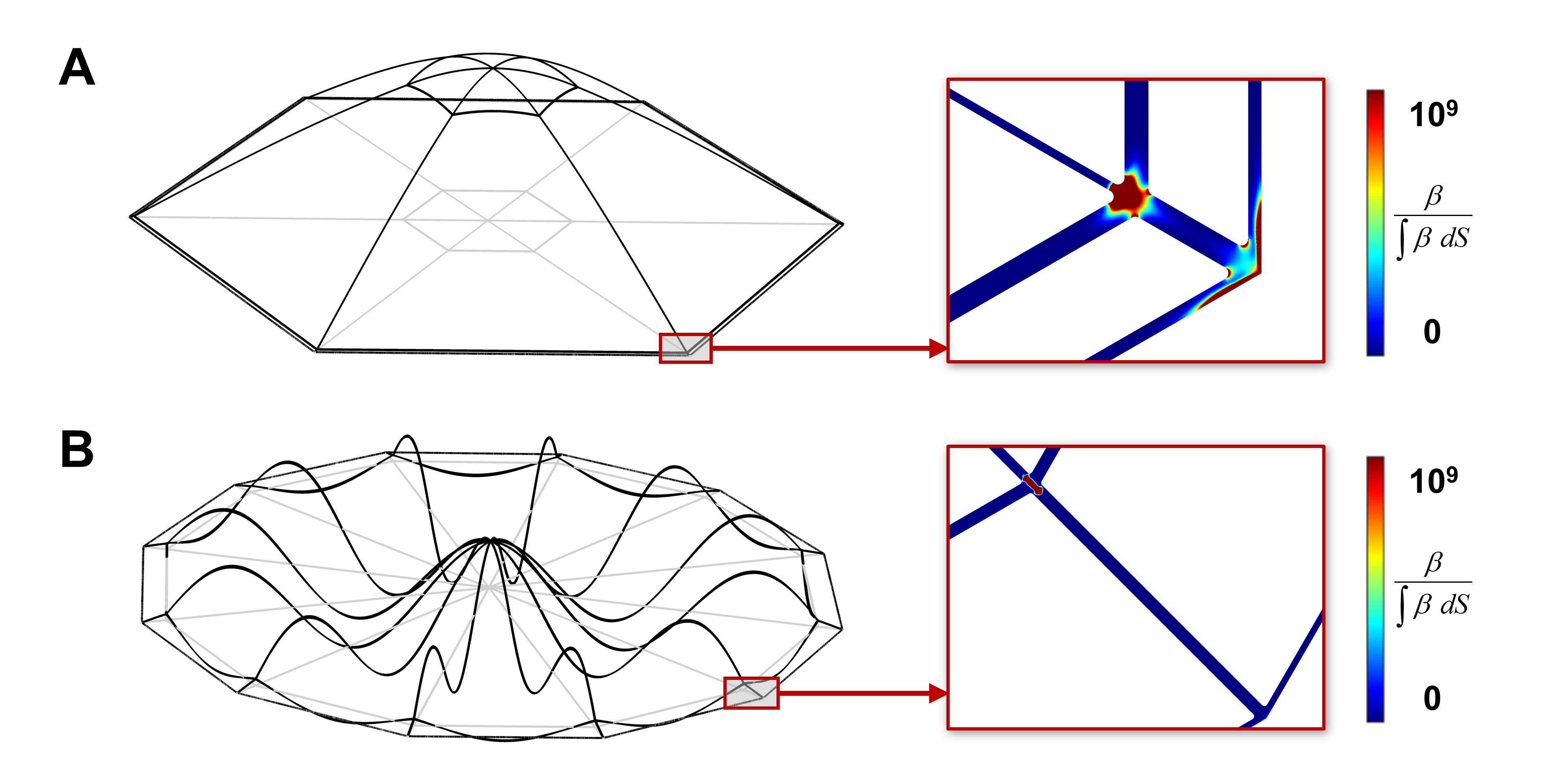}
    \caption{The optimal web-like soft clamping mode for \textbf{A} Design A and \textbf{B} Design B obtained from simulation. The normalized bending loss density around the edge of Design A and Design B shows that the bending loss is concentrated on the boundary and the joints between the lateral and radial beams.}
    \label{fig_S5}
\end{figure}

To analyze the effect of the target mode shapes on the resulting $Q_\mathrm{m}$, we performed two additional optimizations. The first (Design A) focuses on optimizing the $Q_\mathrm{m}$ of the fundamental vibration mode, while the second (Design B) aims to optimize the $Q_\mathrm{m}$ when the out-of-plane deformation at the center of the web is maximized. The optimal Design A, shown in \textbf{Figure \ref{fig_S5}}\textbf{A}, exhibits a small $d$ while pushing the outer lateral beam near the boundary with a large $w_2$. At the same time, the $w_1$ is nearly maximized with the inner lateral beam around the center of the resonator (see \textbf{Table \ref{tab_design_para_AB}}). The result shows a similar strategy to the trampoline design presented in \cite{norte2016mechanical}, using strain engineering. It maximizes the stress of the radial directional beam, which primarily vibrates in the fundamental mode by adding the mass around the center and the boundary. This suggests that the machine learning optimizer is capable of finding similar solutions to physics-based design approaches. 

The optimal Design B increases the number of radial beams (see \textbf{Table \ref{tab_design_para_AB}}), showing that we could achieve a higher $Q_\mathrm{m}$ at higher order modes, unlike uniform string \cite{schmid2011damping} or membrane \cite{yu2012control} resonators. The optimized result adds the lateral beams near the wave propagation node and couples the wave propagation in the radial direction to the lateral direction by the lateral beams' bending motion, as it can be seen in \textbf{Figure \ref{fig_S5}}\textbf{B}. This demonstrates that the optimizer finds soft clamping modes without any pre-information, by alternating the sharp curvature change near the boundary and distributing it to several joint deformations. Note that for both of the designs, the outer lateral beams distribute the bending loss concentrated at the boundaries, similarly to what was recently proposed by fractal \cite{fedorov2020fractal} and hierarchical \cite{beccari2021hierarchical} soft clamping. 

\begin{table}[h]
\centering
 \caption{Optimal design parameters and the corresponding $f_\mathrm{m}$, $Q_\mathrm{m}$, and $S_F$ for Design A and Design B.}
  \label{tab_design_para_AB}
  \begin{tabular}{c c c c c c c c c c}
  \hline
  Design & $N_r$ & $d$(\textmu m) & $w_1$(\textmu m) & $w_2$(\textmu m) & $l_1$(mm) & $l_2$(mm) & $f_\mathrm{m}$(kHz) & $Q_\mathrm{m}$  & $S_F$(aN/$\sqrt{\text{Hz}}$)\\
  \hline
    A & 6 & 1.23 & 3.95  & 3.68  & 0.35  & 1.48 & 63.3 & 45.3 $\cdot 10^6$ & 15.9   \\
    B & 12 & 3.18 & 2.89  & 3.96 & 0.01& 1.40 & 451.5  & 55.4 $\cdot 10^6$ & 60.8    \\
  \hline
 \end{tabular}
\end{table}

\section*{Evolution of the spiderweb nanomechanical resonators designs during the Bayesian  optimization process}

\textbf{Figure \ref{fig_S6}} shows an enlarged view of the design (black) and the vibrational modes (red) obtained at different iterations. The first two designs at iteration 26 and 27 are obtained during the random search phase and the mode is localized in the outer later beam. The following designs at iterations 41 and 51 result from the Bayesian optimization phase and show a localized mode in the inner lateral beam. The last two designs at iterations 78 and 97 show a clear example of the Bayesian optimization \textit{exploring}, resulting in vibration modes different from the previous iterations.

\begin{figure}[h]
  \centering
\includegraphics[width=0.9\textwidth]{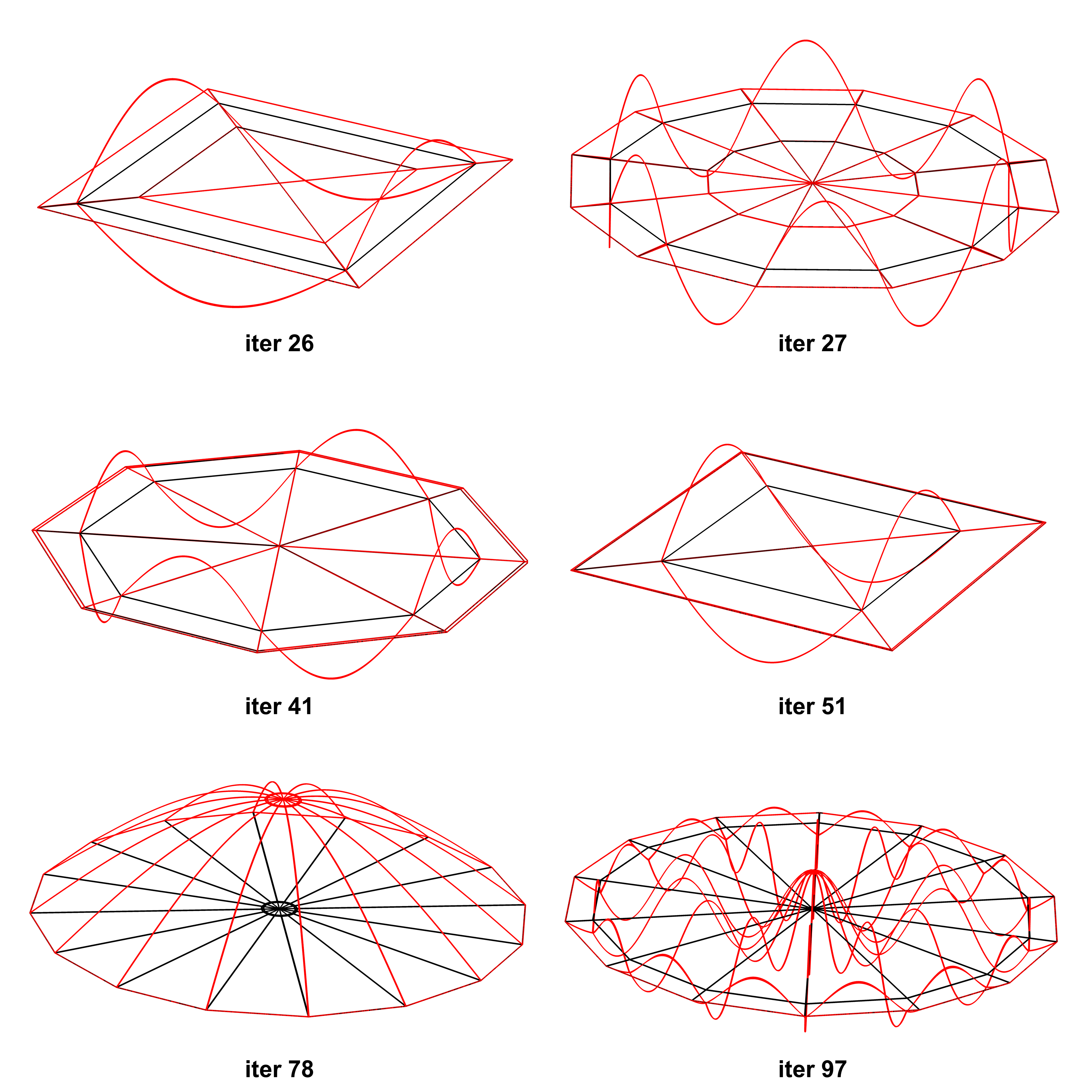}
  \caption{Design and simulated modes shapes at the iterations highlighted by circular markers in \textbf{Figure  3A} in the main text.}
  \label{fig_S6}
\end{figure}

\section*{SEM from tilted view}

\begin{figure}[h]
  \centering
    \includegraphics[width=0.9\textwidth]{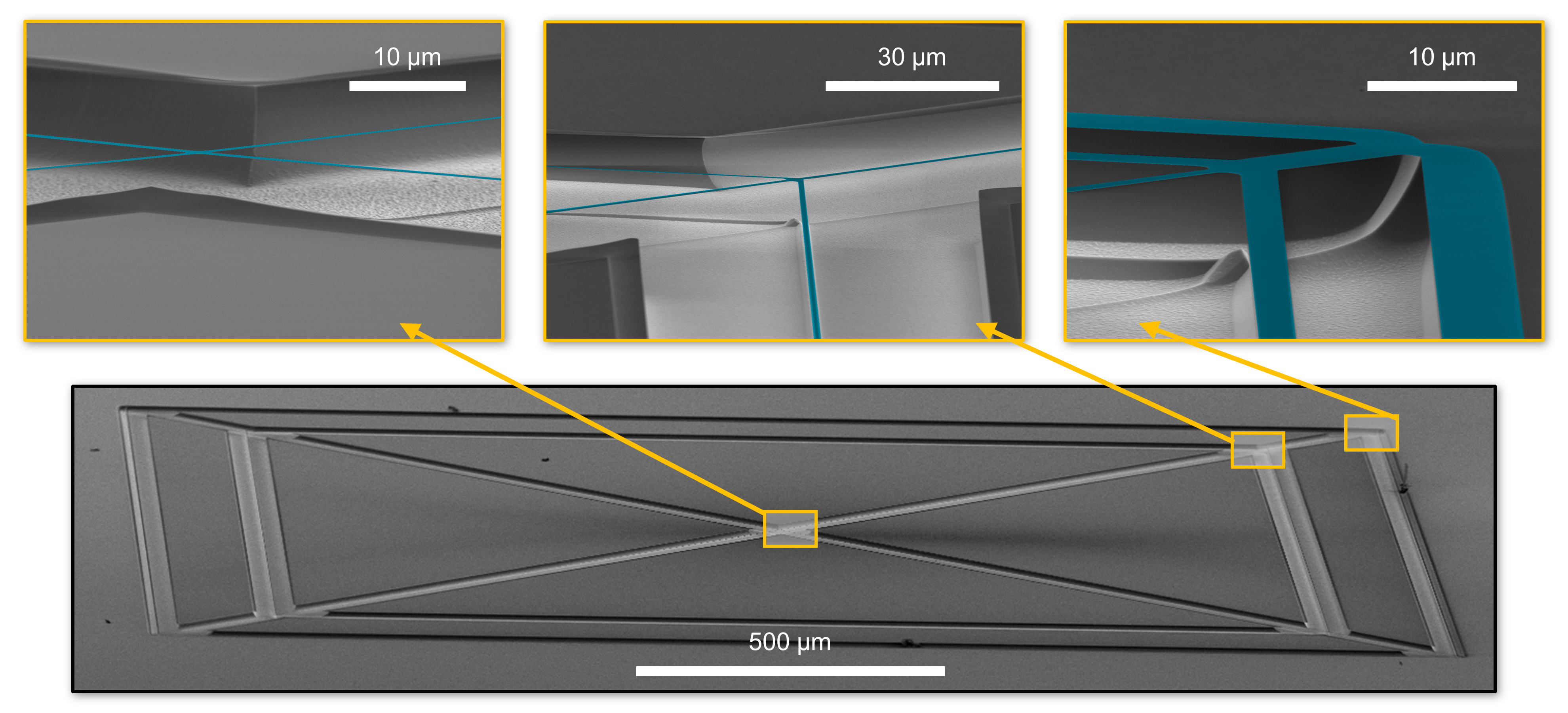}
  \caption{False colored scanning electron microscope images of the optimal design from tilted view. The 3 insets, from left to right, show a zoom at the center, at the joint and at the clamping point.}
  \label{fig_S7}
\end{figure}

The spiderweb nanomechanical resonator, fabricated in high stress silicon nitride, is suspended over the silicon substrate as shown in \textbf{Figure \ref{fig_S7}}. The release step of the suspended nanomechanical resonator, highlighted in blue, is performed from the top, isotropically etching the silicon underneath. It follows that the fixed boundary is etched creating an overhang, as in can be seen in the inset on the top right of \textbf{Figure \ref{fig_S7}}.

\section*{Photothermal effect on Q factor}

\begin{figure}[h]
  \centering
    \includegraphics[width=0.45\textwidth]{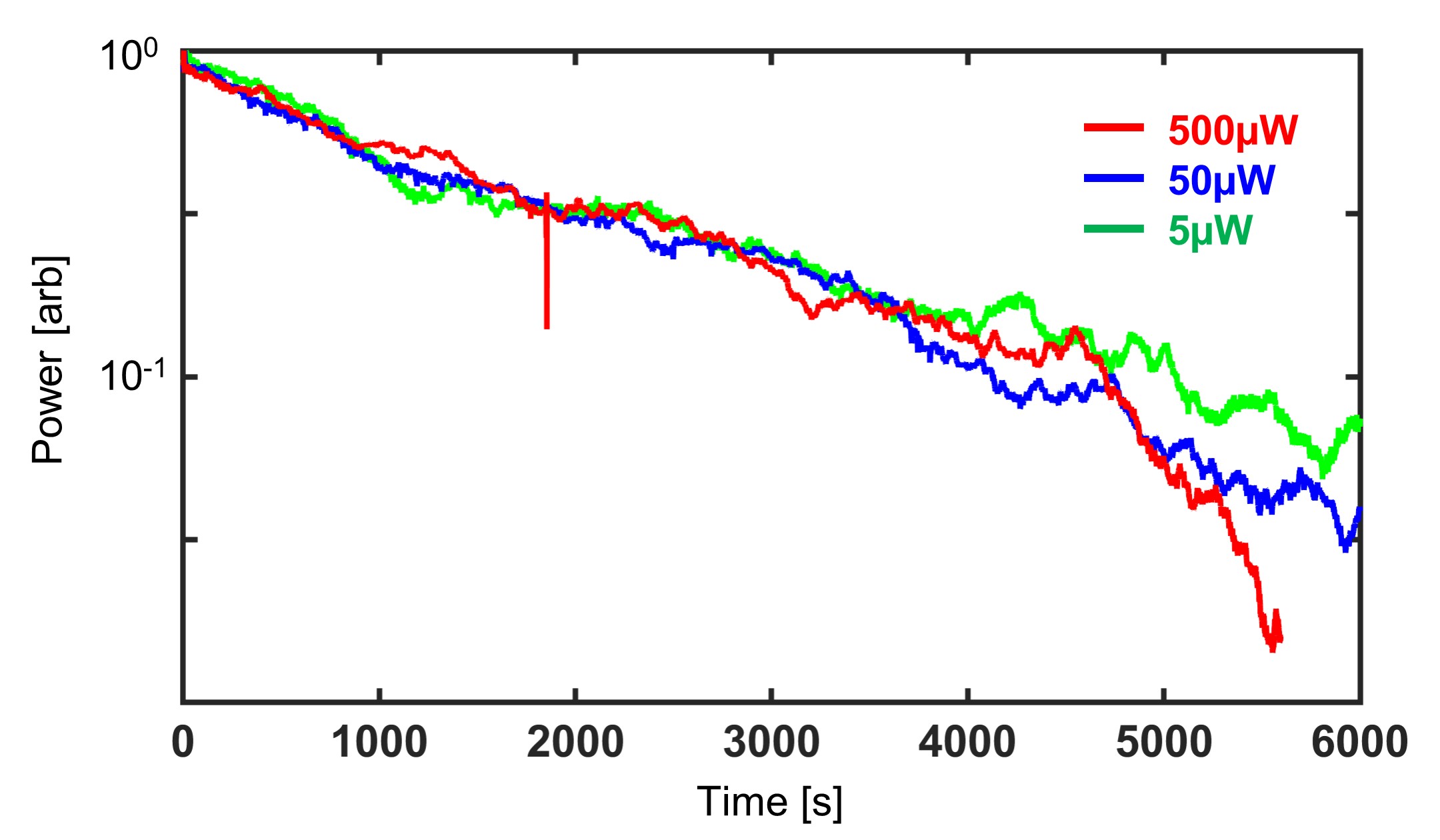}
  \caption{Ringdown measurement of the optimized 3 mm spiderweb nanomechanical resonator excited in its 133.6 kHz mode for three different laser powers. The y axis shows the power acquired from the spectrum analyzer in mW, normalized with respect to the power at time=0 for each curve.}
  \label{fig_S8}
\end{figure}

The interaction between the nanomechanical resonator and the optical field can lead to photothermal effects, which in turn can mask the intrinsic damping rate of the oscillators. Those effects are proportional to the laser power. Therefore when measuring extreme high quality factor, it is important to ensure that the measured decay is not affected by the laser power. To verify it, we performed different ringdown measurements of the mechanical resonance at around 133.6 kHz of the spiderweb nanomechanical resonator shown in the main text in \textbf{Figure 6}\textbf{A}, by varying the laser power incident on the resonator from 500 \textmu W to 5 \textmu W. The result in \textbf{Figure \ref{fig_S8}} shows a comparable decay rate for all the measurements, suggesting that photothermal effects are negligible. The plotted power on the y axis is normalized with respect to the power at the the beginning of the ringdown for each curve to provide a clear comparison. Note that the decaying envelope of the oscillation at carrier frequency is linear until the measured power equals the nanomechanical resonator's thermal fluctuations level leading to large fluctuations of the measured signal. Moreover strong fluctuations caused by unwanted temperature drifts and mechanical vibrations of the setup can bring the measured signal outside the linear region of the interference signal for a short interval of time, resulting in occasional spikes as visible in the curve acquired with a laser power equal to 500 \textmu W at around 1800 seconds.

\section*{Finite element simulation model of the spiderweb design}

\begin{figure}[h]
  \centering
    \includegraphics[width=0.8\textwidth]{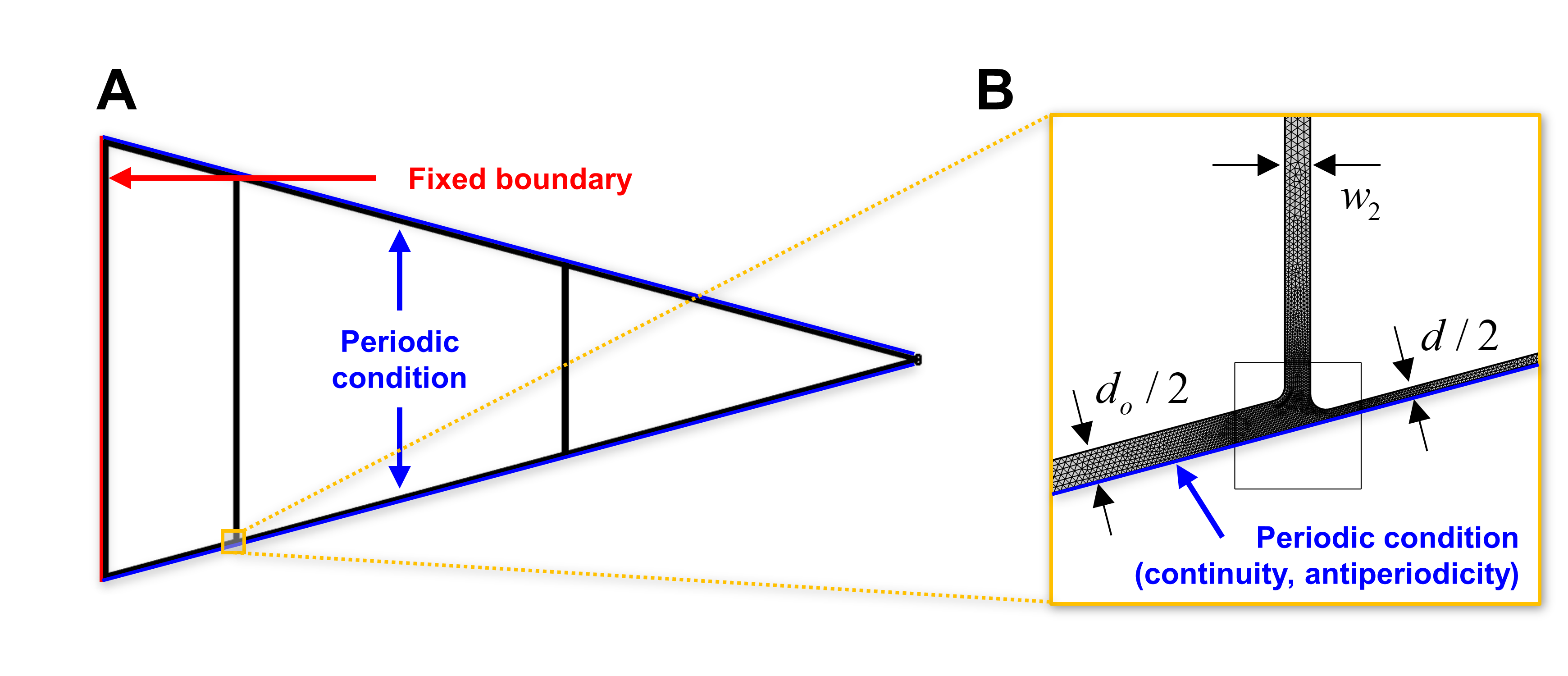}
  \caption{COMSOL finite element simulation layout and mesh, corresponding to the iteration 27 model in the main text. \textbf{A} Top view of the finite element simulated structure showing the periodic condition along the blue line with the fixed boundary condition along the red line. \textbf{B} Zoom-in of the yellow box region in \textbf{A}, showing the dense meshing of the spiderweb nanomechanical resonator with the denser mesh around the joint region.}
  \label{fig_S9}
\end{figure}

As mentioned in the paper, the optimized result in this research was obtained from the finite element analysis performed by COMSOL \cite{comsol}. We used plate elements for two-dimensional structural mechanic simulation dealing with a pre-stressed eigenfrequency analysis, considering the thickness to the width ratio. The plate theory was based on the Mindlin theory with a linear elastic material model. To speed up the simulation, we simulated the partial part of the spiderweb resonator. \textbf{Figure \ref{fig_S9}A} is an example of our iteration 27. The figure shows that it is modeling a portion of the resonator, which depends on $N_r$. The blue lines in the rotational directional edges considered continuity and antiperiodicity periodic conditions for the simulation, and the red line shows the fixed boundary condition. For each simulation, we considered initial in-plane forces ([N/m]) as 1.07 GPa $\cdot$ 50 nm, which is the product of the initial stress and the thickness. We performed the static analysis as the first step, including geometric nonlinearity because of the significant deformation happening from the static analysis \cite{norte2016mechanical}. Note that during the stationary analysis, it is always required to give the periodic condition for the periodic edges even if the antiperiodicity boundary condition is considered for the eigenvalue analysis. After performing the stationary analysis, we performed eigenfrequency simulation, also including geometric nonlinearity. We have set up the maximum number of eigenfrequencies to be a hundred while setting the smallest real part as 100 Hz and the largest real part as 1 MHz. Concerning meshing, we divided the model with small squares around every joint to adjust a more fine mesh to consider the bending loss energy more precisely, as shown in \textbf{Figure \ref{fig_S9}B }. To avoid numerical errors during the optimization, we meshed the upper half of the structure and copied the mesh domain to the lower half of the structure. After the finite element simulation, we calculated the $Q_\mathrm{m}$ based on eq. 1 in the main text. After calculating $Q_\mathrm{m}$ of all the out-of-plane modes, we selected the maximized $Q_\mathrm{m}$ as the performance of the design.

\clearpage

\end{widetext}

\end{document}